\documentclass[nofootinbib,floatfix,superscriptaddress]{elsarticle}
\bibpunct{[}{]}{;}{n}{}{}
\usepackage[utf8x]{inputenc} 
\usepackage[T1]{fontenc}             
\usepackage{lmodern}

\usepackage[greek, english]{babel}
\usepackage{bbm}
\usepackage{bm}
\usepackage{amsmath, amsfonts, amssymb, mathtools}    
\usepackage{graphicx}   
\usepackage{color}      
\usepackage{subfigure}  
\usepackage{epstopdf}
\usepackage{hyperref}   
\usepackage{tabularx}
\usepackage{dcolumn}

\usepackage{physics}
\usepackage{feyn}
\usepackage{slashed} 
\usepackage{tensor}

\newcommand{\moye}[1]{\langle #1 \rangle}
\def\man{\mathcal{M}}
\def\lag{\mathcal{L}}

\def\F{\mathcal{F}}

\def\O{\mathcal{O}}
\def\D{\mathcal{D}}
\def\sl{\slashed}

\def\Psib{\bar{\Psi}}
\def\dag{\dagger}
\def\pa{\partial}

\def\half{\frac{1}{2}}
\def\thalf{\tfrac{1}{2}}

\DeclareMathOperator\sgn{sgn}

\DeclareMathOperator\SU{SU}

\DeclareMathOperator\U{U}

\def\Z{\ensuremath{\mathbb{Z}}}

\def\R{\ensuremath{\mathbb{R}}}

\def\Id{\ensuremath{\mathbbm{1}}}

\usepackage{cancel}
\newcommand{\uds}[1]{\underset{#1}}

\def\lb {\left(}
\def\rb{\right) }
\def\lc{\left[}
\def\rc{\right]}

\def\l.{\left.}
\def\r.{\right.}

\def\beq{\begin{equation}}
\def\eeq{\end{equation}}
\def\bsp{\begin{split}}
\def\esp{\end{split}}
\def\bea{\begin{eqnarray}}
\def\eea{\end{eqnarray}}
\def\beano{\begin{eqnarray*}}
\def\eeano{\end{eqnarray*}}

\newcommand{\eqn}[1]{\begin{align}#1\end{align}}

\newcommand{\pmatr}[1]{\begin{pmatrix}#1\end{pmatrix}}

\def\nn{\nonumber}

\newcommand{\ed}[1]{{\color{red} #1}}

\definecolor{gris}{rgb}{0.4,0.4,0.4}

\DeclarePairedDelimiter\floor{\lfloor}{\rfloor}

\def\nb{N_b}
\def\nf{N}

\def\u{\uparrow}
\def\d{\downarrow}
\def\uf{L}
\def\df{R}
\def\S{s}
\def\ms{m_s}

\def\veps{\varepsilon}
\def\vsig{\mathfrak{s}}
\def\vmu{\mathfrak{v}}
\def\vth{\vartheta}
\def\f0{f_0}
\def\c{c}
\def\cb{c^\dagger}

\def\Fq{F_{q; s}}
\def\fq{f_{q; s}}

\DeclareMathOperator\mQED{QED}

\DeclareMathOperator\QEDt{QED_3}
\DeclareMathOperator\QEDtGN{QED_3-GN}
\DeclareMathOperator\QEDtcHGN{QED_3-cHGN}

\DeclareMathOperator\cHGN{cHGN}
\def\CPo{\text{CP}^1}

\usepackage{wasysym} 

\usepackage{array}
\newcolumntype{L}{>{$}l<{$}}
\newcolumntype{R}{>{$}r<{$}}
\newcolumntype{C}{>{$}c<{$}}

\usepackage{ytableau}
\usepackage{youngtab}
\def\wp{\phantom{.}}

\newcommand\yt[1]{\begin{ytableau}#1\end{ytableau}}
\newcommand{\svdots}{%
  \vbox{
     \baselineskip 2pt \lineskiplimit 0pt
    \hbox {.}\hbox {.}\hbox {.}\kern-0.75pt
  }%
}
\newcommand{\sdots}{%
  \vbox{
     \baselineskip 2pt \lineskiplimit 0pt
     \dots \kern-0.75pt
  }%
}
\ytableausetup{smalltableaux}
\Yboxdim{8pt}
\def\yngc{\Yvcentermath1}

\def\irrepm{
\begin{array}{l}
\overset{\nf-b}{\overbrace{
\begin{array}{c}
\yt{\wp& \none[...] && \none[...] & }
\end{array}
}}\\[-0.3em]
\uds{b}{\underbrace{
\begin{array}{c}
\yt{\wp& \none[...] & }
\end{array}
}}
\end{array}
}

\def\irrepsmin{
\uds{2 \S}{\underbrace{
\begin{array}{c}
\yt{\wp& \none[...] & }
\end{array}
}}
}

\def\irrepmpr{
{\scriptstyle \nf  -b}
\left \{
\!\!\!\!\!
\begin{array}{l}
\left .
\begin{array}{l}
\yt{\wp&\\ \none[\svdots] & \none[\svdots] \\ \wp & \\  }
\end{array}
\!
\right \} {\scriptstyle b}
\\[-0.05em]
\left .
\begin{array}{l}
\yt{\none[\svdots]\\\wp}
\end{array}
\right .
\end{array}
\right .
}

\def\irrepspr{
{\scriptstyle \frac{\nf}{2} + \S}
\left \{
\!\!\!\!\!
\begin{array}{l}
\left .
\begin{array}{l}
\yt{\wp&\\ \none[\svdots] & \none[\svdots] \\ \wp & \\  }
\end{array}
\!
\right \} {\scriptstyle \frac{\nf}{2}-\S}
\\[-0.05em]
\left .
\begin{array}{l}
\yt{\none[\svdots]\\\wp}
\end{array}
\right .
\end{array}
\right .
}

\newcommand\irrepColD[1]{
\left.
\begin{array}[]{l}
\yt{ \wp \\ \none[\svdots] \\ \wp  }
\end{array}
\right \} {\scriptstyle  #1 }
}

\newcommand\irrepColT[1]{
{\scriptstyle  #1 }
\left \{
\begin{array}[]{l}
\yt{ \wp \\ \none[\svdots] \\ \wp  \\  \none[\svdots] \\ \wp }
\end{array}
\right .
}

\newcommand\irrepColDPr[1]{
\left.
\begin{array}[]{l}
\yt{ \wp \\ \none[\svdots]   }
\end{array}
\right \} {\scriptstyle  #1 }
}

\newcommand\irrepColTPr[1]{
{\scriptstyle  #1 }
\left \{
\begin{array}[]{l}
\yt{ \wp \\ \none[\svdots] \\ \wp  \\  \none[\svdots] \\ \wp \\ \wp}
\end{array}
\right .
}

\def\irrepLineN{   
    \uds{N}{\underbrace{
\begin{array}{c}
\yt{\wp& \none[...] & }
\end{array}
}}
}

\def\irrepColN{
    {\scriptstyle N}
\left \{
\begin{array}{l}
\yt{\wp \\ \none[\svdots] \\ \wp  \\  }
\end{array}
\right . 
}

\def\signn{
{\scriptstyle \nf }
\left \{
\begin{array}{c}
\\[-1em]
\yngc
\yt{\wp \\ \none[\svdots]\\ \wp}
\end{array}
\right . 
}

\def\b{b}

\def\evalat{\Bigr|}
\def\lyt{\Upsilon}
\usepackage{comment}

\usepackage{xfrac}
\usepackage{nicefrac}
  
\begin{document}

\begin{frontmatter}
\title{Monopole hierarchy in transitions out of a Dirac spin liquid}
\author[1]{\'Eric Dupuis}

\author[2]{William Witczak-Krempa $^\text{a,}$}
\address[1]{D\'epartement de physique, Universit\'e de Montr\'eal, Montr\'eal (Qu\'ebec), H3C 3J7, Canada}
\address[2]{Centre de Recherches Math\'ematiques, Universit\'e de Montr\'eal; P.O. Box 6128, Centre-ville Station; Montr\'eal (Qu\'ebec), H3C 3J7, Canada}

\date{\today}
\begin{abstract}
Quantum spin liquids host novel emergent excitations, such as monopoles of an emergent gauge field. Here, we study the hierarchy of monopole operators that emerges at quantum critical points (QCPs) between a two-dimensional Dirac spin liquid and various ordered phases. This is described by a confinement transition of quantum electrodynamics in two spatial dimensions ($\QEDt$ Gross-Neveu theories).  Focusing on a spin ordering transition, we get the scaling dimension of monopoles at leading order in a large-$N$ expansion, where $2N$ is the number of Dirac fermions, as a function of the monopole’s total magnetic spin. Monopoles with a maximal spin have the smallest scaling dimension while monopoles with a vanishing magnetic spin have the largest one, the same as in pure $\QEDt$. The organization of monopoles in multiplets of the QCP’s symmetry group $\SU(2) × \SU(N)$ is shown for general N.

\end{abstract}


\end{frontmatter}

\tableofcontents

\section{Introduction}
Anderson first proposed the idea of a quantum spin liquid, an insulator state emerging in frustrated quantum magnets \cite{anderson_resonating_1973}. To formulate this, he used the idea of a resonating valence bond theory which describes a highly entangled state. It was later realized that these kinds of systems can indeed host exotic phases of matter with fractional excitations and emergent gauge fields that evade Landau paradigms. These states motivated the study of gauge theories in a condensed matter context. An important example is the Dirac spin liquid (DSL) which is described by  quantum electrodynamics in $2+1$ dimensions $(\QEDt)$ with $2 \nf$ flavors of gapless  Dirac fermions, with typically ${\nf=2}$ Dirac cones  in quantum magnets. 

The formulation of the  $\QEDt$ model is rather simple, i.e. an abelian gauge field coupled to fermionic matter. Nevertheless, it is a strongly coupled theory with a non-trivial IR limit. The model flows to an interacting fixed point for $\nf$ large enough  while its exhibits a chiral symmetry breaking below some finite number of fermion flavors \cite{pikarski_chiral_1984, vafa_1984_eigenvalue, appelquist_critical_1988, appelquist_phase_2004}. Many recent investigations still explore this dynamical mass generation as well as other aspects of $\QEDt$ \cite{braun_phase_2014,    giombi_conformal_2016, karthik_evidence_2016, dipietro_quantum_2016,  chester_anomalous_2016, giombi_CJ_2016, kotikov_critical_2016, kotikov_critical2_2016}. When the UV divergences of $\QEDt$ are regularized by a lattice, as it is naturally the case in condensed matter systems, one also has to account for the compact nature of the $\U(1)$ gauge field.~\footnote{Throughout the text, the theories we describe are assumed to be compact unless stated otherwise.}  This aspect implies the existence  of topological disorder operators called monopole or instanton operators \cite{borokhov_topological_2003}. 

While monopoles confine the gauge field in a pure gauge theory \cite{polyakov_quark_1977}, a sufficient number of massless matter flavors screening the monopoles prevents their proliferation. The scaling dimensions of monopoles operators determine whether or not monopoles are relevant and destabilize the $\QEDt$ phase. Computations at leading order  \cite{borokhov_topological_2003} and subleading  order  \cite{pufu_anomalous_2014} in $1/\nf$ using the state-operator correspondence indicate a stable theory for $2 \nf  \geq 12 $. This result was recently confirmed with Monte Carlo in the non-compact $\QEDt$ where monopoles are probed using the background field method \cite{karthik_monopole_2018, karthik_numerical_2019}. Conformal bootstrap bounds relating simply and doubly charged monopoles also yield coherent results \cite{chester_towards_2016,  chester_monopole_2018}. To characterize the stability of a DSL, it should also be known which monopole charges are allowed by lattice symmetries. The  first results in this regard were obtained on the square and Kagome lattices  \cite{alicea_monopole_2008, hermele_properties_2008} and were followed by  a comprehensive  analysis of the monopole transformation properties under many lattice symmetries  \cite{song_unifying_2019, song_spinon_2018}. 

Monopole operators have been studied in many other contexts with and without supersymmetry, including non-abelian gauge theories and Chern-Simons Matter theories  \cite{borokhov_monopole_2002, borokhov_monopole_2004, dyer_monopole_2013,  radicevic_disorder_2016, chester_monopole_2018, assel_note_2019}. In fact, monopole operators were first studied in a bosonic theory, the $\CPo$ model. This is the prototype model for deconfined criticality \cite{senthil_quantum_2004, senthil_deconfined_2005, Metlitski_intrinsic_2018, lee_signatures_2019}, i.e.\  QCPs with emergent enlarged symmetry and fractionalized excitations separating classical Landau phases. The monopoles in this model describe a VBS order.  Their scaling dimension have been obtained at leading order \cite{murthy_action_1990, metlitski_monopoles_2008}  and subleading order \cite{dyer_scaling_2015, dyer_erratum_2016} in $1 / \nb$, at order $\O(q^0 \nb^0)$ in the $1/q$ and $1/\nb$  expansion \cite{delaFuente_large_2018}, and with numerical methods on various lattices \cite{ribhu_fate_2013, sreejith_scaling_2015, pujari_transitions_2015}. Another interesting aspect of this model is its conjectured duality with the $\QEDt$-Gross-Neveu model ($\QEDtGN$) with $\nf=2$ fermion flavors \cite{wang_deconfined_2018}. The non-compact realization of this latter model has been studied in Refs.~\cite{janssen_critical_2017, bernhard_deconfined_2018, gracey_fermion_2018, zerf_critical_2018, boyack_deconfined_2019,  benvenuti_easy_2019}. In the compact version of $\QEDtGN$, the scaling dimension of monopole operators  at leading order in $1/\nf$ were found to be the same as in $\QEDt$~\cite{dupuis_transition_2019, proc}. 

The $\QEDtGN$ model also underlies an important aspect of the DSL, which is that this phase has been described as the  parent state for many spin liquids \cite{hermele_algebraic_2005, song_unifying_2019, song_spinon_2018}. Indeed, the $\QEDtGN$ model describes the QPT from the DSL to a chiral spin liquid which is induced by tuning a flavor symmetry preserving a Gross-Neveu interaction in $\QEDt$ \cite{bhattacharjee_kagome_2015}.  By tuning other Gross-Neveu like interactions in $\QEDt$, it is also possible to describe transitions to confined phases. As a flavor-dependent fermion self-interaction is tuned, fermions become gapped and their screening effect is lost, letting monopoles proliferate \cite{ghaemi_neel_2006, lu_unification_2017}.\footnote{This scheme does not apply to $\QEDtGN$ where the symmetric fermion mass generates a Chern-Simons term which prevents monopole proliferation.} This leads to an interesting scenario in the Kagome Heisenberg antiferromagnet whose ground state is putatively described by the DSL with $\nf = 2$ valleys \cite{hastings_dirac_2000}. The confinement of this QSL to a coplanar antiferromagnet is described by $\QEDt$ with  a chiral Heisenberg Gross-Neveu interaction ($\QEDtcHGN$). In this model, a spin-Hall mass is condensed which in turn drives the condensation of monopoles with spin quantum numbers yielding the antiferromagnetic order \cite{hermele_properties_2008, lu_unification_2017}. 

Monopole scaling dimensions in $\QEDtcHGN$ were obtained at leading order in $1/ \nf$ in Ref.~\cite{dupuis_transition_2019}. The minimal monopole scaling dimension obtained is lower than in $\QEDt$. A hierarchy among monopole operators with different quantum numbers was also found, but a proper complete treatment is still lacking. The objective of this paper is to put the hierarchy on a formal footing both qualitatively and quantitatively.  An improved characterization of monopole scaling dimensions  in $\QEDtcHGN$ will yield further analytical results which offer more testing ground for   experimental and numerical explorations of this system. Similarly, a hierarchy of monopole operators was described in Chern-Simons Matter theories for monopoles with varying Lorentz spins.\footnote{Monopole operators are not necessarily Lorentz scalars in Ref.~\cite{chester_monopole_2018} as opposed to the stricter definition provided in Ref.~\cite{borokhov_topological_2003}. } At leading  order  in $1/ \nf$, these operators share the same scaling dimension, but the degeneracy is lifted  by higher order corrections.   This effect was also seen in the conformal bootstrap \cite{chester_monopole_2018}. The hierarchy considered in this work is instead among monopoles with different flavor quantum numbers. The degeneracy lifting is natural as the $\QEDt$ flavor symmetry is partially broken at the QCP, i.e. in the $\QEDtcHGN$ model.
 
The paper is organized as follows. In the following section, we define the $\QEDtcHGN$ theory, and review the role of monopole operators and how different monopole types are distinguished by the fermionic self-interaction in this model.   In Sec.~\ref{sec:scaling}, we obtain the scaling dimensions of monopoles as a function of their total magnetic spin to put the monopole hierarchy on a formal footing.  In Sec.~\ref{sec:large}, the scaling dimensions of monopoles are computed with an analytical approximation valid for large values of the magnetic charge. We interpret these results in Sec.~\ref{sec:irreps} as a degeneracy lifting of monopoles in $\QEDt$ as we organize monopoles in multiplets of the reduced  flavor symmetry group describing $\QEDtcHGN$. In Sec.~\ref{sec:conclusion}, we summarize our results. In  \ref{app:holonomy} and \ref{app:gen-spin_Hall}, more general forms of the gap equations appearing in Sec.~\ref{sec:scaling} are studied to justify the restricted analysis presented in the main text.  In  \ref{app:mu0}, it is shown why a particular region of parameters spaces does not yield solutions of the gap equations. In  \ref{app:reg}, the diverging sums appearing in Sec.~\ref{sec:scaling} are regularized. In  \ref{app:explicit}, the representation of monopoles with minimal magnetic charge for the $N=2$ QCP is explicitly constructed. In  \ref{app:red}, we show the detailed computations yielding the symmetry reduction shown in Sec.~\ref{sec:irreps}. We also discuss how the analysis in this section may be extended to $q>1/2$.

\section{Model \label{sec:model}}

Let us consider $2 \nf$ flavors of massless two-component Dirac fermions, $\psi_A$ where $A = 1,2, \dots, 2 \nf$. In a condensed matter language, these degrees of freedom correspond to the magnetic spin ${\u, \d}$ and $\nf$ nodes in momentum space, typically $\nf=2$ in quantum magnets. These fermions  correspond to the spinons, the  spin-$1/2$ quasiparticles emerging from the fractionalization of a spin$-1$ excitations in a quantum magnet. As crucially noted by Baskaran and Anderson \cite{baskaran_gauge_1988}, such a parton decomposition has a local gauge symmetry, which in turn implies the existence of a gauge field in the low energy description. In particular, we  consider the $\QEDtcHGN$ model, whose action in Euclidean signature is given by  
\eqn{
S =  \int d^3 x \Bigl[ - \Psib \sl{D}_a \Psi  -  \frac{h^2}{2} \lb \Psib \bm{\sigma} \Psi \rb^2     \Bigr]   + \dots \,.
 \label{eq:QFT_transition}
}
The fermions are organized in a flavor spinor, ${\Psi=\pmatr{\psi_1,&\psi_2,&\dots&\psi_{2\nf}}^\intercal}$. 
The gauge  covariant derivative  $\sl{D}_a $ acting on fermions is given by
\eqn{
\sl{D}_a = \gamma^\mu \lb \pa_\mu - i  a_\mu \rb  \,,
} 
where $a_\mu$ is a $\U(1)$ gauge field. The Dirac matrices $\gamma^\mu$  act on Lorentz spinor components and may be written in terms of Pauli matrices $\tau_i$ as ${\gamma^\mu=(\tau_3, \tau_2, - \tau_1)}$.  The $\cHGN$ interaction term has a coupling strength $h$ and is defined with a Pauli matrix vector $\bm \sigma$ acting on the $\SU(2)$ magnetic spin subspace.    The ellipsis denotes the Maxwell free action and the contribution from monopole operators $\man_q^\dag$.

These  topological disorder operators owe their existence to  the  compact nature of the $\U(1)$ gauge field which  implies a $2\pi$ quantization of the magnetic flux. Monopole operators insert integer multiples of the quantum  flux  $4\pi q$ where $2 q \in \Z$. Formally, the charge $q$  may defined  by the action of  the magnetic current operator $j_{\rm top}^\mu(x) = \frac{1}{2 \pi} \epsilon^{\mu \nu \rho} \pa_\nu a_\rho(x)$  on the monopole operator $\man_q^\dag$ that can be developed with the operator product expansion \cite{borokhov_topological_2003}
\eqn{
j_{\rm top}^\mu(x) \man^\dagger_q(0) \sim \frac{q}{4 \pi} \frac{x^\mu}{|x|^3}  \man^\dagger_q(0) + \cdots \,,
\label{eq:flux-op}
}
where the ellipsis denotes less singular terms as ${|x| \to 0}$. The prefactor  corresponds to the magnetic field  of a Dirac monopole with charge $q$.

We can first begin the description of the quantum phase transition (QPT) by analyzing the non-compact $\QEDtcHGN$ model.  For a sufficiently strong coupling strength $h > h_c$, a spin-Hall mass is condensed $\hat n \cdot \moye{\Psib \bm \sigma \Psi} >0$. By using a zeta regularization to find the critical coupling, the effective action at the quantum critical point (QCP) is given by \cite{dupuis_transition_2019}
\eqn{
S_{\rm eff}^c = - \nf \ln \det \lb \sl{D}_a + \bm{\phi} \cdot \bm  \sigma \rb  \,,
\label{eq:S_eff_c}
}
where $\bm \phi$ is an auxiliary vector boson decoupling the $\cHGN$ interaction. 

However, this picture is incomplete. Even if the monopole operators at the QCP are irrelevant and a  $\U(1)_{\rm top}$  global  symmetry emerges in the infrared, the monopole operators will be dangerously irrelevant. While gapless matter in $\QEDt$ may screen the monopoles \cite{borokhov_topological_2003} and prevent the confinement observed in a pure $\U(1)$ gauge theory \cite{polyakov_compact_1975}, the situation changes as fermions are gapped.  Their screening effect is lost following the condensation of the spin-Hall mass which in turn drives the proliferation of spin-polarized monopoles \cite{ghaemi_neel_2006, lu_unification_2017}. This confines the fermions, or in the context of quantum magnets, recombines the spinons that are fractionalized excitations of the underlying spin system. Monopole operators are thus an essential ingredient to properly understand this confinement-deconfinement transition.

A monopole at the QCP is characterized by its scaling dimension  $\Delta_{\man_q}$ that controls the scaling behaviour of the two-point correlation function
\eqn{
\moye{\man_{q}(x)\man_{q}^{\dagger}(0)}\sim \frac{1}{|x|^{2\Delta_{\man_q}}}\,.
}
 Since the model at the QCP is a conformal field theory, the state-operator correspondence can be used to obtain this critical exponent \cite{rychkov_epfl_2017}. More precisely, the correspondence implies that within the set of $4 \pi q$ flux operators, the minimal scaling dimension $\Delta_q$  is equal  to the ground state of an alternate theory, namely the $\QEDtGN$ model defined on $S^2 \times \R$ with  a magnetic flux $\varPhi = 4 \pi q$ piercing the two-sphere. The external gauge field sourcing this flux ${\varPhi = \int \dd A^q }$ may be written as 
\eqn{
A^q(x) = q (1 - \cos \theta ) \dd \phi  \,, \label{eq:Aq}
}
or  ${A^q_\phi = q (1-\cos \theta) / \sin \theta}$ in components notation. The singularity at $\theta =\pi$ can be compensated by a Dirac string which imposes the Dirac condition on the magnetic charge $2 q \in \Z$ as the string must remain invisible.   By computing the free energy of this alternate theory, the scaling dimension $\Delta_q$ was computed at leading order in $1/\nf$ \cite{dupuis_transition_2019}.  This scaling dimension obtained is smaller than in $\QEDt$ and, specifically for the minimal magnetic charge, is given by $\Delta_{1/2} = 2 \nf \times 0.195 + \O(1/\nf^0)$.

As mentioned earlier, spin-polarized monopoles are favored by the spin-Hall mass condensation and yield the order parameter. It is useful to compare how  different types of monopole operators behave at the QCP.  We first precise what is meant by ``types'' of monopoles.  While all monopole operators with a magnetic charge $q$ share the same magnetic properties, they are distinguished by the different possible fermion modes dressings that define supplementary quantum numbers. The fermion occupation also determines Lorentz and gauge properties. For example, a flux operator with a vanishing fermion number is constructed by filling half of the fermion modes \cite{borokhov_topological_2003, dupuis_transition_2019}. Among these fermion modes,  there are $4 |q| \nf$ special fermion zero modes which owe their existence to the topological charge $q$ of the flux operators \cite{atiyah_index_1963}. By filling all negative energy modes and  half of the zero modes, a flux operator with vanishing fermion number \textit{and} a minimal scaling dimension is obtained. The zero modes occupation can be further constrained to select only operators with a vanishing Lorentz spin.  This means that varying which zero modes are dressed  defines a set of distinct $4 \pi q$ flux operators which are Lorentz scalars, and have equal and minimal scaling dimensions. Those are the monopole operators of $\QEDt$ \cite{borokhov_topological_2003}.

The situation is slightly different  in $\QEDtcHGN$. In the mean field theory, monopoles at the QCP are described by a  non-vanishing  spin-Hall mass $\moye{\bm \phi}_{A^q, S^2 \times \R} =  M_q \hat n$ coming from the $\cHGN$ interaction.  This parameter most notably confers a non-vanishing energy to the zero modes. This affects  the scaling dimension of a monopole which is lowered by anti-aligning the spin-Hall mass and the monopole magnetic spin polarization. Monopoles with different ``zero'' modes~\footnote{Since these modes do not have a vanishing energy but nevertheless keep their topological origin and their chiral aspect, we refer to them as ``zero'' modes.} occupation may then have different scaling dimensions, i.e. there is a monopole hierarchy in $\QEDtcHGN$. While the hierarchy was partly explored though this prism in Ref.~\cite{dupuis_transition_2019}, here we provide  a more complete and accurate discussion on this point.  Notably, here the spin-Hall mass is not fixed by the monopole with the lowest scaling dimension, it can instead vary in amplitude and orientation for each type of monopole. In this manner, we characterize the hierarchy among  monopoles operators by obtaining  the scaling dimension as a function of the monopole magnetic spin $\S$  in Sec.~\ref{sec:scaling}. We then show in Sec.~\ref{sec:irreps} how monopoles are organized as irreducible representations of the QCP symmetry group $\SU(2) \times \SU(\nf)$ and how this makes contact with the scaling dimension results.

\section{Scaling dimension with fixed spin  \label{sec:scaling}} 
\subsection{Constraining the monopole magnetic spin}

We now determine the scaling dimension of monopole operators  $\man_{q;\S}$ with magnetic charge $q$ and with magnetic spin $\S$. The minimal scaling dimension in this sector, $\Delta_{q;\S}$, is obtained through the state-operator correspondence  
\eqn{
\Delta_{q;\S} =  \lim_{\beta \to \infty} \Fq \equiv   - \lim_{\beta \to \infty} \beta^{-1} \ln Z_{\S}[A^q]  \,,
\label{eq:del_S}
}
where $\Fq$ is the rescaled free energy. The partition function  and free energy define a ground state for an alternate version of the $\QEDtcHGN$ action at the QCP \eqref{eq:S_eff_c}.  As discussed in the previous section, this alternate model is defined on a compactified spacetime $S^2 \times S^1_\beta$, where the ``time''~\footnote{Here and throughout the text, we put quotes to emphasize that this is not the original time direction on $\R^3$ but rather the real direction obtained with conformal transformation $\R^3 \to S^2 \times \R$.} direction is also taken compact as it is regularized on a ``thermal'' circle with radius $\beta$. Additionally, an external magnetic field $A^q$ \eqref{eq:Aq}  is added to encode the magnetic flux of the monopole operator. Finally, the magnetic spin $s$ of the monopole operator is selected by the inclusion of a Lagrange multiplier field $\mu_S$ in the action 
\eqn{
S^c_{S^2 \times S^1_{\beta}}[A^q] + \beta  \mu_S   \Bigl( \S (\S+1)-    \bm{\mathcal{S}}^2  \Bigr) \,, \label{eq:temp_action}
}
where we introduced the total spin operator  (averaged over time)
\eqn{\bm{\mathcal{S}} = \beta^{-1} \int_{S^2 \times S^1_\beta}\Psib \gamma_0 \tfrac{\bm \sigma}{2} \Psi \,.
}
More explicitly, this operator may be written as ${\bm{\mathcal{S}} = \beta^{-1} \int_{S^2 \times S^1_\beta}\Psi^\dagger \tfrac{\bm \sigma}{2} \Psi}$. The lagrange multiplier equation yields the constraint  $\moye{\bm{\mathcal{S}}^2} = \S(\S+1)$ which sets the magnetic spin of the monopole operator.

Since a monopole operator is dressed with half of the  $4 |q| \nf$ fermion ``zero'' modes, its maximal spin   is ${\S_{\max} =  |q| \nf}$. This corresponds to configurations where only ``zero'' modes with the same spin-$1/2$ polarization are filled. On the other hand, a monopole with a minimal spin $\S_{\min} = 0$ is obtained by dressing  an equal proportion of ``zero'' modes with opposite spins.\footnote{In fact, this is not possible for odd $\nf$  as one ``zero'' mode always remains unmatched, i.e.\  $\S_{\min} = 1/2$. However, this effect is subleading in $1/\nf$ and we only focus on the leading order.} The magnetic spin of a monopole obtained by  filling the Dirac sea and half of the zero modes is thus bounded as 
\eqn{
 0 \leq \S \leq |q| \nf\,. \label{eq:S}
}

\subsection{Free energy at leading order in $1/\nf$}
The interaction  added in Eq.~\eqref{eq:temp_action} to the $\QEDtcHGN$ action to constrain the total monopole spin is quartic in fermions  and can be decoupled with an auxiliary boson field  $\bm \chi$.  As the spin-squared interaction is not diagonal in spacetime $\bm{\mathcal{S}}^2 = \beta^{-2} ( \int_x \Psib \gamma_0 \tfrac{\bm \sigma}{2} \Psi \cdot \int_y \Psib \gamma_0 \tfrac{\bm \sigma}{2} \Psi )  $, we should in principle introduce the boson in the same way.  However, as we only seek to describe the free energy at leading order, we may replace the spin interaction with a diagonal formulation  ${\bm{\mathcal{S}}^2 \to V \beta^{-1} \int_x (\Psib \gamma_0 \tfrac{\bm \sigma}{2} \Psi)^2}$, where ${V = \int_{S^2} \sqrt{g} d^2 x}$ is the area of the two-sphere $S^2$. This will not affect the results as the expectation value is taken to be homogeneous. The auxiliary boson may then  be introduced as the following resolution of the identity
\eqn{
\int \D \bm \chi \exp{- \int \frac{\bm{\chi}^2}{2 V}  } =  \int \D \bm \chi \exp{-  \int \frac{1}{2 V} \lb   \bm \chi - V \sqrt{2 \mu_S\,} \lb \Psib \gamma_0 \tfrac{\bm \sigma}{2} \Psi \rb \rb^2}\,.
\label{eq:identity}
}
For later convenience, we note that the equation of motion for $\bm \chi$ relates the  expectation value of the boson to the spin polarization  as 
\eqn{
 \frac{\moye{\bm \chi}}{ \sqrt{2 \mu_S}}  = V \moye{\Psib \gamma_0 \tfrac{\bm \sigma}{2} \Psi} 
   \,. \label{eq:polarization}
} 
With this auxiliary boson $\bm \chi$ (and the boson $\bm\phi$ decoupling the $\cHGN$ interaction in the original model $\QEDtcHGN$), fermions in the action \eqref{eq:temp_action}  can be integrated out, yielding the following effective action 
\eqn{
\begin{split}
S_{\rm eff}^{\prime c}  =&  -  \nf \ln \det \biggl(  \sl{D}_{a, A^q} -    \sqrt{\tfrac{\mu_S}{2}} \bm \chi  \cdot  \gamma_0 \bm \sigma  + \bm \phi \cdot \bm \sigma    \biggr) +  \beta \mu_S \S (\S+1)  +   \int\frac{\bm \chi^2}{2 V}   \,.
\end{split}
\label{eq:eff_2}
} 
We can now compute the free energy \eqref{eq:del_S} using a large-$\nf$ expansion corresponding to a saddle point expansion of the effective action. The leading order  will be obtained by computing the saddle point value of the effective action
\eqn{
\Fq^{\rm L.O.} = \nf  \Fq^{(0)} = \frac{1}{\beta} S_{\rm eff}^{\prime c}\big|_{\text{Saddle pt.}} \,.
}

  We take the saddle point configurations to be homogeneous. The two auxiliary bosons are $\SU(2)_{\rm Spin}$ vectors~\footnote{By  $\SU(2)_{\rm Spin}$ vectors, we only refer to the spin flavor group without specifying transformation properties under time reversal. } and one of them may be oriented along $\hat z$ without loss of generality. We orient the first auxiliary boson $\bm \phi$ along a general unit vector $\hat n$,
\eqn{
\moye{\bm \phi} = M_q \hat n\,, \label{eq:Mq}
}
 where $M_q>0$ is a spin-Hall mass.\footnote{The spin-Hall mass could be written as $M_{q;s}$ since its value will depend on the magnetic spin $\S$. We omit this index for simplicity.} The second boson can then be written simply as
\eqn{
\moye{\bm{\chi}} = P_z \hat z  = \sqrt{2\mu_S} \ms \hat z\,,
\label{eq:chi}
}
where the spin polarization $\ms$ \eqref{eq:polarization} can be positive, negative or zero. As for the dynamical gauge field, gauge invariance requires its expectation value to vanish ${\moye{a_\mu} = 0}$.\footnote{In fact, the gauge field may have a non-trivial holonomy  on the ``thermal'' circle $\int_0^\beta d \tau \moye{a_0} \neq 0$. However, in the zero ``temperature'' limit, it is sufficient to take $\moye{a_0} = 0$. See  \ref{app:holonomy}} 

These mean field ansatz are inserted in the effective action. The free energy can be further simplified by recognizing that the spin variables scale as $\nf$ since their magnitude can be formulated as a fraction of the total number of fermion ``zero'' modes \eqref{eq:S}. With this in mind, a subleading term in the total spin charge ${\S (\S+ 1) = \S^2 + \O(\nf^1)}$ is dropped and the total spin and  spin polarization  ${\S, \, \ms \to \nf \S,\, \nf \ms}$ are rescaled. The rescaled total spin is thus bounded from above by ${s_{\max} = q}$, i.e. $s_{\max} = 1/2$ for the minimal magnetic charge. The free energy then becomes 
\eqn{
\Fq^{(0)} = \fq + \frac{\mu'}{\ms} \bigl(\S^2   + \ms^2 \bigr) \,, \label{eq:free}
}
where $\fq$ is the determinant operator 
\eqn{
\fq =  -  \beta^{-1} \ln \det \bigl( \sl{D}_{-i\mu' \sigma_z, A^q} + M_q  \hat n \cdot \bm \sigma \bigr) \,, \label{eq:fq}
} 
and where we defined 
\eqn{
\mu' =\mu_S  \ms\,. \label{eq:mupr}
}

The saddle point parameters must solve the gap equations and minimize the free energy. Before determining them, we need to reexpress the free energy by  developing the determinant operator. In $\QEDtcHGN$ \cite{dupuis_transition_2019}, the basis of spinor monopole harmonics can be used to diagonalize the determinant operator. The procedure is a simple generalization of the pure $\QEDt$ case \cite{borokhov_monopole_2002, pufu_anomalous_2014}. Here, the formulation is a bit more involved  since the spin-Hall mass and the spin polarization are not necessarily along the same axis.  In what follows,  we will suppose this is the case and write $\bm M_q = M_q \hat z$, where we recall that $M_q > 0$. This assumption is motivated by Ref.~\cite{dupuis_transition_2019} where it is found that the monopole with the minimal scaling dimension can be interpreted as a monopole with anti-aligned  spin-Hall mass condensate and spin polarization. Using this assumption, solutions for both $\vartheta =0$ and $\vartheta = \pi$ are found in what follows. These two solutions yield opposite spin polarizations $m_s$, which supports the idea that auxiliary bosons should anti-align. Another analytical solution exists for $\vartheta = \pi/2$, but it yields a larger scaling dimension and can thus be discarded as it does not correspond to a global minimum of the free energy.  The intution that the auxiliary bosons should anti-align is confirmed in \ref{app:gen-spin_Hall} as no other solutions are found when taking a  general orientation of the spin-Hall mass $\bm{M_q} = M_q (\sin \vartheta \cos \varphi,  \sin \vartheta \sin \varphi, \cos \vartheta)$. We thus proceed with the simplification $\bm M_q = M_q \hat z$. 

The determinant operator \eqref{eq:fq} can be diagonalized by introducing  spinor monopole harmonics $S^{\pm}_{q, \ell, m}$.  These functions diagonalize generalized total spinor operator  $J^2_q \to j_{\pm} (j_{\pm} +1)$, where $j_{\pm} = \ell \pm 1/2$.  The azimuthal and magnetic quantum numbers, respectively  $\ell \in \{|q|, |q|+1, \dots\}$ and $ m \in \{-\ell, -\ell+1, \dots \ell\}$, define the eigenvalues of $L_q^2$ and $L_q^z$ which are diagonalized by monopole harmonics \cite{wu_dirac_1976}  which serve as components of the spinor monopole harmonics. For minimal angular momentum $\ell = |q|$, only the $S^{\pm}_{q, |q|,m}$ spinor exists and it corresponds to a zero mode of the Dirac operator.  In the $j=\ell-1/2$ basis $(S^+_{q;\ell-1,m}, S^-_{q;\ell,m} )^\intercal$, the Dirac operator becomes a matrix with c-number entries \cite{borokhov_topological_2003}.  As for the spin-Hall mass, its contribution is diagonal in this basis as noted in Ref.~\cite{dupuis_transition_2019}. The resulting diagonal determinant operator therein is adapted by shifting the Matsubara frequency  to account for the presence of the spin chemical potential
\eqn{
\begin{split}
 \fq   =& - \dfrac{1}{\beta} \sum_{\sigma = \pm  1} \sum_{n \in \Z} \biggl[  d_{q}  \ln \bigl( \omega_n - i  \mu' \sigma  + i \sigma M_q \bigr)  + \!\! \sum_{\ell = q+1}^\infty d_\ell \ln  \bigl( \lb \omega_n - i \mu' \sigma \rb^2 + \veps_\ell^2 \bigr)  
  \biggr] \,,
  \end{split}
  \label{eq:free0}
  }
  where  $\veps_\ell$ is the energy and $d_\ell$ is the degeneracy   
  \eqn{
  \veps_\ell = \sqrt{\ell^2 -q^2   + M_q^2}\,, \quad d_\ell = 2\ell\,, \label{eq:en_den}
  }
  and $\omega_n$ are the fermionic Mastubara frequencies
\eqn{
\omega_n = \frac{2\pi}{\beta} (n + 1/2)\,, \quad n \in \Z\,.
}
Note that the energy $\veps_\ell$ is dimensionless as we choose units where the radius of the two-sphere  $\sqrt{V/ (4\pi)}$ is equal to one.  Also, we have supposed that the magnetic charge is positive $q>0$.  In the end, the monopole scaling dimension is independent of the sign of the charge. Taking the sum over Mastubara frequencies\footnote{We may define $\tilde{M}_q = M_q - \mu'$ in the first term and $\tilde \mu = \mu \sigma$ in the second term, and directly read results from Ref.~\cite{dupuis_transition_2019}}, Eq.~\eqref{eq:free0} is simplified to
  \eqn{
  \begin{split}
\fq =& - \beta^{-1}  \biggl[   d_{q} \ln \bigl( 2 \lc 1+ \cosh(\beta (M_q - \mu') )\rc \bigr) \\
&+ \sum_{\ell = q+1}^\infty  2 d_\ell \ln \bigl( 2 \lc \cosh(\beta \veps_\ell) + \cosh(\beta \mu') \rc \bigr) \biggr] \,.
\end{split}  
\label{eq:f0}  
  }
  
  \subsection{Solving the gap equations}
  We now obtain the gap equations by varying the free energy with respect to the original saddle point parameters $M_q, \mu_S, P_z$
\eqn{
\pa_{M_q}\fq  = 0\,, \label{eq:gapp1}\\
\thalf \ms \pa_{\mu'} \fq  + \S^2  = 0 \,, \label{eq:gapp2}\\
\sqrt{\tfrac{\mu_S}{2}}  \pa_{\mu'} \fq  + P_z = 0 \label{eq:gapp3}\,.
}
The last gap equation could be solved with ${\mu_S  = P_z = 0}$, but this yields unphysical results (see  \ref{app:mu0}). Instead, if we take  $\mu_S \neq 0$ and $P_z \neq 0$, the third gap equation can be written as 
\eqn{
\pa_{\mu'} \fq  = - 2 \ms\,, \label{eq:third}
}
where we used Eq.~\eqref{eq:chi}. The LHS can be developed explicitly as 
\eqn{
\begin{split}
\pa_{\mu'} \fq =  d_{q}  \biggl( \frac{\sinh(\beta (M_q - \mu') )}{1+ \cosh(\beta (M_q- \mu') )} \biggr) -2\sum_{\ell = q+1}^\infty   d_\ell \biggl( \frac{\sinh(\beta \mu')}{\cosh(\beta \veps_\ell) + \cosh(\beta \mu')} \biggr)\,. \label{eq:pa_mupr}
 \end{split}
}
By taking $\mu'$ as 
\eqn{
\mu' =  M_q +  \beta^{-1} \ln \lb \frac{1+ 2 \ms/d_q}{1- 2 \ms/d_q} \rb \,, \label{eq:ansatz}
}
the sum in Eq.~\eqref{eq:pa_mupr} vanishes at  leading order in $1/\beta$ since $\mu' < \veps_{q+1}$ while the first term in Eq.~\eqref{eq:pa_mupr} yields the required result \eqref{eq:third}. Inserting this in the second gap equation \eqref{eq:gapp2}, we obtain 
\eqn{
 \ms^2 = \S^2\,.
 } 
 This means that the polarization is maximized. We turn to the remaining gap equation for $M_q$. The derivative of the determinant operator  with respect to $M_q$ is given by
\eqn{
\begin{split}
\pa_{M_q} \fq =  - d_{q} \biggl( \frac{\sinh(\beta (M_q- \mu') )}{1+ \cosh(\beta (M_q- \mu') )} \biggr)
 - 2 M_q \sum_{\ell = q+1}^\infty \frac{  d_\ell \veps_\ell^{-1} \sinh(\beta \veps_\ell)}{\cosh(\beta \veps_\ell) + \cosh(\beta \mu')} \,.
 \end{split}
\label{eq:pa_m}
}
Inserting in this expression the result for $\mu'$  \eqref{eq:ansatz}, the first gap equation \eqref{eq:gapp1} becomes ${2  \ms   -2 M_q \sum_{\ell} d_{\ell} \epsilon_{\ell}^{-1} = 0}$. A positive solution for $M_q$ can only be found for $\ms < 0$. This shows, as mentioned above, that the spin polarization and the spin-Hall mass should be anti-aligned.    Taking $\ms = - \S$, the gap equation then becomes 
\eqn{
-2 \S -2 M_q \sum_{\ell=q+1}^{\infty} d_{\ell} \epsilon_{\ell}^{-1} = 0\,. \label{eq:gap_M}
}
This equation can be solved for any allowed spin ${0 \leq \S \leq d_q /2}$. We discuss its solutions later on.  

We first turn to the computation of the free energy and the scaling dimension. Using the solution for $\mu'$ \eqref{eq:ansatz}, the determinant operator \eqref{eq:fq} at leading order in $1/\beta$ is given by $- 2 \sum_\ell d_\ell \veps_\ell$. The rest of the free energy \eqref{eq:free} is reexpressed using $\mu' = M_q + \O(1/\beta)$ and $\ms = -\S$\,\footnote{
Combining Eqs. \eqref{eq:chi} and \eqref{eq:mupr}, we may note that $P_z =  \sqrt{2 \mu' \ms}$, which, since $\mu' >0$ and $\ms < 0$, is imaginary.
As $\bm \chi$ is an auxiliary boson introduced as a resolution of the identity \eqref{eq:identity}, an imaginary expectation value poses no problem: A gaussian integral shifted in the complex plane yields the same result.
}. Taking the zero ``temperature'' limit of the resulting free energy, we obtain the scaling dimension at leading order in $1/\nf$
\eqn{
\Delta_{q; \S}  = -   2 \nf \biggl( \S  M_q +  \sum_{\ell = q+1}^{\infty} d_\ell \veps_\ell \biggr) + \O(1/\nf^0)  \,. \label{eq:scaling}
}

\subsection{Scaling dimension and spin-Hall mass}
 The last result \eqref{eq:scaling} shows how monopoles with the largest spin $\S$  have a minimal contribution of the spin-Hall mass to their scaling dimension.  The only remaining parameter is the spin-Hall mass $M_q$ as other fields were evaluated. This mass can be obtained by solving numerically a regularized version of the gap equation \eqref{eq:gap_M}. Inserting this result in a regularized version of Eq.~\eqref{eq:scaling}, the monopole scaling dimension with spin $\S$ is found. Both regularized expressions are shown in  \ref{app:reg}. Here, we simply show their solutions for multiple values of $\S$. 

\begin{figure}[ht]
\centering
{
\includegraphics[width=\linewidth]{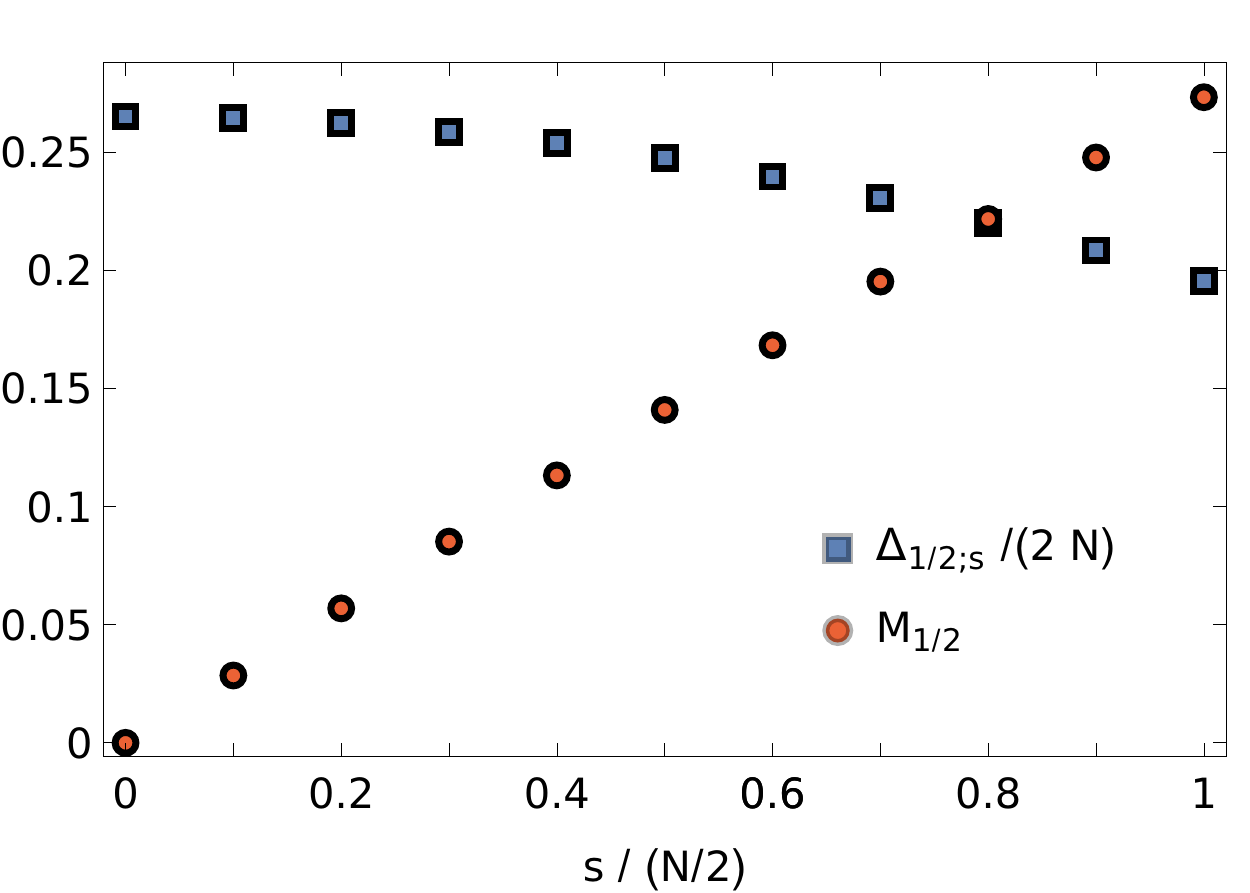}
}
\caption{Scaling dimension $\Delta_{q;\S}$ per number of fermion flavors $2 \nf$ and spin-Hall mass $M_q$ as a function of the monopole spin $\S$  for a minimal magnetic charge $q=1/2$. The spin is expressed as a fraction of its maximal value $s_{\max} = \nf/2$. Here, the spin $s$ is not rescaled by $N$. \label{fig:scaling_mass}}
\end{figure}

\begin{figure}[ht]
\centering
\subfigure[]
{
\includegraphics[width=0.4\linewidth]{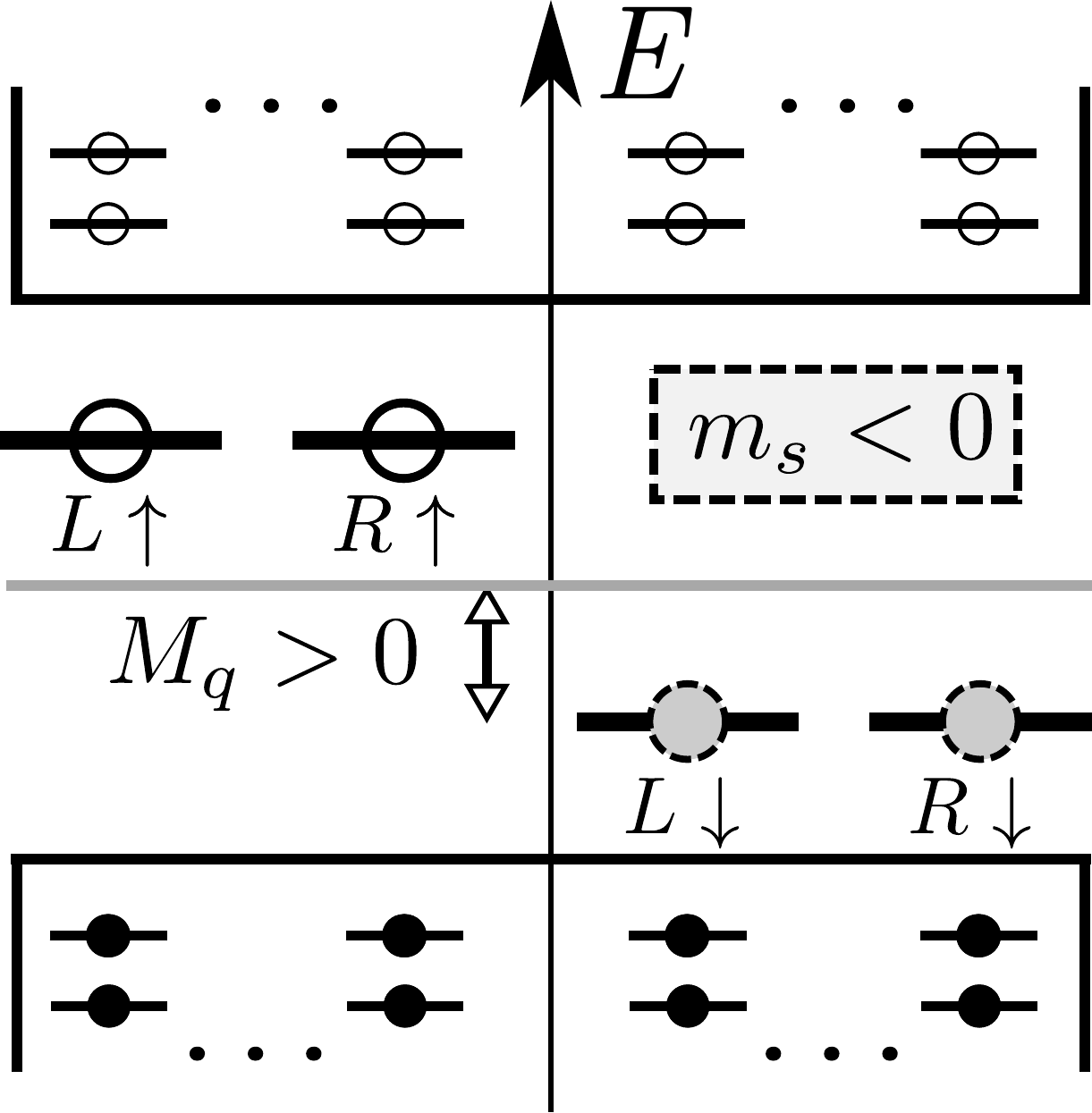}
}
\;\;\;\; \quad
\subfigure[]
{
\includegraphics[width=0.4\linewidth]{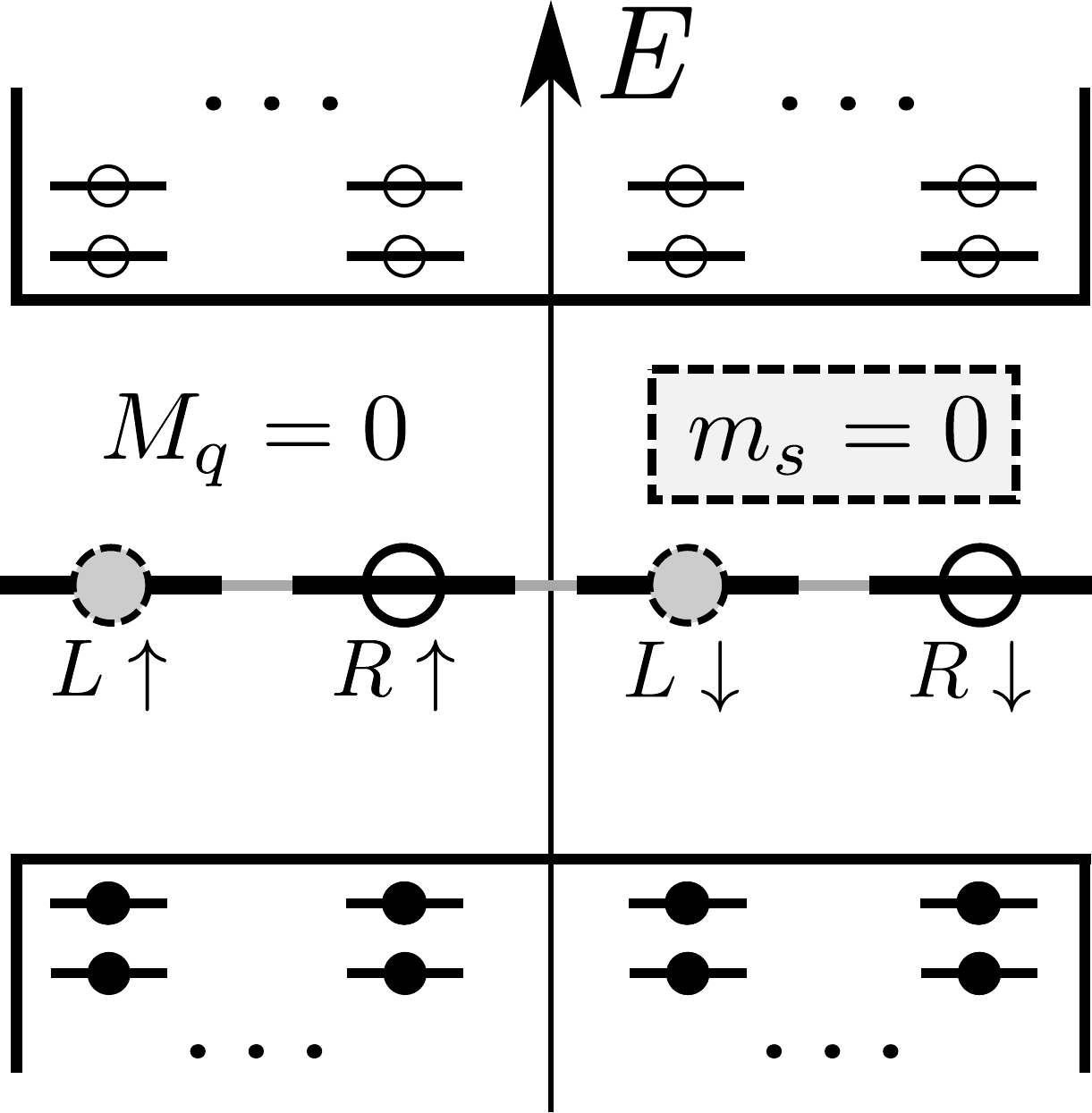}
}
\caption{Schematic representation of the fermion modes dressing of monopole operators with ${\nf = 2}$ valleys ${v = L, R}$ and a minimal magnetic charge $q=1/2$. (a)  The $\S = 1, \ms=-1$ monopole is described by a non-vanishing spin-Hall mass $\moye{\Psib \bm \sigma \Psi} \propto M_q \hat z$ where $M_q>0$. (b) A spin singlet monopole ($s=m_s=0$) is described by a vanishing spin-Hall mass $M_q=0$. Here, the monopole polarized along the $L$ valley is shown. \label{fig:pol_mass}}
\end{figure}

The spin-Hall mass $M_q$ and the scaling dimension  $\Delta_{q;\S}$  for the minimal magnetic charge ${q=1/2}$  are shown as a function of the total magnetic spin  $\S$ in Fig.~\ref{fig:scaling_mass}.
For the maximal spin ${\S_{\max}=d_q /2}$, the gap equation and the minimal scaling dimension  in the $4 \pi q$ sector of $\QEDtcHGN$ found in Ref.~\cite{dupuis_transition_2019} are retrieved, that is ${\Delta_{q; \S_{\max}} = \Delta_{q}^{\QEDtcHGN}}$. More explicitly, the scaling dimension is given by
\eqn{ 
\Delta_{q; \S_{\max}}
= - \nf \Bigl( d_q M_q + 2 \sum_\ell d_\ell \veps_\ell \Bigr)\Big|_{\text{Saddle pt.}}  + \O(\nf^0)\,,
}
where the spin-Hall mass is evaluated at its saddle point value found by solving the gap equation \eqref{eq:gap_M} for $s=s_{\max}$. The last expression for $\Delta_{q;s_{\max}}$ yields the minimal scaling dimension $\Delta_{q}^{\QEDtcHGN}$ found in Ref.~\cite{dupuis_transition_2019}.  For a minimal spin $\S_{\min} = 0$, the mass at saddle point vanishes.   In turn, this means that the leading order scaling dimension corresponds to the monopole scaling dimension in pure $\QEDt$, ${\Delta_{q; \S_{\min}} = \Delta_q^{\QEDt}}$, or, more explicitly,  
\eqn{
\Delta_{q;0} =   - 2 \nf \sum_{\ell} d_{\ell} \veps_\ell|_{M_q=0} + \O(\nf^0)\,.
} 
The scaling dimension ranges between these values for intermediate values of the spin $ \S_{\min} < \S < \S_{\max}$ 
\eqn{
\Delta_q^{\QEDtcHGN} \leq \Delta_{\man_{q}}^{\QEDtcHGN} \leq \Delta_q^{\QEDt}\,, \quad \text{L.O. in } 1/\nf\,. \label{eq:hierarchy}
} 
This result includes the effect of fermion occupation on the mass $M_q$. This aspect was neglected in  Ref.~\cite{dupuis_transition_2019} where the mass was considered fixed in orientation and amplitude, defined as to yield the smallest possible lower bound for scaling dimensions in the $4 \pi q$ sector. 
The operator corresponding to this minimal scaling dimension was dubbed ``spin down'' monopole and other monopoles were obtained by modifying its ``zero'' modes occupation.  Here, by finding an optimal spin-Hall mass parameter in each spin sector, a smaller upper boundary on monopole scaling dimensions in $\QEDtcHGN$ \eqref{eq:hierarchy} is found.  This is schematized in Fig.~\ref{fig:pol_mass} for the case $q=1/2$ and $\nf=2$. 

\begin{figure}[ht]
\centering
{
\includegraphics[width=\linewidth]{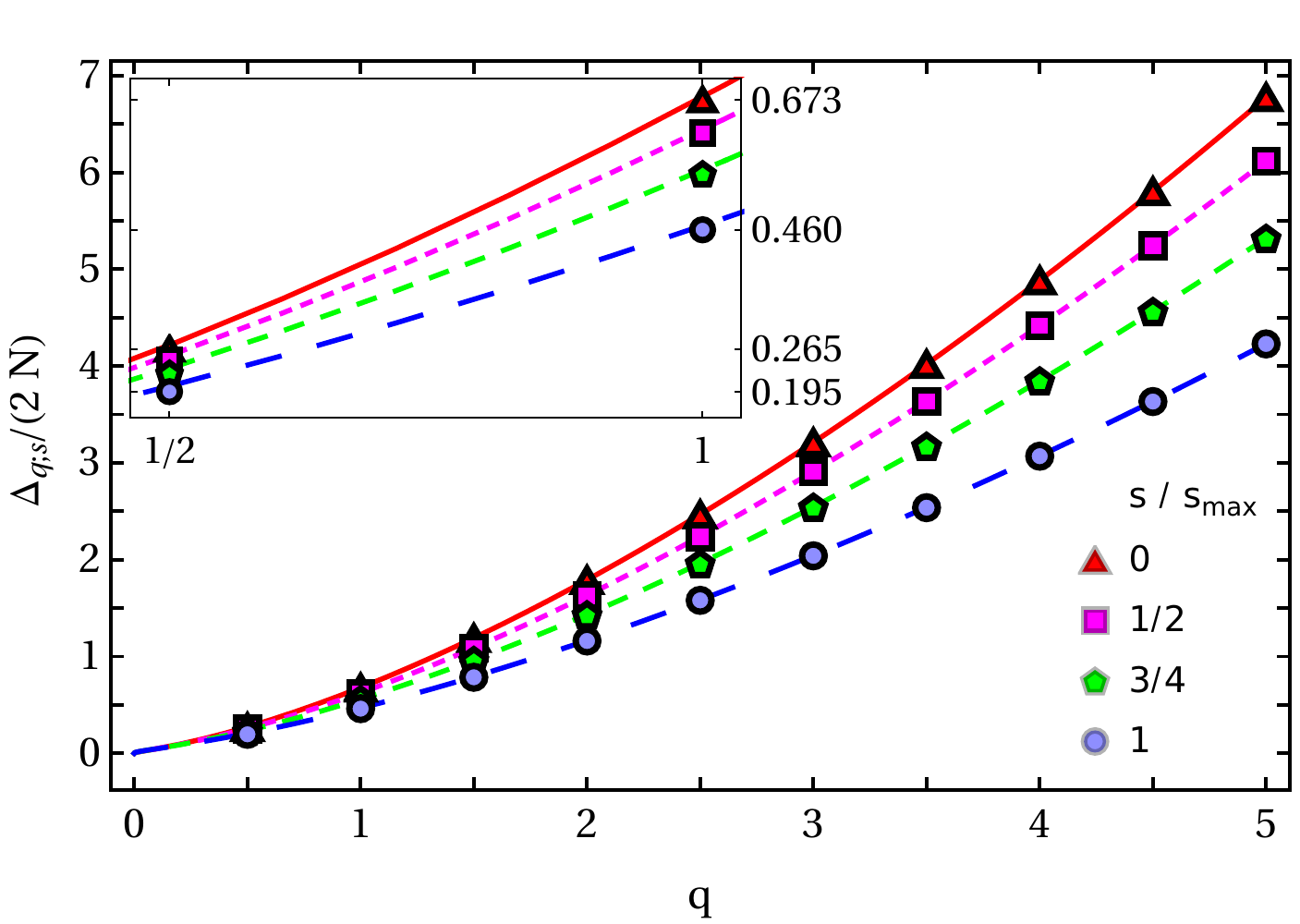}
}
\caption{Scaling dimension $\Delta_{q;\S}$ per number of fermion flavors $2 \nf$ as a function of $q$ for $s/s_{\max} = \{0, 1/2, 3/4, 1\}$ where $s_{\max} = q \nf$. The lines correspond to analytical approximations of the scaling dimensions obtained through a large-$q$ expansion and shown in Table~\ref{tab:large}.\label{fig:scaling2}}
\end{figure}

This hierarchy is characterized explicitly for various values of  spin and magnetic charge. Monopole scaling dimensions obtained numerically are shown in Fig.~\ref{fig:scaling2}. Every $q=1/2$ monopole has a smaller scaling dimension than monopoles with larger magnetic charge $q$. This is not necessarily the case for monopoles with larger magnetic charges, e.g.    $\Delta_{4;s_{\min}} < \Delta_{5;s_{\max}}$. Analytical approximations for the scaling dimensions obtained with a large-$q$ expansion and shown in Table~\ref{tab:large} are also plotted in Fig.~\ref{fig:scaling2}.  There is a good agreement with the numerical results even for small values of the magnetic charge. 

\section{Scaling dimensions for large-$q$ \label{sec:large}}
The scaling dimension of monopole operators may also be approximated by an analytical expression obtained with a large-$q$ expansion. The expansion was presented in Ref.~\cite{dupuis_transition_2019} for the monopole with maximal spin and minimal scaling dimension. Building on this result, analytical approximations were also proposed for other types of monopoles, but these results again neglected the backreaction of the ``zero`` modes occupation on the spin-Hall mass. Here, we show the proper analysis for any magnetic spin $s$.

To start this computation, we may read the unregularized free energy (at zero temperature) off the scaling dimension \eqref{eq:scaling} with the relation ${\Delta_{q;s} =\nf F_{q;s}^{(0)} + \dots}$ By changing the summation index   $\ell \to \ell + q + 1$,  the free energy  becomes
\eqn{
\frac{F_q^{(0)}}{2}  
= -  4 q M_q  \tilde{s}   - 4\sum_{\ell = 0}^\infty (\ell + q +1) \sqrt{(\ell + q +1)^2 -q^2 + M_q^2}  \,,
\label{eq:F0_unreg_2}
}
where we introduced  $\tilde s \equiv s / s_{\max}$ in order  to factorize the dependence on $q$. The saddle point equation is then given by
\eqn{
 - 4 q  \tilde{s}  - 4 M_q \sum_{\ell = 0}^\infty \frac{ \ell + q + 1  }{\sqrt{\lb \ell + q + 1 \rb^2 + M_q^2- q^2}} = 0  \,.
\label{eq:cond}
}
Inserting the following mass squared ansatz
\eqn{
M_q^2 = 2 \chi_0 q + \chi_1 + \O(q^{-1})  \,,
\label{eq:ansatz_m2}
}
the free energy \eqref{eq:F0_unreg_2} and the gap equation \eqref{eq:cond}  can be expanded in powers of $1/q$, respectively 
\eqn{
\begin{split}
\half F_q^{(0)}
=& - 2 \sqrt{2} q^{3/2} \lb  \zeta_{-1/2} + \tilde s \chi _0^{1/2}  \rb  - \frac{q^{1/2}}{\sqrt{2}} \bigg( \!   \left(\chi _0^2+\chi _1\right) \zeta_{1/2} \\
&- 6  \chi_0 \zeta_{-1/2} + 5 \zeta_{-3/2} + \tilde s \chi_1 \chi_0^{-1/2} \bigg)  + \O(q^{-1/2}) \,. \label{eq:F0g_0}
\end{split}
}
and 
\eqn{
0 =& \; 4 q^{1/2} \lb  \zeta_{1/2} + \tilde  s \chi_0^{-1/2} \rb 
 - q^{-1/2} \Big( \lb   \chi_1  +  \chi _0^2\rb \zeta_{3/2} \nn \\
& + 2  \chi_0 \zeta_{1/2} - 3  \zeta_{-1/2}  + \tilde s \chi_1 \chi _0^{-3/2} \Big)  +\O(q^{-3/2})  \,,
\label{eq:gapg}
}
where $\zeta_n \equiv \zeta(n, 1 + \chi_0)$  is the zeta function used to regularize ${\sum_{l=0} (l+(1+\chi_0))^{-n}}$.  The gap equation at leading order yields a transcendental condition  defining $\chi_0$ while $\chi_1$ is determined by a linear condition at next-to-leading order
\eqn{	
 \zeta_{1/2} + \tilde s \chi_0^{-1/2}  &= 0  \,,
\label{eq:chi0}\\
\chi_1 + \frac{ 2  \chi_0^{3/2} \lb
 \chi_0^2 \zeta_{3/2}
 +2  \chi_0 \zeta_{1/2} 
 -3   \zeta_{-1/2} 
 \rb }{\tilde s +  \chi_0^{3/2} \zeta_{3/2}} &=0  \,.
\label{eq:chi1}
}
We solve these equations by fixing $s$ and by finding numerically the values of $\chi_0$ and $\chi_1$. These coefficients yield the value of the spin-Hall mass and are inserted in the free energy \eqref{eq:F0g_0}.\footnote{As noted in last section, the gap equation \eqref{eq:cond} for $s=0$ is solved for a vanishing spin-Hall mass $M_q=0$. In this case, we must simply take $\chi_0=\chi_1=0$.} The resulting monopole scaling dimensions at leading order in $1/\nf$ for various magnetic spin $s$ are shown in Table~\ref{tab:large}.
\begin{table}[ht]\caption{Analytical approximation of the monopole scaling dimensions for $s/s_{\max} = \{0, 1/2, 3/4, 1\}$ obtained in the large-$q$ expansion. \label{tab:large}}
\centering
\begin{tabular}{c|c}
$\displaystyle \frac{s}{s_{\max}} $ &  $\displaystyle \frac{\Delta_{q; s}}{2\nf} + \O(q^{-1/2}) + \O(\nf^{-1})$\\[0.75em] \hline
$1$ & $0.356 q^{3/2} + 0.111 q^{1/2}$\\
$3/4$ & $0.455 q^{3/2} + 0.103 q^{1/2}$\\
$1/2$ & $0.528 q^{3/2} + 0.096 q^{1/2}$\\
$0$ & $0.588 q^{3/2} + 0.090 q^{1/2}$
\end{tabular}
\end{table}
The lines in Fig.~\ref{fig:scaling2} giving the monopole scaling dimensions against $q$ are plotted using these analytic approximations.

Note that the absence of the $\O(q^0)$ term in $\Delta_{q;s}$ is expected. In $2+1$-dimensional CFTs with a $\U(1)$ global charge $q$, the $q^0$ contribution in the large-$q$ expansion of the scaling dimension is predicted to be universal. It must thus be independent of the number of fermion flavors $2N$ defining our specific model ~\cite{Hellerman_on_2015, delaFuente_large_2018}, i.e. the $\O(q^0 N^1)$ in the large-$N$ and large-$q$ expansion of $\Delta_{q;s}$ must vanish.
  
\section{Hierarchy as degeneracy lifting \label{sec:irreps}}
 The hierarchy shown in Eq.~\eqref{eq:hierarchy} is now analyzed from the point of view of symmetry. Our non-perturbative analysis does not depend on a large-$N$ expansion. The multiplet organization of monopoles at the QCP is obtained by showing how monopoles in the DSL reorganize as the flavor symmetry of $\QEDt$ is broken to the flavor symmetry of  $\QEDtcHGN$. To make contact with the scaling dimensions obtained in the last section, it is important to recognize that the spin-Hall mass is a $\SU(2)_{\rm Spin}$ vector.

The $\cHGN$ interaction $\delta \lag \sim (\Psib \bm\sigma \Psi)^2$ inducing the confinement-deconfinement transition breaks down the flavor symmetry of $\QEDt$ as
\eqn{
 \SU(2 \nf)  \to \SU(2) \times \SU(\nf)\,.
 \label{eq:symmetry_breaking}
 } 
 While monopoles in $\QEDt$ are all related by $\SU(2 \nf)$ rotations and share the same scaling dimension, this is not the case in $\QEDt-\cHGN$. The hierarchy of monopole operators in $\QEDt-\cHGN$ observed in the previous section may be explained as a degeneracy lifting of monopoles in $\QEDt$.
 
 Monopoles in  $\QEDt$ are organized as irreducible representations (irreps) of the flavor symmetry group $\SU(2 \nf)$. We focus our attention of monopoles with a minimal magnetic charge ${q=1/2}$. This is the simplest case as  monopole operators are then automatically Lorentz scalars  \cite{borokhov_topological_2003}. The case  with a larger magnetic charge is briefly discussed in  \ref{app:red}. It is useful to first define a bare monopole operator $\man_{\rm Bare}^\dag$ creating a $2\pi$ magnetic flux background and filled only with negative energy modes. A  monopole operator is then obtained by adding in half of the  $2 \nf$ zero modes creation operators $c^{\dag}_{I_{i}}$   
\eqn{
\man^\dag_{I_1 \dots I_N} = c^\dag_{I_1} \dots c^{\dag}_{I_{N}} \man^\dag_{\rm Bare}\,, \quad I_i \in \{1, 2, \dots, 2 \nf\}\,.  \label{eq:mon_gen}
}
Given the antisymmetric commutation relations between the fermionic creation operators, the expression above clearly shows how $q=1/2$ monopoles  in $\QEDt$ form the rank-$\nf$ completely antisymmetric irrep of $\SU(2\nf)$.
\subsection{Multiplets at the QCP for $\nf=2$}
We first discuss the monopole hierarchy for a finite $\nf$ situation to provide some intuition.  While the discussion is centered on symmetries and not dynamics, we do need to assume that the QCP still exists at finite $\nf$.  We focus on the case with $\nf=2$ valleys ${v= L,R}$ which is the most relevant to quantum magnets. Monopole operators then have two zero modes creation operators and may be expressed in the following form
\eqn{
c^\dag A(c^{\dag})^{\intercal}  \man^\dag_{\rm Bare}\,, \label{eq:QED3_2}
}
where $A$ acts on vectors in flavor space  ${c^\dag =  \big( c_{\u, L}^\dag,\,  c_{\u, R}^\dag,\,  c_{\d, L}^\dag,\,  c_{\d, R}^\dag \big)}$. At the $\QEDtcHGN$ QCP, monopoles are organized as triplets \cite{hermele_properties_2008, song_unifying_2019}
\eqn{
\bm{\man}_{\rm Spin}^{\dag}  = c^\dag \lb \sigma_y \bm \sigma  \otimes  \mu_y \rb (c^{\dag})^{\intercal} \, \man_{\rm Bare}^{\dag}\,, \label{eq:triplets-a} \\
\bm{\man}_{\rm Nodal}^{\dag} = c^\dag \lb \sigma_y \otimes  \mu_y \bm \mu \rb (c^{\dag})^{\intercal} \, \man_{\rm Bare}^{\dag}\,,
\label{eq:triplets-b}
}
where $\bm \sigma$ and $\bm \mu$ are Pauli matrices vectors respectively acting on magnetic spin and nodal subspaces (in our notation, $\mu_z = \ket{\uf}\bra{\uf} - \ket{\df} \bra{\df}$ and  $\sigma_z$ has the usual definition $ \ket{\u}\bra{\u} - \ket{\d} \bra{\d}$). In particular, the spin triplet forms the order parameter $\hat n \cdot  \moye{i \bm{\man}_{\rm Spin}^{\dag }}$ for coplanar antiferromagnetic phase of  the Kagome heisenberg  antiferromagnet \cite{hermele_properties_2008, lu_unification_2017}.  

\begin{figure}[ht]
\centering
\subfigure[]
{\includegraphics[width=0.4\linewidth]{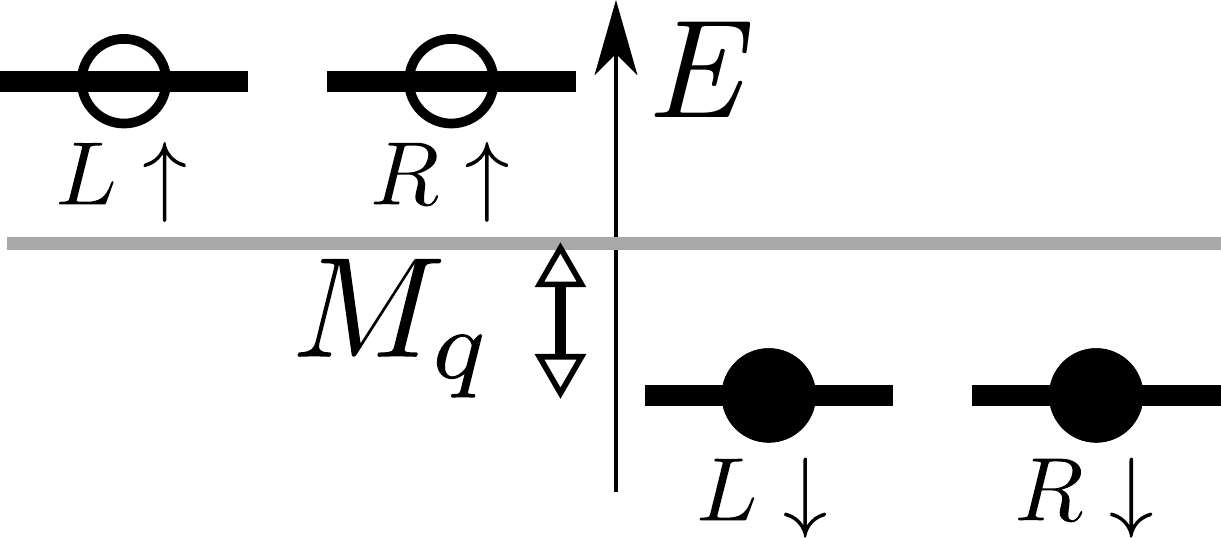}
}\qquad
\subfigure[]{
\includegraphics[width=0.4\linewidth]{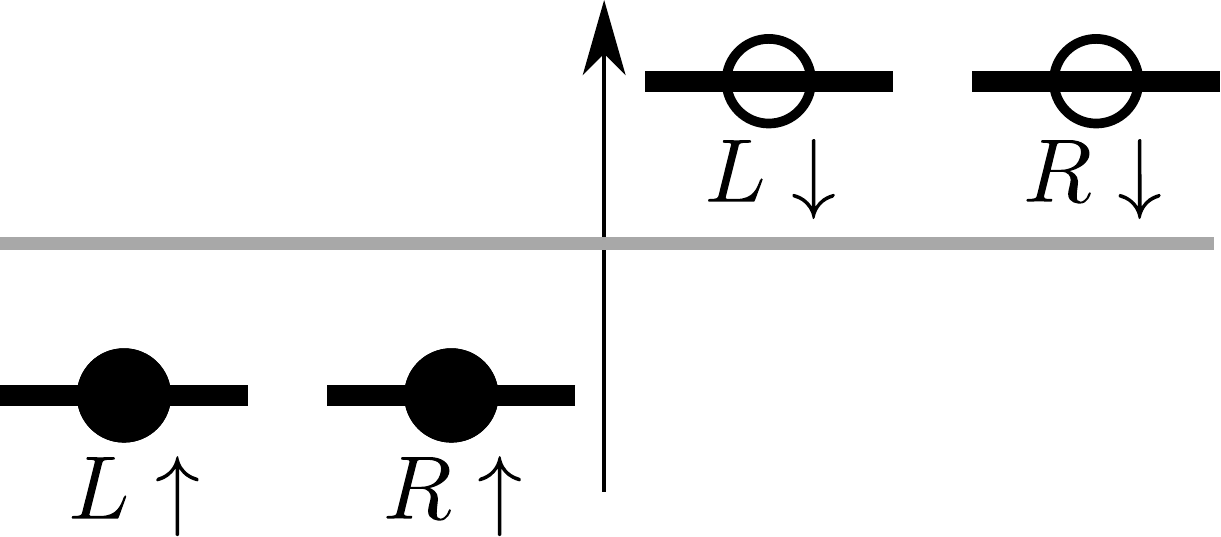}
}\,
\caption{Schematic representation of the fermion zero modes dressing of monopole operators with ${\nf = 2}$ valleys ${v = L, R}$ and a minimal magnetic charge $q=1/2$. (a)  The $\S = 1, \ms=-1$ monopole with spin-Hall mass $\moye{\Psib \bm \sigma \Psi} \propto M_q \hat z$ can be rotated to (b) the $\S = 1, \ms=-1$ monopole with spin-Hall mass $\moye{\Psib \bm \sigma \Psi} \propto -M_q \hat z$. \label{fig:spin_flip}}
\end{figure}

More formally, monopoles in  $\QEDt$ \eqref{eq:QED3_2} form the ${\text{rank-}2}$ completely antisymmetric irrep, that we note by its dimension $\bm 6$. Following the symmetry reduction \eqref{eq:symmetry_breaking}, this irrep is reduced in irreps $(\bm m, \bm n)$ with dimension $m \times n$ of the QCP subgroup ${\SU(2)_{\rm Spin} \times \SU(2)_{\rm Nodal}}$  
\eqn{
\bm 6 \to \lb \bm 3, \bm 1\rb \oplus \lb \bm 1 , \bm 3 \rb\,. \label{eq:reduction_2}
}
The spin and nodal triplet, respectively $(\bm 3, \bm 1)$ and $(\bm 1, \bm 3)$, have total magnetic spin given by $S=1$ and $S=0$. They are the finite-$\nf$ analogues of  ${q=1/2}$ monopoles  with $\S_{\max} = 1$ and $\S_{\min}=0$, respectively, described in Sec.~\ref{sec:scaling}
\eqn{
 \man_{1/2; \S_{\max}}^\dag \to \lb \bm 3, \bm 1\rb\,, \quad  \man_{1/2; \S_{\min}}^\dag \to \lb \bm 1, \bm 3\rb\,.
 } 
 As we perform a $\SU(2)_{\rm Spin}$ rotation, the ``zero'' modes occupation is modified  and so does the orientation of the spin-Hall mass. For example, a spin flip exchanges spin down and spin up ``zero'' modes, changing the polarization sign of a monopole, i.e.\ $\man^\dag_{q; \S, \ms} \to \man^\dag_{q; \S, - \ms}$. With this transformation, the sign of the spin-Hall mass also changes ${\moye{\Psib \sigma_z \Psi} \to - \moye{\Psib \sigma_z \Psi}}$, which leaves the scaling dimension unchanged.  The spin flip is schematically shown  in Fig.~\ref{fig:spin_flip}. 
In  \ref{app:explicit}, we explore the explicit realization of monopoles suggested in Eqs.~(\ref{eq:triplets-a} -\ref{eq:triplets-b}) and we also obtain a representation of the  $\SU(2) \times \SU(2)$ generators. Acting with these on the monopoles, it can be seen that these operators indeed form a reducible representation of this group. We also build explicitly the rotation shown in Fig.~\ref{fig:spin_flip}, and the rotation that sends the quantization axis in the $x-y$ plane. 
 
 \subsection{Multiplets at the QCP for general $\nf$}
 We now give a description of the monopole hierarchy for general $\nf$. Again, our starting point is the organization of monopoles in $\QEDt$. As noted above, these monopoles form the rank-$\nf$  antisymmetric tensor of $\SU(2 \nf)$ \eqref{eq:mon_gen}. In terms of a Young tableau, this may be written as a single column of $\nf$ boxes
\eqn{
 \signn \,. \label{eq:sign}
}
The organization of monopole operators at the $\QEDtcHGN$ QCP can then be understood by finding how this $\SU(2 \nf)$ irrep reduces as a representation  of the subgroup $\SU(2) \times \SU(\nf)$. 

In the last section, we discussed the case $\nf=2$ where the rank-$2$ completely antisymetric irrep $\bm 6$ of $\SU(4)$ is reduced to the representation $ \lb \bm 3, \bm 1 \rb \oplus \lb \bm 1, \bm 3 \rb $  of ${\SU(2) \times \SU(2)}$. The following Young tableaux schematize this  reduction  \eqref{eq:reduction_2} 
\eqn{
 \yngc  \yng(1,1)_{\bm 6} &\to \Bigl( \yngc \yng(2)_{\bm 3}, \yng(1,1)_{\bm 1} \Bigr) \oplus \Bigl( \yngc \yng(1,1)_{\bm 1}, \yng(2)_{\bm 3} \Bigr)\,, \label{eq:decomp_4}
   }   
   where the bold subscripts indicate the respective irrep's dimension. The reduction can also be written explicitly  for general $\nf$. The rank-$\nf$ completely antisymmetric irrep \eqref{eq:sign} is reduced as 
   \eqn{ 
\signn
&\to
\bigoplus_{b=0}^{\floor{\nf/2}} \Biggl(\;  \irrepm , \;  \irrepmpr \; \Biggr) \,,
\label{eq:reduction}
}
This  is coherent with results in Ref.~\cite{itzykson_unitary_1966} where this reduction is obtained up to $\nf=8$,\footnote{Using the notation employed in Ref.~\cite{itzykson_unitary_1966}, Eq.~\eqref{eq:reduction} can be reexpressed as $(1^{\nf}) \to {\displaystyle \bigoplus_{b=0}^{\floor{\nf/2}}}  \Bigl( (\nf-b, b), (2^b, 1^{\nf-2b}) \Bigr)$.} or with the output of a recent symbolic computation  package for representation theory \cite{feger_lieart_2019}. This result was stated in Ref.~\cite{chester_monopole_2018} where the organization of flux operators in Lorentz symmetry multiplets was discussed. We give the proof of this result \eqref{eq:reduction}  in  \ref{app:red}. 

The previous notation highlights well that the RHS of Eq.~\eqref{eq:reduction} is simply the sum over all pairs of conjugate Young tableaux with $\nf$ boxes,  $(\lyt_\nu, \lyt_{\tilde \nu})$, where the first diagram $\lyt_\nu$ has a maximum of two rows since it  must be a $\SU(2)$ irrep. This plays a key role when proving Eq.~\eqref{eq:reduction}.  This total number of boxes also corresponds to the number of fermion zero modes dressing the monopole. Indeed, the monopole may be noted as $\man^\dagger_{\sigma_1 v_1, \sigma_2 v_2, \dots ,\sigma_N v_N}$. The $\SU(2)_{\rm Spin}$ irrep encodes symmetry relations between spin indices $\sigma_i$, and the same goes for valley indices $v_i$ and the $\SU(N)_{\rm Nodal}$ irrep.  In the simplest case where $b=0$, the irrep becomes
  \eqn{
\biggl( \; \irrepLineN ,\;  \irrepColN \; \biggr) \,.
\label{eq:b0}
    }
    The corresponding monopole simply transforms symmetrically in its spin indices and antisymmetrically in its valley indices. That is, the monopole is a valley singlet and a spin multiplet with maximal spin. For other values of $b$, the irreps describe mixed symmetries between the indices. Every box represents a spin or valley index which is antisymmetrized with its column neighbours and symmetrized with its row neighbours. The antisymmetry of Eq.~\eqref{eq:reduction}'s  LHS is realized in each of the RHS irreps through the following prescription : i) Match pairs of antisymmetrized spins (columns with two boxes) with pairs  of symmetrized valleys (rows with  two boxes) ii) Match the remaining symmetrized spins  with the remaining antisymmetrized valleys.  

To gain a better physical intuition, it is useful to label the Young tableaux in Eq.~\eqref{eq:reduction}  with the magnetic spin $\S$ instead of a number of boxes $b$. The dimension of the $\SU(2)_{\rm Spin}$ irrep  is $\nf -2 b +1$.  This number should naturally be identified with $2 \S +1$.   Also,  we can eliminate every column of two boxes for the $\SU(2)_{\rm Spin}$ irreps which correspond to antisymmetrized pairs of spins. Ignoring those ``bounded'' spin indices by removing their corresponding boxes obscurs the underlying zero modes dressing, but it puts the $\SU(2)_{\rm Spin}$ irreps in a more familiar form. The remaining boxes correspond to the ``free'' spin indices that form the spin-$s$ multiplet. With this, the monopole representation at the QCP can be written as 
    \eqn{
    \bigoplus_{s = (\nf\, \mathrm{mod}\, 2)/2}^{\nf/2} \Biggl(\;  \irrepsmin , \;  \irrepspr \; \Biggr)\,. 
    \label{eq:irreps_s}
    }
     For the maximal spin $s=N/2$, the  valley irrep is a singlet, as noted earlier in Eq.~\eqref{eq:b0}.
    
    \subsubsection{Degeneracy in each magnetic spin sector}    
    The degeneracy of monopoles at the QCP for each magnetic spin sector is found in what follows. The do so, we compute the dimension of the reduced irreps in     Eq.~\eqref{eq:irreps_s}, which is given by the product of the spin and valley irreps' dimensions. The dimension of the spin irrep is the usual spin degeneracy $2s + 1$. 
    As for the valley irrep, it can first be observed that it is generated through a tensor product of completely antisymmetric tensors 
    \eqn{
\begin{split}
\irrepColT{\frac{N}{2}+s} \!\! \otimes  \!\!  \irrepColD{\frac{N}{2}-s} 
\;
=
\;
\irrepspr
 \,\, \oplus\,\, 
 \irrepColTPr{\frac{N}{2}+s+1} \!\! \otimes \!\! \irrepColDPr{\frac{N}{2}-s-1} \,.
\end{split}
\label{eq:tensor}
}
 The valley irrep's dimension can be obtained by computing the dimensions of the other representations in this relation. Each composite antisymmetric tensor has a dimension given by a binomial factor.  Specifically, the dimension of the $\SU(N)$ irrep corresponding to a single column of $b$ boxes is $\binom{N}{b}$.   The tensor product's dimension is simply the product of these binomial factors. The dimension of the valley irrep is then
\eqn{
\dim \lb   \irrepspr  \rb = \binom{N}{\frac{N}{2}+s}\binom{N}{\frac{N}{2}-s} -  \binom{N}{\frac{N}{2}+s+1}\binom{N}{\frac{N}{2}-s-1}\,. 
\label{eq:dim_sub}
}
 We can then obtain the total degeneracy of monopoles in each magnetic spin sector 
\eqn{
 \Omega_s = (2 \S + 1) \times \lc \binom{N}{\frac{N}{2}+s}\binom{N}{\frac{N}{2}-s} -  \binom{N}{\frac{N}{2}+s+1}\binom{N}{\frac{N}{2}-s-1} \rc \,.
 \label{eq:dim_sub_irreps}
}
In \ref{app:red}, we confirm this result with a method appropriate for general Young tableaux.  We also show that the total dimension of the reduced irreps is equal to the original antisymmetric irrep in $\SU(2N)$, that is $\sum_s \Omega_s =  \binom{2N}{N}$. 

We briefly reformulate the last result. The factor $2s+1$  comes from the possible spin polarizations of the monopole. As for the valley sector, we take as a starting point two antisymmetric tensors   $A_{v_1,v_2,\dots , v_{N/2+s}}\tilde A_{v'_1,v'_2,\dots , v'_{N/2-s}}$  that assign valley indices to the $N$ zero modes. This corresponds to the tensor product in the LHS of Eq.~\eqref{eq:tensor}.  The $2s$ supplementary valley indices in the first tensor can be attributed to the $2s$ ``free'' spin indices which give the monopole its magnetic polarization.  Indeed, these ``free'' spin indices form a symmetric multiplet and must be matched with antisymmetrized valley indices to yield an antisymmetric state. On the other hand, the remaining $N-2s$ ``bounded'' spin indices (the boxes eliminated in Eq.~\eqref{eq:irreps_s}) form antisymmetrized spin pairs that should be matched with symmetrized valley pairs. Thus, every configuration with at least one pair of antisymmetrized valley indices  matched to an antisymmetrized spin pair should be removed. Such configurations can be written as  $A_{v'_1, v_1, v_2, \dots , v_{N/2+s}}\tilde A_{v'_2,v'_3, \dots , v'_{N/2-s}}$, which is the tensor product in the RHS of Eq.~\eqref{eq:tensor}. It then follows that distinct spin pairs are also related by antisymmetrized valley indices, which yields the overall antisymmetric state.

\subsubsection{Reduction for $N=3$}

Using the result in Eq.~\eqref{eq:reduction}, we can retrieve the reduction of monopoles at the QCP for $N=3$ that was discussed in Ref.~\cite{proc}. In this case, the flavor symmetry breaking is ${\SU(6) \to \SU(2)_{\rm Spin} \times \SU(3)_{\rm Nodal}}$. Monopoles in $\QEDt$ are organized as the rank-$3$ antisymmetric representation  $\bm{20}$ of $\SU(6)$. At the QCP, this is decomposed as 
\eqn{
 \yngc  \yng(1,1,1)_{\bm{20}} &\to \Bigl( \yngc \yng(3)_{\bm 4}, \yng(1,1,1)_{\bm 1} \Bigr) \oplus \Bigl( \yngc \yng(2,1)_{\bm 2}, \yng(2,1)_{\bm 8} \Bigr)\,, \label{eq:20}
   }  
   which agrees with Ref.~\cite{itzykson_unitary_1966}. The total dimension of the reduced irreps is ${4 \times 1 + 2 \times 8 = 20}$, as required. The largest $\SU(2)_{\rm Spin}$ spin is $s_{\max} = 3/2$ since $q=1/2$ and $N=3$ (see Eq.~\eqref{eq:S}). It corresponds to a spin quadruplet, the $\bm 4$ in Eq.~\eqref{eq:20}. This irrep represents monopoles with the lowest scaling dimension. As noted before, the fact that $N$ is odd yields a non-vanishing minimal with $s_{\min} = 1/2$, the  $\bm 2$ in Eq.~\eqref{eq:20}. As three ``zero'' modes must be filled, the lowest polarization that can be obtained is by paring two modes in a singlet, leaving one remaining spin that forms the doublet.

\section{Conclusion \label{sec:conclusion}}
We characterized the hierarchy of monopole operators in $\QEDtcHGN$. Using the state-operator correspondence, we obtained the scaling dimensions of monopoles as a function of their magnetic spin at leading order in $1/\nf$. The spin-Hall mass parameter generated by  the critical fermion self-interaction allows to lower the scaling dimension of monopoles, and this effect is more pronounced  for monopoles with larger magnetic spins.  The minimal scaling dimension identified in Ref.~\cite{dupuis_transition_2019} corresponds to monopoles with a maximal magnetic spin. Monopoles with a vanishing spin have the  largest scaling dimension which is the same as in $\QEDt$.  This hierarchy is natural from the point of view of symmetry as monopoles at the QCP are organized as irreps of $\SU(2) \times \SU(\nf)$ which are labeled by the magnetic spin. These irreps  were obtained explicitly for $q=1/2$ monopoles at the QCP  by finding the branching rules of $\SU(2\nf) \to \SU(2) \times \SU(\nf)$ to reduce the monopole irrep formed in $\QEDt$. This also allowed to obtain the  degeneracies remaining at the QCP following the degeneracy lifting. It would be interesting to improve the analysis for $q>1/2$ that was began in  \ref{app:red}.

With the many recent numerical investigations of $\QEDt$ with Monte Carlo  \cite{karthik_monopole_2018, yang-xu_monte_2018, meng_monte_2019, karthik_numerical_2019} and conformal bootstrap \cite{chester_towards_2016, chester_monopole_2018}, it would be interesting to see similar investigations in the $\QEDtGN$-like models.

\section*{Acknowledgements}

  We thank Jaume Gomis and Sergue\"i Tchoumakov for useful discussions.
{\'E.D.} was funded by an Alexander Graham Bell CGS from NSERC. {W.W.-K.} was funded by a Discovery Grant from NSERC, a Canada Research Chair, a grant from the Fondation Courtois, and a ``\'Etablissement de nouveaux chercheurs et de nouvelles chercheuses universitaires'' grant from FRQNT.

\appendix 
\section{Holonomy of the gauge field \label{app:holonomy}}
Let us also consider the holonomy of the gauge field on the ``thermal''  circle. 
\eqn{
\alpha = \frac{1}{V} \int_{S^2 \times S^1_{\beta}} d^3 x \sqrt{g}\,   a_0\,.
}
In our mean field ansatz, we now leave open the possibility of a non-trivial expectation value of the gauge field 
\eqn{
\moye{a_0} =\beta^{-1} \alpha\,.
}
 Modifying  Eq.~\eqref{eq:free0} accordingly, the determinant operator becomes 
\eqn{
\begin{split}
 \fq   =& - \beta^{-1} \sum_{\sigma = \pm  1} \sum_{n \in \Z} \biggl[  d_{q}  \ln \bigl( \omega_n + \beta^{-1}\alpha - i  \mu' \sigma  + i \sigma M_q \bigr) \\
   &+ \sum_{\ell = q+1}^\infty d_\ell \ln  \Bigl( \lb \omega_n + \beta^{-1}\alpha  - i \mu' \sigma \rb^2 + \veps_\ell^2 \Bigr)  
  \biggr] \,.
  \end{split}
  }
 Taking the sum over the mastubara frequencies,  the same logic in passing from Eq.~\eqref{eq:free0} to Eq.~\eqref{eq:f0}  is used
   \eqn{
  \begin{split}
\fq =& - \beta^{-1}  \biggl[   d_{q} \ln \bigl( 2 \lc \cos (\alpha)+ \cosh(\beta (M_q - \mu') )\rc \bigr)\\ 
&+ \sum_{\ell = q+1}^\infty \sum_{\sigma = \pm 1}  d_\ell \ln \bigl( 2 \lc \cosh(\beta \veps_\ell) + \cosh(\beta ( \sigma \mu' + i \beta^{-1}\alpha)) \rc \bigr) \biggr] \,.
\end{split}  
  }
  The gap equation for $a_0$ is 
  \eqn{
0= \beta \pdv{\fq}{\alpha} =& 
  - d_q \dfrac{\sin(\alpha)}{\cos(\alpha) + \cosh(\beta(M_q - \mu'))}  \\
  &+ \sum_{\ell = q+1}^\infty \sum_{\sigma = \pm 1}  \dfrac{i  \sinh(\beta ( \sigma \mu' + i \beta^{-1}\alpha))}{\cosh(\beta \veps_\ell) + \cosh(\beta ( \sigma \mu' + i \beta^{-1}\alpha))} \,.
  }
This  vanishes for $\alpha = 0, \pi$. Inserting this solution in the determinant operator, we obtain
   \eqn{
  \begin{split}
\fq^\pm =& - \beta^{-1}  \biggl[   d_{q} \ln \bigl( 2 \lc \pm 1 + \cosh(\beta (M_q - \mu') )\rc \bigr) \\
&+ \sum_{\ell = q+1}^\infty  d_\ell \ln \bigl( 2 \lc \cosh(\beta \veps_\ell) \pm \cosh(\beta \mu' ) \rc \bigr) \biggr] \,.
\end{split}  
  }
  With this, we can proceed to solve the remaining gap equations to find the other mean field parameters in these two cases for $M^\pm_q, \mu^\pm_S, \Sigma^\pm$. At leading order in $1/\beta$, the effective action  evaluated at both these saddle points is the same. The partition function obtained by summing over these two saddle points then yields an extra factor of $2$, $Z \approx 2 e^{-\beta \nf \Fq^{(0)}}$ where $\Fq^{(0)}$ was found in the main text. We could directly ignore this as $\ln 2 = \O(\beta^0 \nf^0)$ and  the factor does not contribute to the leading order results
  \eqn{
   Z_\S[A^q]  = \exp{-\beta \nf \Fq^{(0)} + \O(\beta^1 \nf^0, \nf^1 \beta^{0})  }\,.
   } 
   
   However, it is more interesting to remark that this factor is cancelled with a proper normalization.   When computing the scaling dimension $\Delta_{q;\S}$, we should actually use the normalized partition function 
 \eqn{
 \Delta_{q;\S} = -\beta^{-1} \lim_{\beta \to \infty} \ln  \lb \dfrac{Z_\S[A^q]}{Z_0[0]} \rb\,.
 }
 Normally, this vacuum parition function is not mentioned, as the free energy usually vanishes, leaving a trivial normalization factor $Z_0[0]=1$. However, on the ``thermal'' circle, the non-trivial holonomy also contributes to this partition function.  By setting $q=0$ (and $\mu'=0$) in the gap equation for $a_0$, the two solutions for holonomies $\alpha=0, \pi$ remain. In this case, $Z_0[0] = 2$, which exactly cancels the extra factor in $Z_\S[A^q]$.

\section{General spin-Hall mass\label{app:gen-spin_Hall}}
In this section, we find the monopole scaling dimension using a more general ansatz than the one employed in Sec.~\ref{sec:scaling}. We let the auxiliary bosons have different orientations. We keep $\bm \chi$ along $\hat z$ while the spin-Hall mass is oriented more generally as $\bm M_q = M_q \hat n$. The more general determinant operator in Eq.~\eqref{eq:fq} then  becomes
\eqn{
 \fq
  =& - \beta^{-1}\sum_{n \in \Z}
  \Bigg [ 
 d_{q} \ln \det \lc  -i  (\omega - i \mu' \sigma_z)   + \bm M_q \cdot \bm \sigma  \rc \nn \\  
 &+ \sum_{\ell=q+1}^\infty d_\ell \ln \det \lc -i \bm{O}_{q, \ell} \lb \omega - i \mu' \sigma_z  + i \bm{P}_{q, \ell} \rb +  \bm M_q \cdot \bm \sigma  	\rc   \Bigg]   \,,
\label{eq:det}
}
where the matrices $\bm{O}_{q, \ell}$ and  $\bm{P}_{q, \ell}$ are given by \cite{borokhov_topological_2003, dupuis_transition_2019}
\eqn{
\begin{split}
\bm{O}_{q, \ell} = \frac{1}{\ell} \pmatr{-q  & - \sqrt{\ell^2 -q^2} \\ - \sqrt{\ell^2 -q^2} &  q } \,, \\
\bm{P}_{q, \ell} = \frac{\sqrt{\ell^2 -q^2}}{\ell} \pmatr{
 \sqrt{\ell^2 -q^2}	& -q  \\
-q  	& -  \sqrt{\ell^2 -q^2}
}
\end{split}
\,.
}
 The spin-Hall mass $\bm M_q$ may be parameterized by its norm $M_q$ and two angles $(\vartheta, \varphi)$ for its orientation $\hat n$
\eqn{
M_q \hat n = M_q \lb \sin \vartheta \cos \varphi, \sin \vartheta \sin \varphi, \cos \vartheta \rb\,.
}
The determinant operator can be diagonalized in the magnetic spin subspace
\eqn{
\begin{split}
\fq    =&
-\beta^{-1}\Bigg( d_q \ln \lc 2 \lb 1+ \cosh(\beta \sqrt{M_q^2 \sin^2 \vartheta + \lb M_q \cos \vartheta - \mu' \rb^2} )\rb \rc \\
&+  2 \sum_{\sigma = \pm 1} \sum_{\ell=q+1}^\infty d_\ell \ln  \lc 2 \cosh \lb \frac{\beta}{2} \epsilon_{\ell, \vartheta,\sigma} \rb \rc \Bigg)\,,
\end{split}
\label{eq:F_app}
}
where
\eqn{
\epsilon_{\ell, \vartheta, \sigma} = \sqrt{ \lb  \sqrt{ \ell^2 - q^2 + M_q^2 \cos^2 \vartheta} + \sigma \mu' \rb^2 + M_q^2 \sin^2 \vartheta}  \,.
}
For convienience, we may write this as 
\eqn{
\epsilon_{\ell, \vartheta, \sigma}  \equiv \sqrt{ \lb  \veps_{\ell, \vartheta} + \sigma \mu' \rb^2 + M_q^2 \sin^2 \vartheta}\,,
}
where
\eqn{
 \veps_{\ell, \vartheta} = \sqrt{\ell^2 -q^2 + M_q^2 \cos^2 \vartheta}\,.
}
For $\vartheta = 0, \pi$, this corresponds to the eigenvalue $\veps_{\ell}$ defined in the main text.
 
We note that the determinant operator \eqref{eq:F_app} is independent of $\varphi$ which indicates an azimutal symmetry. By setting $\vartheta=0$,  the determinant operator used in the main text \eqref{eq:f0} is retrieved.  As shown in Eq.~\eqref{eq:free}, the full free energy expression is given by  $\Fq^{(0)} = \fq + \frac{\mu'}{\ms} \bigl(\S^2   + \ms^2 \bigr)$. Inserting Eq.~\eqref{eq:F_app} in this last expression, it is found that the free energy is invariant under 
\eqn{
\vartheta \to \pi - \vartheta\,,  \quad \ms \to - \ms \,,
\label{eq:sym}
}
where the last transformation also implies  $\mu' \to -\mu'$ \eqref{eq:mupr}. This means that $\vartheta = \pi$ and $\ms = \S$ is a solution with the same free energy as the $\vartheta = 0$ and $\ms = - s$ solution found in the main text. For this second solution, the spin polarization and the spin-Hall mass are still anti-aligned.

For later convenience, we write explicitly  the free energy. We may already take the large-$\beta$ limit for the non-zero modes as $\ell^2 -q^2$ is order $\O(\beta^0)$ which lets us take $\log \lc 2 \cosh \lb \beta \epsilon_{\ell, \vartheta,\sigma} /2 \rb \rc  \to \beta \epsilon_{\ell, \vartheta,\sigma} / 2$. In this limit, the free energy is given by 
\eqn{
\begin{split}
\Fq^{(0)} =& -\beta^{-1}  d_q \ln \lc 2 \lb 1+ \cosh(\beta \sqrt{M_q^2 \sin^2 \vartheta + \lb M_q \cos \vartheta - \mu' \rb^2} )\rb \rc \\
&+ \frac{\mu'}{\ms} \bigl(\S^2   + \ms^2 \bigr)  -  \sum_{\sigma = \pm 1} \sum_{\ell=q+1}^\infty d_\ell  \epsilon_{\ell, \vartheta,\sigma}\,.
\end{split}
\label{eq:free_gen}
}

\subsection{Gap equations}
The  gap equations are obtained by varying $F_{q;s}^{(0)}$ with respect to  the original saddle point parameters $M_q, \mu_S, P_z, \vartheta$.  In this more general case, the gap equations for $\mu_S$ and $P_z$, 
\eqn{
\thalf \ms \pa_{\mu'} \fq  + \S^2  = 0 \,, \\
\sqrt{\tfrac{\mu_S}{2}} \lb   \pa_{\mu'} \fq  +  2 m_s \rb = 0\,,
} 
can still be combined  to yield the condition  $m_s^2 = s^2$ if $\mu_s \neq 0$. We are then left with a system of three gap equations 
\eqn{
\pa_{M_q}\fq  &= 0\,, \\ 
\pa_{\mu'} \fq -2 m_s &= 0\,,\\
\pa_{\vartheta}\fq  &= 0 \,,
}
where the second equation is the gap equation for $P_z$ divided by $\sqrt{\mu_S/2}$. The explicit expression for the gap equations is given by
\eqn{
 - \lb \frac{ M_q-\mu ' \cos \vartheta}{ M_q \cos \vartheta - \mu'} \rb  C 
- \frac{M_q}{2} \sum_\sigma \sum_\ell d_{\ell} \left( \frac{ \veps_{\ell, \vartheta} + \cos^2 \vartheta \mu' \sigma}{\veps_{\ell, \vartheta} \, \epsilon_{\ell, \vartheta, \sigma}} \right) &= 0  \label{eq:appGap1}\,,
 \\
  C -2 m_s - \half \sum_\sigma \sum_\ell d_{\ell}  \, \sigma  \lb \frac{\veps_{\ell, \vartheta} + \mu' \sigma }{\epsilon_{\ell, \vartheta, \sigma} } \rb  &= 0 \label{eq:appGap2}\,, \\ 
 -  \lb  \frac{\mu ' M_q \sin \vartheta}{M_q \cos \vartheta - \mu' } \rb C
+  \frac{M_q^2 \mu' \sin \vartheta \cos \vartheta}{2} \sum_\sigma \sum_\ell d_{\ell} \frac{ \sigma }{\veps_{\ell, \vartheta}}\frac{1}{\epsilon_{\ell, \vartheta, \sigma}} &= 0 \,,
\label{eq:appGap3}
}
 where $C$ is defined as  
\eqn{
C = \frac{d_q  \lb M_q \cos \vartheta - \mu' \rb \tanh \left(\frac{1}{2} \beta  \sqrt{\mu^{\prime 2}-2 \mu ' M_q \cos \vartheta+M_q^2}\right)}{ \sqrt{\mu^{\prime 2}-2 \mu ' M_q \cos \vartheta+M_q^2}} \label{eq:CC}\,.
}

\subsection{Analytical solutions for $\vartheta \in \{0, \pi / 2, \pi\}$}
We first focus on the cases $\vartheta \in \{0, \pi/2, \pi\}$. For these angles, the sum over non-``zero'' modes in the $\vartheta$ gap equation \eqref{eq:appGap3} does not contribute as it is proportional to $\sin \vartheta \cos \vartheta \to 0$.  The sum in the second gap equation \eqref{eq:appGap2}  also vanishes if we suppose that $\mu' < \veps_{\ell}$ for $\vartheta = 0, \pi$ and $\mu' \sim \O(1/\beta)$ for $\vartheta = \pi/2$. In this case, $(\veps_{\ell, \vartheta} + \mu' \sigma ) /\epsilon_{\ell, \vartheta, \sigma} \to 1$ and the two terms in the sum over $\sigma$ cancel each other.  Later on, we see that this assumption does allow to find a solution. The gap equations then become
\eqn{
 - \lb \frac{ M_q-\mu ' \cos \vartheta}{ M_q \cos \vartheta - \mu'} \rb  C 
- \frac{M_q}{2} \sum_\sigma \sum_\ell d_{\ell} \left( \frac{ \veps_{\ell, \vartheta} + \cos^2 \vartheta \mu' \sigma}{\veps_{\ell, \vartheta} \, \epsilon_{\ell, \vartheta, \sigma}} \right) &= 0 \label{eq:appGap1pr}  \,,
 \\
  C  - 2 m_s &= 0  \label{eq:appGap2pr} \,, \\ 
 -  \lb  \frac{\mu ' M_q \sin \vartheta}{M_q \cos \vartheta - \mu' } \rb C
&= 0 \label{eq:appGap3pr} \,.
}
The gap equation for $\vartheta$  \eqref{eq:appGap3pr} is solved trivially for $\vartheta = 0, \pi$. If $\vartheta = \pi/2$, it requires $M_q = 0$. As for the second equation derived from $\pa F_q^{(0)} / \pa P_z$ \eqref{eq:appGap2pr},    it is solved with the following  $\mu'$
\eqn{
\mu' = M_q \cos \vartheta + \beta^{-1}  \ln \lb \frac{1+2\ms / d_q}{1-2\ms / d_q} \rb\,. \label{eq:mupr_ansatz_2}
}
Using all previous results, the remaining gap equation becomes
\eqn{
- 2 \sgn(\cos \vartheta) \ms -2 M_q \sum_{\ell} d_{\ell}\veps_{\ell}^{-1} =0\,,
}
where $\sgn(0) = 0$. For $\vartheta = \pi/2$, the first term vanishes and the gap equation is solved since it was established that $M_q=0$ for this angle. For $\vartheta = 0, \pi$ and $\ms = - \sgn(\cos \vartheta) \S$, we retrieve the same gap equation as in the main text \eqref{eq:gap_M}. Putting together all the previous results, we can also find back the scaling dimension in Eq.~\eqref{eq:scaling}.  As for $\vartheta  = \pi/2$, the only term that contributes to the free energy \eqref{eq:free} at leading order in $1/\beta$ is the sum over non-``zero'' modes. Thus, the free energy becomes  
\eqn{
\Fq^{(0)}\bigr|_{\vartheta = \pi/2}  = - 2 \sum_\ell d_\ell \veps_{\ell}|_{M_q=0} + \O(1/\beta)\,.
}
This is the free energy we would obtain in $\QEDt$. The scaling dimension for $\vartheta = \pi / 2$ is thus larger than for $\vartheta = 0, \pi$.  This solution should thus be discarded as it is not a global minimum.

\subsection{Numerical study for $0 < \vartheta < \tfrac{\pi}{2}$}
We found two minima of free energy for $\vartheta = 0, \pi$  and a maximum for $\vartheta = \pi/2$.  Other values of $\vartheta$ cannot be solved so simply. This is because the sum over non-``zero'' modes contributions doesn't simplify like it does for $\vartheta \in \{0, \pi/2, \pi\}$. In this case, we resort to solving the gap equations numerically. 

We search the root of the three gap equations (\ref{eq:appGap1} - \ref{eq:appGap3}) yielding the solution for $M_q, \mu'$ and $\vartheta$ (the solution for the polarization is already known, $m_s = - 2 s \sgn(\cos \vartheta)$).  Had we not taken the large-$\beta$ limit for non-``zero'' modes starting from Eq.~\eqref{eq:free_gen}, the sums would also contain a factor $\tanh \lb \beta \epsilon_{\ell, \vartheta, \sigma} /2\rb$. We reinstate this factors in our numerical analysis as we take finite beta. We will set $q=1/2$ and $s = s_{\max}/2$, but the situation is similar for other magnetic charges and magnetic spins. 

\begin{figure}[ht!]
\centering
{
\subfigure[\label{fig:num_sol}]
{\includegraphics[width=0.75\linewidth]{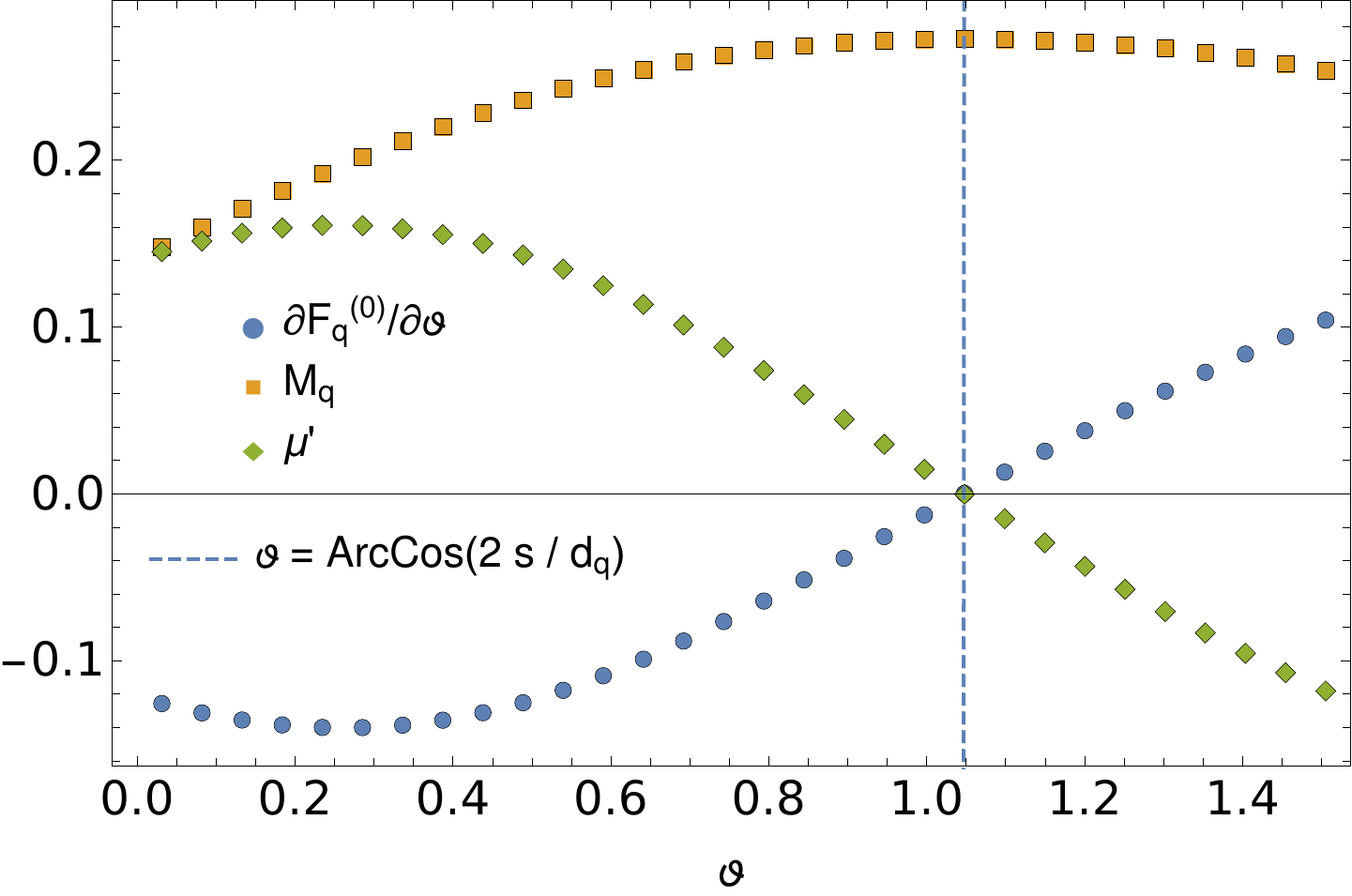}}
\subfigure[\label{fig:num_sol_0}]
{\includegraphics[width=0.75\linewidth]{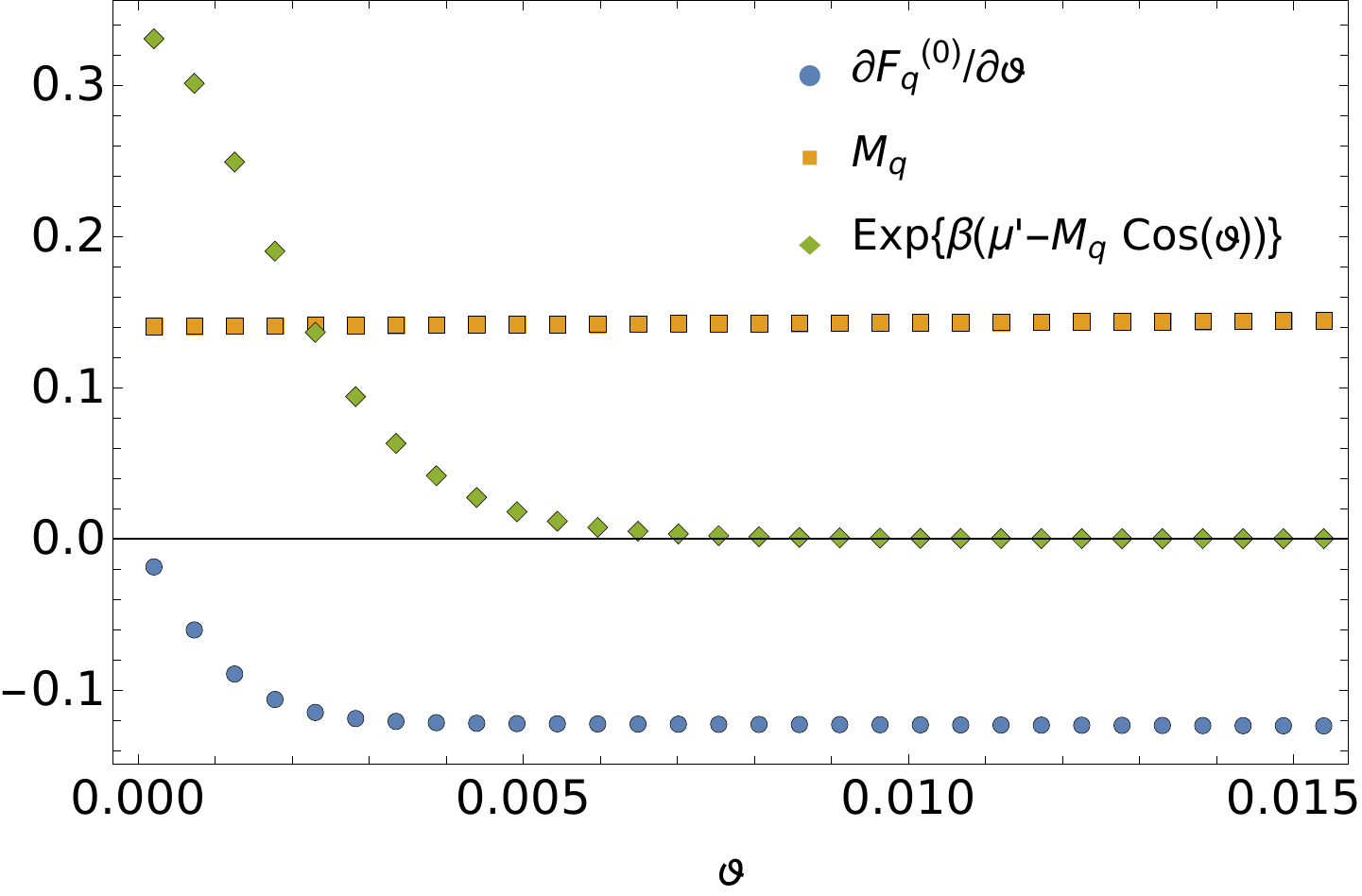}}
}
\caption{Numerical investigation of possible solutions to the gap equations for a) $0 < \vartheta < \pi/2$ and b) near $\vartheta = 0$. We set $\beta = 10^4$. Values of $M_q$ and $\mu'$ are found by solving Eqs.~(\ref{eq:appGap1}-\ref{eq:appGap2}) at fixed $\vartheta$. $\pa F_q^{(0)}/ \pa \vartheta$ which yields the LHS of the remaining gap equation becomes $0$ for $\vartheta = 0$ and $\vartheta = \arccos(2 s / d_q)$. 
}
\end{figure}

To solve the gap equations, we first seek a solution of the first two gap equations with a fixed value for $\vartheta$. We then insert the solutions for $M_q$ and $\mu'$ and the fixed value of $\vartheta$ in the last gap equation to see if it is satisfied, i.e. $\pa F_q^{(0)} / \pa \vartheta$ should vanish. As shown in Fig.~\ref{fig:num_sol}, the only would-be solution in the range $0 < \vartheta < \pi /2$ occurs when $\mu' = 0$. However, this contradicts the assumption that $\mu_S \neq 0$ unless we take $m_s = 0$. The full treatment of the four gap equations, without assuming $\mu_S \neq 0$, yields the same solution. In the case where $\mu'=0$, the gap equation for $\mu_S$ yields the condition $m_s \cos \vartheta = - 2 s^2 / d_q$. As we enforced $m_s = - s$ in the range $0 < \vartheta <  \pi/2$, the numerical solution indeed occurs at $\vartheta = \arccos (2 s / d_q)$, as shown in Fig.~\ref{fig:num_sol}. However, as argued in the main text and further in App.~\ref{app:mu0}, this solution has divergences in the second derivatives of the free energy and should be discarded. 

A closer look at the situation near $\vartheta = 0$ is shown in  Fig.~\ref{fig:num_sol_0}. The solution obtained in the main text is recovered in this limit as $M_q \to  0.14 $  and $\exp{\beta (\mu' - M_q \cos \vartheta)} \to 1/3$ which are the expected values for $q=1/2$ and $s= s_{\max} /2$ (see Fig.~\ref{fig:scaling_mass} and Eq.~\eqref{eq:mupr_ansatz_2} with $m_s = -s$).

\begin{figure}[ht]
\centering
{
{\includegraphics[width=0.75\linewidth]{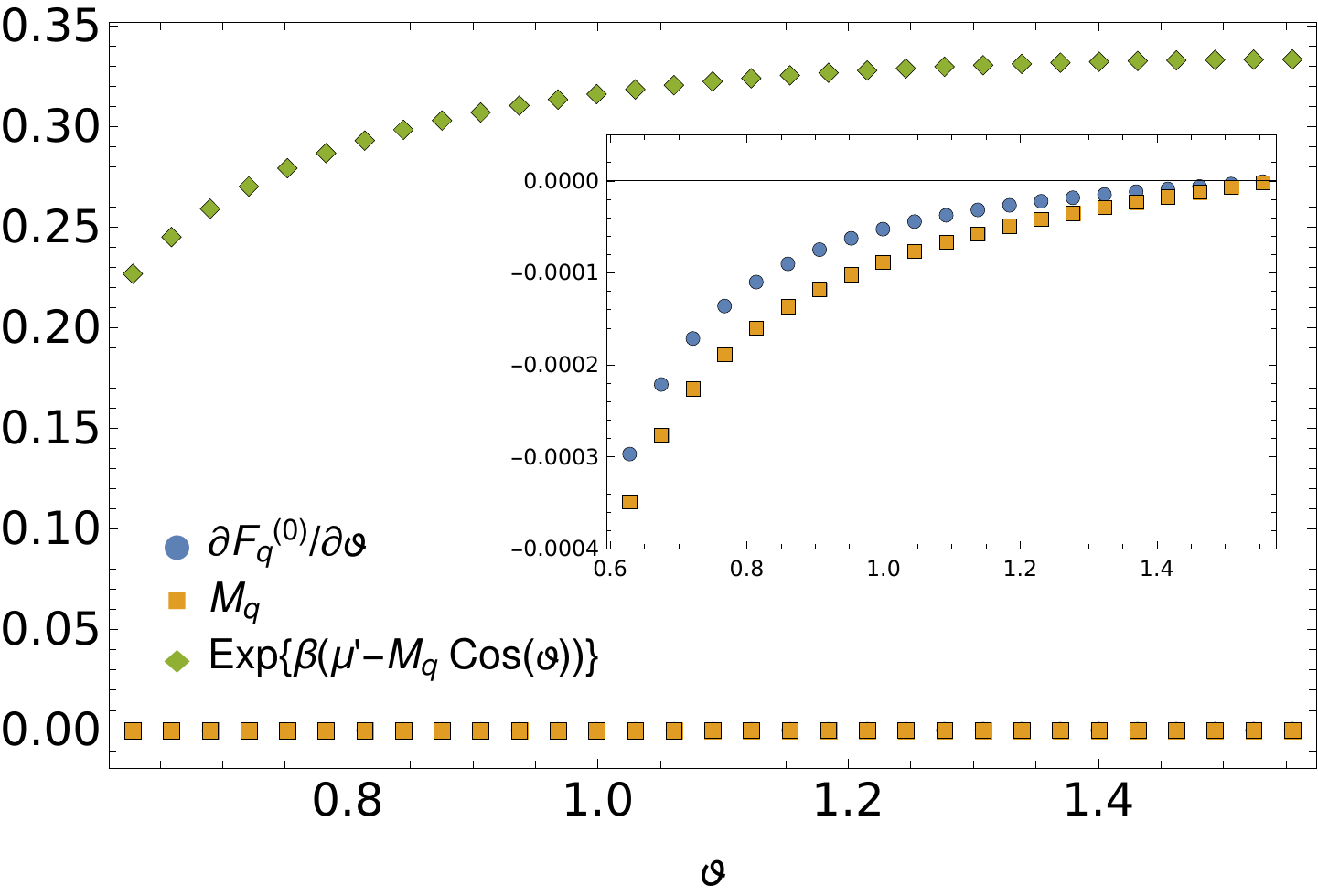}}
}
\caption{Numerical investigation of possible solutions to the gap equations for near $\vartheta = \pi/2$.    Values of $M_q$ and $\mu'$ are found by solving Eqs.~(\ref{eq:appGap1}-\ref{eq:appGap2}) at fixed $\vartheta$.  We set $\beta = 10^4$.  The solution at $\pi/2$ where $\pa F_q^{(0)}/ \pa \vartheta \to 0$ lies in a different region of parameters' space than the one considered in Fig.~\ref{fig:num_sol}.  Here, $M_q \to 0$.
\label{fig:num_sol_pi2}
}
\end{figure}

As for the $\vartheta = \pi/2$ solution, Fig.~\ref{fig:num_sol} near $\pi/2$ is a bad starting point. Seeking solutions in another part of the parameters' space as shown in Fig.~\ref{fig:num_sol_pi2}, we do recover the $ \vartheta = \pi/2 $ solution for which $M_q \to 0$, and, again, $\exp{\beta (\mu' - M_q \cos \vartheta)} \to 1/3$ .

We found no other solutions numerically. This justifies our assumption in the main text where we worked only with $\vartheta = 0$.

\section{Gap equations and $\mu'=0$ \label{app:mu0}}
In the main text, we obtained three gap equations (\ref{eq:gapp1}-\ref{eq:gapp3})
for $M_q$, $\mu_S$ and $P_z$, respectively.  The third one can be solved for $\mu_S = P_z = 0$. This also means that $\mu'=0$. The first gap equation then simply becomes
\eqn{
-d_q  -  2 M_q \sum_{\ell = q+1}^{\infty} d_{\ell} \veps^{-1}_{\ell} = 0\,,
}
which can be written as 
\eqn{
- 2 \S_{\max} -  2 M_q \sum_{\ell = q+1}^{\infty} d_{\ell} \veps^{-1}_{\ell} = 0\,.
}
This has the same form as the gap equation found in the main text \eqref{eq:gap_M} yields as a solution the maximal possible spin-Hall mass. As for the free energy, it also takes the form obtained in the main text for a maximal spin 
\eqn{
F_{q;\S}^{(0)} = - 2  M_q \S_{\max} - 2 \sum_{\ell=q+1} d_{\ell} \veps_{\ell} + \O(\beta^{-1})\,.
}

This solution with $\mu'=0$ thus seems to indicate a smaller scaling dimension that the one proposed in the main text where $\S_{\max} \to \S$.  However, by inspecting the second derivatives of the free energy at this saddle point, divergences are found. The third gap equation can be derived with respect to $\mu_S$ 
\eqn{
\frac{\pa^2 F_{q;\S}^{(0)}}{\pa \mu_S \pa P_z}
=
\frac{1}{2 \sqrt{\mu_S}} \pa_{\mu'}  f_{q;\S} + \frac{1}{2} \ms  \sqrt{\frac{\mu_S}{2}}\pa_{\mu'} \pa_{\mu'} f_{q;\S}\,.
}
Developing this and taking the large-$\beta$ limit, this becomes 
\eqn{
\frac{\pa^2 F_{q;\S}^{(0)}}{\pa \mu_S \pa P_z}
=
\frac{1}{2 \sqrt{\mu_S}}\,,
}
which is singular.  This shows how this solution has a bad behavior and should be  ignored.

\section{Regularized gap equation and scaling dimension \label{app:reg}}
The free energy used to find the monopole scaling dimension has a diverging sum  $\sum_{\ell=q+1}^{\infty} d_\ell \veps_\ell$, where the degeneracy and energy are defined in Eq.~\eqref{eq:en_den}. By obtaining the first two orders in the $1/\ell$ expansion of the summand 
\eqn{
d_\ell \veps_\ell = 2\ell^2 + (M_q^2-q^2) + \O(\ell^{-2})  \equiv d_\ell \veps^{\rm div}_\ell + \O(\ell^{-2})\,,
}
the diverging sum can be rewritten as 
\eqn{
 \sum_{\ell = q +  1}^\infty d_\ell \veps_\ell 
 &=  \sum_{\ell = q +  1}^\infty d_\ell (\veps_\ell - \veps_{\ell}^{\rm div}) +   \sum_{\ell = q +  1}^\infty d_\ell \veps^{\rm div}_\ell  \,.
}
In this expression, the first sum is  convergent 
\eqn{
 \sum_{\ell = q +  1}^\infty d_\ell (\veps_\ell - \veps_{\ell}^{\rm div}) 
 =
 \sum_{\ell = q +  1}^\infty  \lc d_\ell \veps_{\ell} - \half d_\ell^2 - (M_q^2 - q^2) \rc  \,,
}
while the second sum is divergent 
\eqn{
\sum_{\ell = q +  1}^\infty d_\ell \veps^{\rm div}_\ell 
=
2\sum_{\ell = q +  1}^\infty \left [  \ell^{2(1-s)} +   \Bigl( \half - s \Bigr) \lb M_q^2- q^2 \rb \ell^{-2s} \right ] \biggr |_{s=0}\,.
}
This divergent sum may be continued analytically to the Hurwitz zeta function $\sum_{k=0}^{\infty} (k+a)^{-s} = \zeta(s,a)$ \cite{NIST:DLMF}
\eqn{
\sum_{\ell = q +  1}^\infty d_\ell \veps^{\rm div}_\ell 
&=  2 \sum_{\ell=0}^{\infty} (\ell+(q+1))^{2} + \lb M_q^2- q^2 \rb \sum_{\ell=0}^{\infty} (\ell+(q+1))^0  \nn \\
&=  2  \zeta(-2,q+1) + \lb M_q^2- q^2 \rb \zeta(0,q+1)
}
Replacing the zeta functions with their polynomial expressions, we obtain 
\eqn{
\sum_{\ell = q +  1}^\infty d_\ell \veps^{\rm div}_\ell 
=
\half (2q+1) \lb \frac{q(q-2)}{3} - M_q^2 \rb \,.
}
Using these results, we obtain the regularized version of the gap equation \eqref{eq:scaling}
\eqn{
\begin{split}
\frac{\Delta_{q; \S}}{2\nf}  
=&- 2 s  d_q M_q  +(2 q +1) \lb M_q^2 -  \frac{q (q-2)}{3} \rb\\
 &-  \sum_{\ell  = q + 1}^\infty \Big[ 2 d_\ell \veps_\ell -  d_\ell^2 - 2 \lb M_q^2- q^2 \rb \Big]\,,
\end{split}
\label{eq:F0_reg}
}
where the spin-Hall mass  $M_q$ is determined by the regularized version of the gap equation \eqref{eq:gap_M} 
\eqn{
-  2  s d_q +  2 M_q \biggl( 2 q + 1   -    \sum_{\ell = q + 1}^\infty   \left [  d_\ell \veps_\ell^{-1} -   2 \right ] \biggr) = 0\,.
}

  \section{Representation of $q=1/2$ monopoles for $\nf=2$ \label{app:explicit}}
\begin{figure}[ht]
\centering
{
\includegraphics[width=0.5\linewidth]{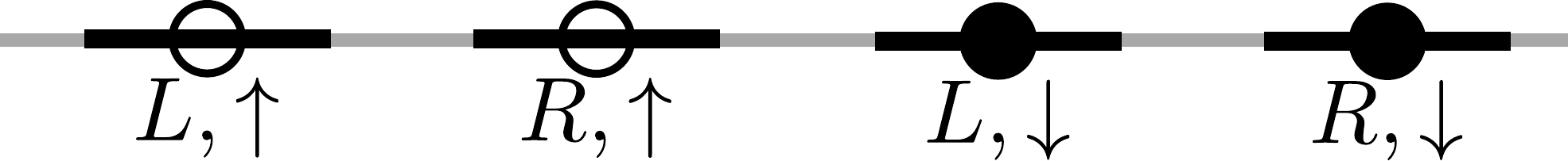}
}
\caption{Schematic representation of the spin down monopole with the fermion zero modes occupation.\label{fig:mon_down}}
\end{figure}

We can generally write a monopole operator as
\eqn{
\man_{I}^\dag = D^\dag_{I} \man_{\rm Bare}^{\dag}\,,
}
where $D^\dag_{I}$ is an operator that creates half of all zero modes available. For $\nf=2$ and $q=1/2$, $D_I^\dag$ corresponds to a zero modes creation operators bilinear $c^\dag A_I(c^{\dag})^{\intercal}$ as  in Eq.~\eqref{eq:QED3_2}. Taking the $z$ axis as the spin quantization axis, monopoles in the helicity basis are written as follows 
\eqn{
D_\d^\dag =  \half  c^\dag \lc \frac{1-\sigma_z}{2} \otimes i \mu_y  \rc (c^{\dag})^{\intercal}
 =
 \dfrac{1}{2} \lb \cb_{\uf \d} \cb_{\df \d} - \cb_{\df \d} \cb_{\uf \d} \rb \,, \\
 D_\u^\dag =   \half  c^\dag \lc \frac{1+\sigma_z}{2} \otimes i \mu_y  \rc (c^{\dag})^{\intercal}
 =
 \dfrac{1}{2} \lb \cb_{\uf \u} \cb_{\df \u} - \cb_{\df \u} \cb_{\uf \u} \rb \,,\\
D_{\u \d}^\dag =  \half  c^\dag \lc \frac{\sigma_x}{\sqrt 2} \otimes i \mu_y  \rc (c^{\dag})^{\intercal} = \frac{1}{2}\lc  \frac{\lb \cb_{\uf \u} \cb_{\df \d} +  \cb_{\uf \d} \cb_{\df \u} \rb}{\sqrt 2} - \lb \uf \leftrightarrow \df \rb \rc\,.
}
For example, the spin down monopole acting on the vacuum $D_\d^\dag  \man_{\rm Bare}^{\dag} \ket{0}$ may schematically be represented as shown in Fig.~\ref{fig:mon_down}.

 We may reorganize the monopoles in the real vector basis used in the main text in Eqs. \eqref{eq:triplets-a} and \eqref{eq:triplets-b}
\eqn{
\pmatr{  \man_\d - \man_\u \\  - i \lb  \man_\d + \man_\u  \rb \\  \sqrt{2} \man_{\u \d}}
&=
\half  c^\dag \lc \sigma_y  \bm{\sigma} \otimes  \mu_y  \rc (c^{\dag})^{\intercal} \man_{\rm Bare}^\dag \,. 
\label{eq:real_vector}
}
The same can be done for monopoles in the nodal triplet by simply exchanging spin and valley indices $\u, \d\, \leftrightarrow \, \uf, \df$.

\subsection{$\SU(2) \times \SU(2)$ generators}
Generators $T_a$ of the $\SU(4)$ algebra may be realized with one zero mode creation operator and one zero mode destruction operator
\eqn{
T_a = c^\dag \O_a c\,.
}
For example, we may identify the raising spin operator, which is part of the $\SU(2)_{\rm Spin} \otimes \SU(2)_{\rm Nodal}$ subalgebra of $\SU(4)$, as 
\eqn{
\vsig_+ = \half  c^\dag \lb (\sigma_x + i \sigma_y) \otimes \Id \rb  c  = \cb_{\uf \u} \c_{\uf \d}  +  \cb_{\df \u} \c_{\df \d}  \,.
}
Under the action of this operator, the spin down monopole transforms in other monopoles in spin triplet while the monopoles in the nodal triplet are annhiliated 
\eqn{
\vsig_+ \vsig_+ \man_{\d} = \sqrt 2 \vsig_+ \man_{\u \d} = 2 \man_\u \,,\\
\vsig_+ \man_\df = \vsig_+ \man_{\uf \df} =  \vsig_+ \man_{\uf} =0\,. 
}
The factors involved are the usual total magnetic spin eigenvalue $\sqrt{\S(\S+1)}$. One may also define the lowering spin and   the azimutal spin number operators, respectively given by
\eqn{
\vsig_- = \vsig_+^\dagger\,, \quad \vsig_z = \half  c^\dag \lb \sigma_z \otimes \Id \rb  c   =\half ( c_{L \u}^\dag  c_{L \u} - c_{L \d}^\dag  c_{L \d}  +  \lb \uf \leftrightarrow \df \rb )\,.
}
It is simple to show that these operators obey the $\SU(2)$ algebra commutation relations $[\vsig_+, \vsig_-] = 2 \vsig_z$. Again, analogous valley operators $\bm{\vmu}$ may be constructed by the exchange $\u, \d\, \leftrightarrow \, \uf, \df$. The action of these $\SU(2) \times \SU(2)$ generators $\{ \vsig_-, \vsig_z, \vsig_+, \vmu_-, \vmu_z, \vmu_+\}$ on the monopole operators is shown in Table~\ref{tab:triplets}.  It is readily seen that the monopole spin and nodal triplets form a reducible representation of $\SU(2) \times \SU(2)$.  
\begin{table}[h]
\caption{The monopoles form a reducible representation $(\bm 3, \bm 1) \oplus (\bm 1,  \bm 3)$ of $\SU(2) \times \SU(2)$ (helicity basis). \label{tab:triplets}}
\centering
\begin{tabular}{C||CCC|CCC}
								& \man^\d 		&\man^\u 		& \man^{\u \d} 		& \man^{\df} 		& \man^{\uf} 	& \man^{\uf \df}\\ \hline 
 \frac{1}{\sqrt 2}\vsig_{+} 	& \man^{\u \d} 	& 0 			& \man^\u 			& 0 				& 0 			& 0 \\
 \vsig_z 						& -\man^\d 		& \man^\u 		& 0 				& 0 				& 0 			& 0 \\
 \frac{1}{\sqrt 2}\vsig_{-} 	& 0 			& \man^{\u \d} 	& \man^\d 			& 0 				& 0				& 0 \\ 
 \frac{1}{\sqrt 2} \vmu_+ 		& 0 			& 0 			& 0 				& \man^{\uf \df} 	& 0 			& \man^{\uf} \\
 \vmu_z 						& 0 			& 0 			& 0 				& -\man^{\df} 		& \man^{\uf} 	& 0 \\
 \frac{1}{\sqrt 2} \vmu_+ 		& 0 			& 0 			& 0 				& 0 				&\man^{\uf \df}	& \man^{\df} \\
\end{tabular}
\end{table}

\subsection{Rotation of spin monopoles}
We may explicitly show the rotation of spin monopoles mentioned in the main text. To do so, we first rexpress monopoles in the real vector representation \eqref{eq:real_vector} in the following basis
\eqn{
\pmatr{1\\0\\0} = \frac{1}{\sqrt 2} \lb \man_\d - \man_\u \rb\,, \quad 
\pmatr{0\\1\\0} = \frac{-i}{\sqrt 2} \lb \man_\d + \man_\u \rb\,, \quad  
\pmatr{0\\0\\ 1} = \man_{\u \d}\,.
}
In this real basis, the angular momentum operators take the form $(J^I)_{jk} = \epsilon_{Ijk}$. Considering an $\SU(2)$ transformation of the fermions 
\eqn{
\Psi &\to e^{i \bm \theta \cdot  \bm \sigma} \Psi\,,
}
we can find the transformation of the vector representations. The fermion bilinears and the monopoles transform a bit differently 
\eqn{
\Psib \sigma_i \Psi &\to \Psib e^{-i \bm \vth \cdot  \bm \sigma}  \sigma_i e^{i \bm \vth \cdot  \bm \sigma} \Psi =  R_{ij} \Psib \sigma_j \Psi\,,\\
c^\dag (i \sigma_y \sigma_i) (c^{\dag})^{\intercal} &\to 
c^\dag  (i \sigma_y) e^{i \bm \vth \cdot  \bm \sigma^\intercal} \sigma_i e^{-i \bm \vth \cdot  \bm \sigma^\intercal} (c^{\dag})^{\intercal} 
=
\tilde{R}_{ij} c^\dag (i \sigma_y \sigma_j) (c^{\dag})^{\intercal} \,,
} 
where $\tilde{R}_{ij} = R_{ij}|_{\vartheta_{x,z} \to - \vartheta_{x,z}}$ and we used the fact that $(\bm \vth \cdot \bm \sigma) \sigma_y = - \sigma_y (\bm \vth \cdot \bm \sigma^\intercal)$. We will perform these rotations explicitly on the spin-Hall mass $\Psib \sigma_z \Psi$ and the spin down monopole $\man_\d$. In the real vector basis, they are expressed as 
\eqn{
\Psib \sigma_z \Psi = \pmatr{0\\0\\1}\,, \quad \man_\d =\frac{1}{\sqrt{2}} \pmatr{- i \\ 1 \\ 0}\,.
}
We can compare how these vectors rotate along the $y$ axis. In this case, they are transformed by the same rotation matrix
\eqn{
R_{\vth_y} = \tilde R_{\vth_y} = \pmatr{\cos \vth_y & 0 &  \sin \vth_y \\ 0 & 1 & 0 \\ - \sin \vth_y &0 & \cos \vth_y}\,.
}

We first consider a rotation  $\vth_y = \pi/2$ that sends the quantization axis in the $x-y$ plane. The spin-Hall mass is rotated to $\Psib \sigma_x \Psi$
\eqn{
R_{\vth_y = \pi/2} \lb \Psib \sigma_z \Psi \rb 
= \pmatr{0 & 0 &  1 \\ 0 & 1 & 0 \\ - 1 &0 & 0} \pmatr{0\\0\\1} = \pmatr{1\\0\\0} = \Psib \sigma_x \Psi\,.
}
As for  the spin down monopole, it is rotated to a combination of all the monopoles (in the helicity basis) of the spin triplet 
\eqn{
R_{\vth_y = \pi/2} \man_\d
= \frac{1}{\sqrt 2}\pmatr{0 & 0 &  1 \\ 0 & 1 & 0 \\ - 1 &0 & 0} \pmatr{-i\\1\\0}  
= \frac{1}{\sqrt 2} \pmatr{0\\1\\i} 
&= \frac{1}{2} \lb \man_\d + \man_\u \rb - \frac{1}{\sqrt 2} \man_{\u \d} \,.
\label{eq:rot_mon}
}
This is an eigenstate of the angular momentum oriented along $\hat x$, $(J_x)_{ij} = \epsilon_{1ij}$. Indeed, $J_x  R_{\pi/2} \man_\d =  -  R_{\pi/2} \man_\d$. This monopole operator now creates a state with $S_x=-1$ which minimizes the rotated spin-Hall mass $\Psib \sigma_x \Psi$.

 The $\vth_y = \pi$ rotation  is more simple. In this case
\eqn{
R_{\vth_y = \pi} \Psib \sigma_z \Psi =  \pmatr{-1 & 0 & 0 \\ 0 & 1 & 0 \\ 0 &0 & -1} \pmatr{0\\0\\1} =  - \Psib \sigma_z \Psi\,, \\
R_{\vth_y = \pi} \man_\d = \pmatr{-1 & 0 & 0 \\ 0 & 1 & 0 \\ 0 &0 & -1}  \frac{1}{\sqrt{2}} \pmatr{- i \\ 1\\0} =  \frac{1}{\sqrt{2}} \pmatr{ i \\ 1\\0}  = \man_\u\,.
}
Thus, the mass is shifted $\Psib \sigma_z \Psi \to - \Psib \sigma_z \Psi $ while the spin down monopole is rotated to the spin up monopole $\man_\d \to \man_\u$, just as expected. This situation was shown in Fig.~\ref{fig:spin_flip}

\subsection{Computing the spin-Hall energy}
We may also explicitly compute the energy of the spin-Hall mass term for the state $\ket{\psi_I} = \man_I^\dag \ket{0}$ in the three situations considered above. i) For the spin down monopole, the state is  $\man^\dagger_\d \ket{0}  = \ket{\d \d} \equiv \ket{\S=1, \ms = -1}$. The energy of the spin-Hall mass term oriented along $\hat z$ for this state is 
\eqn{
M_q (\sigma_z \otimes \Id + \Id \otimes \sigma_z) \ket{\d \d} = - 2 M_q \ket{\d \d}\,.
}
ii) After the $\pi/2$ rotation, the related state may be written using \eqref{eq:rot_mon} 
\eqn{
R_{\vth_y = \pi/2} \man^\dag_\d \ket{0}  = \frac{\ket{\d \d} + \ket{\u \u} - \ket{\d \u} - \ket{\u \d}}{2}\,,
}
while the spin-Hall mass becomes oriented along $\hat x$. Its energy contribution to this state remains $-2 M_q$
\eqn{
M_q &(\sigma_x \otimes \Id + \Id \otimes \sigma_x) 
\Bigl( R_{\vth_y = \pi/2} \man^\dag_\d \ket{0} \Bigr)\\
=&
M_q
\frac{\lc
\lb \ket{\u \d} + \ket{\d \u} - \ket{\u \u} - \ket{\d \d} \rb +
 \lb \ket{\d \u} + \ket{\u \d} - \ket{\d \d} - \ket{\u \u} \rb
\rc}{2} \nn 
 \\
=& - 2 M_q \Bigl( R_{\vth_y = \pi/2} \man^\dag_\d \ket{0} \Bigr)\,.
}
iii) Finally, after the $\pi$ rotation, the state and the action of the spin-Hall mass are as expected 
\eqn{
R_{\vth_y = \pi} \man^\dag_\u \ket{0}  =\ket{\u \u}\,, \quad   - M_q (\sigma_z \otimes \Id + \Id \otimes \sigma_z) \ket{\u \u} = - 2 M_q \ket{\u \u}\,.
}

\section{General reduction problem \label{app:red}}     
 \subsection{A relation with the permutation group}
    There is an ambiguity when discussing the reduction $\SU(2\nf) \to \SU(2) \times \SU(\nf)$ as the ${\SU(2) \times \SU(\nf)}$ subgroup of $\SU(2\nf)$ is not uniquely defined. The subgroup  ${\SU(2) \times \SU(\nf)}$ describing the $\QEDtcHGN$ model is not the same as the one obtained by the chain ${\SU(2\nf) \supset \SU(\nf) \times \SU(\nf) \supset  \SU(2) \times \SU(\nf)}$. The two subgroups have different branching rules. In this regard, it is useful to consider a more general reduction problem
\eqn{
\SU(M N) \to \SU(M) \times \SU(N)\,, \label{eq:gen-symmetry_breaking}
}
where $M$ and $N$ are integers, and our case corresponds to $M=2$ and $N=\nf$. 

We note in passing that this embedding in a larger symmetry group is also used to find the multiplicity of flux operators that transform as Lorentz scalars. For magnetic charges larger than the minimum $q=1/2$, the number of zero modes dressing a monopole \eqref{eq:mon_gen} is $4 |q| \nf$ rather than $2\nf$. However, starting with requirement that half of zero modes should be filled, it is natural to first build a rank-$2 |q| \nf$ completely antisymmetric irrep of $\SU(4 |q| \nf)$. As an intermediate step to reduce this to the real symmetry group $\SU(2) \times \SU(2 \nf)$, one can consider  the reduction $\SU(4 |q| \nf) \to \SU(2 |q|) \times \SU(2 \nf)$ \cite{borokhov_topological_2003, dyer_monopole_2013, chester_monopole_2018} which corresponds to setting $M=2 |q|$ and $N = 2\nf$.

The generators of $\SU(M N)$  may be parameterized by taking Kronecker products of the $\SU(M)$ and $\SU(N)$ generators
\eqn{
\SU(MN) : \quad  T^{MN}_{a} \in \{T^M_{a'} \otimes \Id,\, \Id \otimes T^{N}_{a''},\, T^M_{a'} \otimes  T^N_{a''}  \}\,,
 } 
 where $a \in \{1,\dots, (MN)^2-1 \}$, $a' \in \{1,\dots, (M)^2-1\}$ and $a'' \in \{1,\dots, (N)^2-1\}$. The subgroup $\SU(M) \times \SU(N)$ is completely defined by the set of unbroken generators, which are 
\eqn{
\SU(M) \times \SU(N) : \quad  \{T^M_{a'} \otimes \Id,\, \Id \otimes T^{N}_{a''}\}\,. \label{eq:subgroup}
} 
 This represents the reduced symmetry group of the $\QEDtcHGN$ QCP. While the spin-Hall term  $\Psib \bm \sigma \Psi$  transforms as a vector under the first generators  $T^M_{a'} \otimes \Id$, the $\cHGN$ interaction term $(\Psib \bm \sigma \Psi)^2$ is a scalar built from this vector and thus transforms trivially. Once the unbroken generators are specified, it follows that  the fundamental representation of $\SU(MN)$ simply reduces as the fundamental of $\SU(M)$ and $\SU(N)$
 \eqn{
 \yng(1) \to \lb  \yng(1),  \yng(1)\rb \,. \label{eq:decomp_fundamental}
 }
 This may be understood with rank-1 tensors as a $\SU(MN)$ index $\alpha \in \{1, \dots, MN \}$ can be decomposed in one $\SU(M)$ index  $\sigma \in \{1, \dots , M\}$ and one $\SU(N)$ index $v \in \{1, \dots , N\}$
 \eqn{
 h_\alpha = h_{(\sigma v)} \equiv f_\sigma g_v\,.\label{eq:decomp_rank1}
 }
 
 We may also define rank-$2$ tensors in this manner,  $h_{\alpha \beta}  = f_{\sigma \eta} g_{v w}$. Defining symmetrized and anti-symmetrized tensor respectively as ${t_{\{i,j\}} = \half \lb t_{ij} + t_{ji} \rb}$ and   ${\half t_{[i,j]} = \half \lb t_{ij} - t_{ji} \rb}$, we may write explicitly the RHS of \eqref{eq:decomp_4} as a tensor whose simplified form corresponds to the LHS   of \eqref{eq:decomp_4} 
\eqn{
f_{\{\sigma,\eta\}} g_{[v,w]} + f_{[\sigma,\eta]} g_{\{v,w\}} 
= \half \lb f_{\sigma \eta} g_{vw} - f_{\eta \sigma} g_{wv} \rb = h_{[\alpha, \beta]}\,.\label{eq:decomp0}
}
This method is difficult to implement as the rank of the tensors is increased, i.e.\ the number of boxes in the Young diagrams is increased. However, this  specific example has the merit of showcasing an important property of the reduction of interest \eqref{eq:gen-symmetry_breaking}:  The decomposition is  independent of $M$ and $N$. Indeed, the only information needed to show  Eq.~\eqref{eq:decomp0} is the  $\SU(MN)$ index decomposition \eqref{eq:decomp_rank1}  which characterizes the reduction studied (\ref{eq:gen-symmetry_breaking} , \ref{eq:subgroup}).  This can be understood by inspecting the relation between the reduction  $\SU(MN) \to \SU(M) \times \SU(N)$ and the finite permutation group of $f$ objects $S_f$ \cite{itzykson_unitary_1966}. 

This independence also manifests by the fact that general reductions of  $\SU(MN) \to \SU(M) \times \SU(N)$ may be built by taking products of the  fundamental representation  \cite{itzykson_unitary_1966}. For example, by taking the product of the  fundamental representation reduction  \eqref{eq:decomp_fundamental} with itself
\eqn{
\yngc \yng(1) \otimes \yng(1) \to \lb \yng(1)\otimes \yng(1), \yng(1) \otimes \yng(1) \rb\,, \label{eq:red2}
}
one finds can write the reduction for $ (1) \otimes (1) = (2) \oplus (1^2)$. By also developing the RHS, the reduction  in Eq.~\eqref{eq:decomp_4} can be found.  Given that this is the reduction of a reducible representation $(2) \oplus (1^2)$, the associations  made between the LHS and RHS of \eqref{eq:red2} are not straightforward.   However, this shows again that this decomposition is independent of the indices $M$ and $N$ as they were not involved in the computation\footnote{This general procedure may however include diagrams with too many rows and that are not allowed in either $\SU(M)$ or $\SU(N)$. These must simply be removed.} With this procedure, it is clear that a Young tableau in the $\SU(MN)$ side  with $f$ boxes generally reduces to Young tableaux of $\SU(M) \times \SU(N)$ with the same number $f$ of boxes. Those properties hint at a the importance of the permutation group of $f$ objects $S_f$ in this reduction problem \eqref{eq:gen-symmetry_breaking}.

If we define  $\lyt_\lambda$  as a certain diagram $\lambda$ among Young tableaux with $f$ boxes, then the general reduction \eqref{eq:gen-symmetry_breaking} may be written as
\eqn{
\lyt_\lambda \to \bigoplus_{\nu, \rho} c_{\nu \rho \lambda} (\lyt_\nu, \lyt_\rho)\,, \label{eq:CFP}
}
where  $c_{\nu \rho \lambda}$ is a coefficient of fractional parentage (CFP). 
The relation with $S_f$ manifests itself through these CFPs: For each Young tableau $\lyt_\lambda$, there is a corresponding irrep $D_\lambda$ of $S_f$, and the CFP of $\lyt_\lambda$  in $(\lyt_\nu, \lyt_\rho)$ is the Clebsch-Gordan coefficient (CGC) of $D_\lambda$  in the decomposition of the direct product representation  $D_\nu \otimes  D_\rho$~\cite{itzykson_unitary_1966}
\eqn{
 D_\nu \otimes D_\rho = \bigoplus_{\lambda} c_{\nu \rho \lambda} D_\lambda \,.
 \label{eq:direct}
}
Since the characters $\chi$ of a direct product is the product of the characters 
\eqn{
\chi_{\nu \otimes \rho} = \chi_{\nu} \chi_\rho\,, \label{eq:characters}
}
these CGCs are simply given by
 \eqn{
c_{\nu \rho \lambda} = \frac{1}{\dim(S_f)} \sum_r p_r \chi_\nu(\mathcal{C}_r) \chi_\rho(\mathcal{C}_r) \chi_\lambda^*(\mathcal{C}_r)\,, \label{eq:CGC}
}
where $\chi_\nu(\mathcal{C}_r)$ is the character of $D_\nu$ and $p_r$ is the number of group elements in a conjugacy class $\mathcal{C}_r$.

\subsection{Clebsch-Gordan coefficients of the sign irrep}
We seek to reduce the rank-$\nf$ completely antisymmetric irrep of $\SU(2 \nf)$ which is a Young diagram with $f=\nf$ boxes. More precisely, this irrep corresponds to a single column of $\nf$ boxes \eqref{eq:sign}. This diagram corresponds to the sign irrep of $S_{\nf}$ that we note $D_{\rm sign}$. To find out how the rank-$\nf$ completely antisymmetric irrep of $\SU(2 \nf)$ reduces, we must therefore find the CGCs  $c_{\nu \rho \, \text{sign}}$ that give the contribution of  $D_{\rm sign}$ in the reduction of $ D_\nu \otimes D_\rho$.

Let $D_\nu$ be an irrep of $S_{\nf}$ and $D_{\tilde \nu}$ its conjugate.  Diagrammatically, these Young tableaux are the transposed of each other (e.g.\ the two diagrams in the RHS of Eq.~\ref{eq:reduction}). As the conjugate irrep $D_{\tilde{\nu}}$ is simply the direct product of the irrep $\nu$ with the sign irrep,  ${D_{\tilde{\nu}} = D_{\rm sign} \otimes D_\nu}$, its character is simply the product of their characters \eqref{eq:characters}, i.e.\ ${\chi_{\tilde{\nu}} = \chi_{\rm sign} \chi_{\nu}}$. Using this, the  product of characters that define $c_{\nu \rho \, \text{sign}}$ \eqref{eq:CGC} may be rewritten as
\eqn{
\chi_{\nu}(\mathcal{C}_r) \chi_{\rho}(\mathcal{C}_r) \chi^*_{\rm sign}(\mathcal{C}_r) 
=
\chi_{\nu}(\mathcal{C}_r)  \chi_{\rm trivial}(\mathcal{C}_r)
\chi^*_{\tilde{\rho}}(\mathcal{C}_r)\,. 
\label{eq:res}
}
To obtain this, we also used that, for any equivalence class $\mathcal{C}_r$,  the characters of the sign irrep are $\pm 1$, implying that \  ${|\chi_{\rm sign}|^2=1 = \chi_{\rm trivial}}$, and more generally that the characters of the permutation group are real, \ ${\chi_{\tilde{\rho}} = \chi^*_{\tilde{\rho}}}$. This relation \eqref{eq:res}   implies that  $c_{\nu \rho \,  \rm sign}  
=  c_{\nu \, \text{trivial}\, \tilde{\rho}}$. The latter CGC gives the decomposition of $D_{\tilde{\rho}}$ in $D_\nu \otimes D_{\rm trivial} = D_\nu$. Obviously, the coefficient is only non-``zero'' if $\tilde{\rho} = \nu$, which in turn means that ${c_{\nu \rho \,  \rm sign}  = \delta_{\rho, \tilde{\nu}}}$.  More explicitly, this means that only pairs of conjugate irreps have a contribution from the sign irrep in their direct product decomposition  
\eqn{
D _{\nu} \otimes D _{\rho} = \delta_{\rho, \tilde{\nu}} D _{\rm sign}  \oplus \dots \,. \label{eq:CGC_final}
}
This result \eqref{eq:CGC_final} implies that the CFPs  \eqref{eq:CFP} are equal to  one for every pair of irrep  and its conjugate
\eqn{
\lyt_\lambda \to \bigoplus_{\nu}  (\lyt_\nu, \lyt_{\tilde{\nu}})\,. \label{eq:CFP2}
}
 In the case of interest where  $M=2$ and $N =\nf$, we must exclude irreps of $\SU(2) \times \SU(\nf)$ where tableaux in the $\SU(2)$ side have more than two rows since they are not include in $\SU(2)$. In the end, this corresponds exactly to the reduction we announced in Eq.~\eqref{eq:reduction}.  In  \ref{app:dimensions}, we explicitly check that the dimensions of these diagrams match.

\subsection{Dimensions of the reduced irreps \label{app:dimensions}}
We explicitly check the dimensions of irreps in the reduction of monopoles shown in the main text \eqref{eq:reduction}
   \eqn{ 
\lb \signn \rb_{\SU(2N)}
&\to
\bigoplus_{b=0}^{\floor{\nf/2}} \lc \;  \Biggl( \irrepm \Biggr)_{\SU(2)} , \;  \Biggl( \irrepmpr \Biggr)_{\SU(N)} \; \rc \,.
}
\subsubsection*{Dimensions of the $\SU(2) \times \SU(N)$ irreps }
The $\SU(2)$ subdiagram simply has dimension $\nf-2\b+1$
\eqn{
\dim  \; \Biggl( \irrepm \Biggr)_{\SU(2)}  = \dim \Biggl( \uds{N - 2\b}{\underbrace{
\begin{array}{c}
\yt{\wp& \none[...] & }
\end{array}
}} \Biggr)_{\SU(2)} = \nf-2\b+1\,,
}
where we removed columns of two boxes which transform trivially in $\SU(2)$. 
 The $\SU(\nf)$ diagrams requires more work. The dimension $F/H$ of such a diagram is found using the factor over hooks rule \cite{georgi_lie_1999}. The $\SU(\nf)$ Young tableau's boxes can be labeled as 
\eqn{
\ytableausetup{boxsize=2.95em}
{\scriptstyle \nf  - \b}
\left \{
\begin{array}{l}
\left .
\begin{array}{l}
\yt{\scriptstyle \nf &\scriptstyle \nf+1 \\ \none[\svdots] & \none[\svdots] \\ \scriptstyle \nf+1-\b &\scriptstyle \nf+2-\b  \\  }
\end{array}
\right \} {\scriptstyle \b}
\\[-0.1em]
\left .
\begin{array}{l}
\yt{ \scriptstyle \nf -\b \\\none[\svdots]\\ \scriptstyle \b}
\end{array}
\right \} {\scriptstyle \nf- 2\b}
\end{array}
\right .\,.
}
Then, the numerator $F$ is the product of all the box labels in the Young tableau above. It can be decomposed as the product of labels in the first column and in the second 
\eqn{
F=F_L \times F_R =\dfrac{(\nf)!}{\b!} \times \dfrac{(\nf+1)!}{(\nf+1-\b)!}\,.
} 
As for the denominator $H$, the length of the hooks for each box must be multiplied. It is useful to decompose it in three sections
\eqn{
H = h_A \times h_B \times h_C = (\nf-2\b)! \times \frac{\lb \nf-\b+1 \rb!}{\lb \nf-2\b+1 \rb!} \times  \b!  \,, 
\quad 
\ytableausetup{smalltableaux}
{\scriptstyle \nf  -\b}
\left \{
\begin{array}{l}
\left .
\begin{array}{l}
\yt{B & C \\ \none[\svdots] & \none[\svdots] \\ B & C \\  }
\end{array}
\right \} {\scriptstyle \b}
\\[-0.1em]
\left .
\begin{array}{l}
\yt{ A \\\none[\svdots]\\ A}
\end{array}
\right \} {\scriptstyle \nf- 2\b}
\end{array}
\right .
\,.
} 
Putting all together, this becomes 
\eqn{
\begin{split}
\dim \Biggl( \irrepmpr \Biggr)_{\SU(N)}
&= \dfrac{\dfrac{(\nf)!}{\b!} \times \dfrac{(\nf+1)!}{(\nf-\b+1)!}}{(\nf-2\b)! \times \frac{\lb \nf-\b+1 \rb!}{\lb \nf-2\b+1 \rb!} \times  \b! } \\
&= \frac{\lb \nf-2\b + 1 \rb \lb \nf+1 \rb}{\lb \nf-\b+1 \rb^2} \times \binom{\nf}{\b}^2\,.
\end{split}
}
The total dimension of the $\SU(2) \times \SU(N)$ irrep is 
\eqn{
\begin{split}
\dim  \Biggl(  \irrepm \Biggr)_{\SU(2)} \times \dim \Biggl( \irrepmpr \Biggr)_{\SU(N)}\\
  = 
 \frac{\lb \nf-2\b + 1 \rb^2 \lb \nf+1 \rb}{\lb \nf-\b+1 \rb^2} \times \binom{\nf}{\b}^2\,.
 \end{split}
}
Replacing $\b$ with $N/2 -s$, it can be verified that this result indeed corresponds to Eq.~\eqref{eq:dim_sub_irreps}.

\subsubsection*{Total dimension}
The dimension of the $\SU(2N)$ irrep should match the dimension of the $\SU(2) \times \SU(N)$ representation to which it is reduced 
\eqn{
\begin{split}
&\dim \Biggl( \signn  \Biggr)_{\SU(2 \nf)} = \\
&\sum_{\b=0}^{\floor{\nf/2}} \dim \Biggl(\irrepm \Biggr)_{\SU(2)} \times \dim \Biggl( \irrepmpr\Biggr)_{\SU(\nf)} \,.
\end{split}
\label{eq:dim_reduction}
}

Summing over the index $\b$, we obtain the RHS of Eq.~\eqref{eq:dim_reduction}.  The LHS of Eq.~\eqref{eq:dim_reduction} is the dimension of the rank-$\nf$ antisymmetric irrep of $\SU(2\nf)$ which is $\binom{2\nf}{\nf}$. It is left to show  that 
\eqn{
\binom{2\nf}{\nf} &=  \sum_{\b=0}^{\floor{\nf/2}}   \frac{\lb \nf-2\b + 1 \rb^2 \lb \nf+1 \rb}{\lb \nf-\b+1 \rb^2} \times \binom{\nf}{\b}^2\,.\label{eq:dimension_reduction}
}
The  RHS of Eq.~\eqref{eq:dimension_reduction} with even $\nf = 2 x$ with $x \in \Z^+$  can be simplified to the expected result
\eqn{
\begin{split}
\sum_{\b=0}^{x}   \frac{\lb \nf-2\b + 1 \rb^2 \lb \nf+1 \rb}{\lb \nf-\b+1 \rb^2} \times \binom{\nf}{\b}^2\evalat_{\nf=2x} &=
\frac{\pi^{\frac{3}{2}} \left(2 \, x\right)!^{2} \left(x - \frac{1}{4}\right)! \left(x - \frac{3}{4}\right)!}{ \left(x - \frac{1}{2}\right)!^{3} x!^{3} \Gamma\left(\frac{3}{4}\right) \Gamma\left(\frac{1}{4}\right)}\\
 &=
\dfrac{(4x)!}{(2x)!^2} = \binom{2\nf}{\nf}\evalat_{\nf=2x}\,.
\end{split}
}
For odd $\nf$, we could not show \eqref{eq:dimension_reduction} analytically, but it was confirmed numerically up to $\nf=10^4+1$.

\subsection{Monopoles with larger magnetic charges}
The results in the present section directly apply to the study of monopole with a magnetic charge larger than the minimum $q=1/2$. As we take a magnetic charge higher than the minimum, $q>1/2$, more fermion zero modes become available. However, filling half of those $2d_q \nf$ zero modes will generate monopole with non vanishing Lorentz spins. This is still a good starting point:  We define $4 \pi q$  flux operators with vanishing fermion number by generalizing Eq.~\eqref{eq:mon_gen}
\eqn{
\Phi^\dag_{I_1 \dots I_N} = c^\dag_{I_1} \dots c^{\dag}_{I_{N}} \man^\dag_{\rm Bare}\,, \quad I_i \in \{1, 2, \dots, 2 d_q \nf\}\,.  \label{eq:mon_gen_q}
}
These flux operators form the rank-$d_q \nf$ completely antisymmetric tensor of $\SU(2 d_q \nf)$. In $\QEDt$,  it reduces in irreps of the symmetry group  $\SU(2)_{\rm rot} \times \SU(2 \nf)$. Consider the following chain 
\eqn{
\SU(2 d_q \nf) \supset \SU(d_q) \times \SU(2 \nf) \supset \SU(2) \times \SU(2 \nf)\,.
}
Then, a first step in reducing the antisymmetric irrep of $\SU(2 d_q \nf)$ is to first consider the reduction 
\eqn{
\SU(2 d_q \nf) \to  \SU(d_q) \times \SU(2 \nf)\,.
}
This is just a subcase of the general reduction considered above \eqref{eq:gen-symmetry_breaking} with ${M=d_q}$ and ${N  = 2 \nf}$. The reduction is then simply 
\eqn{
(1^{d_q \nf}) \to  \bigoplus_{\nu} \lb \lyt^q_\nu , \lyt^q_{\tilde \nu} \rb\,,
}
where $\lyt^q_{\nu}, \lyt^q_{\tilde \nu}$ are pairs of conjugate Young tableaux with $d_q \nf$ boxes, and $\lyt^q_\nu$ has at most $d_q$ rows. To obtain irreps in $\QEDtcHGN$, this should then be reduced as 
\eqn{
\SU(d_q) \times \SU(2 \nf) \to \SU(2) \times \SU(2 \nf)\,.
}
Of course, we only need to know how the treat the reduction of the first subalgebra $\SU(d_q) \to \SU(2)$. At this point, $\SU(2)$ singlets may be selected and represent the monopole operators, i.e. flux operators with vanishing fermion number and that transform as Lorentz scalars
\eqn{
\SU(2 d_q \nf) &\to  \SU(2) \times \SU(2 \nf)\,,\\
(1^{d_q \nf}) &\to  \bigoplus_{\nu \in V_0} \Omega_{\nu} \lb \bm 1 , \lyt^q_{\tilde \nu} \rb\,, \quad V_0 =  \{\nu \;|\; (\lyt^q_\nu)_{\SU(d_q)} \to \Omega_\nu \bm 1_{\SU(2)} \oplus \dots \}\,,
}
where $\Omega_\nu$ is a degeneracy to be determined. Then, we can proceed to the reduction of the flavor symmetry irreps with  $\nu \in V_0$ just as we did in the main text for monopoles with $q=1/2$ 
\eqn{
 \SU(2 \nf) &\to \SU(2) \times \SU(\nf)\,,\\
  \lyt^q_{\tilde \nu}  &\to \qquad \;\;  \dots \,,
}
where the ellipsis indicates the various diagrams involved in the reduction. To our knowledge, there is no way to systematically obtain the reduction for general $q$ like we did for general $\nf$ and $q=1/2$. We will simply examine the case $\nf =2$. We consider the reduction given by 
\eqn{
\SU(4 d_q) &\to  \SU(d_q) \times \SU(4)  \to \SU(d_q) \times \SU(2) \times \SU(2)\,.
}

For the second smallest magnetic charge $q=1$, the reduction is given by
\eqn{
\SU(8) &\to \SU(2) \times \SU(4)\,,\\
\yngc \yng(1,1,1,1)_{\bm{70}} &\to \yngc \lb \yng(2,2)_{\bm{1}}, \yng(2,2)_{\bm{20}'} \rb 
\oplus \yngc \lb \yng(3,1)_{\bm{3}} , \yng(2,1,1)_{\bm{15}} \rb 
\oplus  \yngc \lb \yng(4)_{\bm{5}} , \yng(1,1,1,1)_{\bm{1}} \rb \,.
}
Keeping only the Lorentz singlet, we the decompose the multiplet $\bm{20}'$ of $\SU(4)$ in irreps of $\SU(2) \times \SU(2)$
\eqn{
\SU(4) 		&\to \SU(2) \times \SU(2) \,, \\
\bm{20}' 	&\to (\bm 1, \bm 1) \oplus (\bm 1, \bm 5)  \oplus (\bm 3, \bm 3) \oplus (\bm 5, \bm 1) \,.
}
The first two irreps shows there is accidental degeneracy. One could continue for larger  values of $q$, e.g.  $q=3/2$
\eqn{
\SU(12) &\to \SU(3) \times \SU(4)\,, \\
\yngc \yng(1,1,1,1,1,1) &\to  \yngc \lb \yng(2,2,2), \yng(3,3) \rb \oplus  \yngc \lb \yng(3,2,1) , \yng(3,2,1) \rb \oplus \yngc  \lb \yng(3,3) , \yng(2,2,2) \rb \oplus \yngc \lb \yng(4,2) , \yng(2,2,1,1) \rb \,,
}
and so on. In these other cases, the first subalgebra must also be reduced to $\SU(2)$ in order to select Lorentz scalars.

\bibliographystyle{apsrev}

\begin{thebibliography}{67}
\expandafter\ifx\csname natexlab\endcsname\relax\def\natexlab#1{#1}\fi
\expandafter\ifx\csname bibnamefont\endcsname\relax
  \def\bibnamefont#1{#1}\fi
\expandafter\ifx\csname bibfnamefont\endcsname\relax
  \def\bibfnamefont#1{#1}\fi
\expandafter\ifx\csname citenamefont\endcsname\relax
  \def\citenamefont#1{#1}\fi
\expandafter\ifx\csname url\endcsname\relax
  \def\url#1{\texttt{#1}}\fi
\expandafter\ifx\csname urlprefix\endcsname\relax\def\urlprefix{URL }\fi
\providecommand{\bibinfo}[2]{#2}
\providecommand{\eprint}[2][]{\url{#2}}

\bibitem[{\citenamefont{Anderson}(1973)}]{anderson_resonating_1973}
\bibinfo{author}{\bibfnamefont{P.}~\bibnamefont{Anderson}},
  \bibinfo{journal}{Materials Research Bulletin} \textbf{\bibinfo{volume}{8}},
  \bibinfo{pages}{153 } (\bibinfo{year}{1973}), ISSN \bibinfo{issn}{0025-5408},
  \urlprefix\url{http://www.sciencedirect.com/science/article/pii/0025540873901670}.

\bibitem[{\citenamefont{Pisarski}(1984)}]{pikarski_chiral_1984}
\bibinfo{author}{\bibfnamefont{R.~D.} \bibnamefont{Pisarski}},
  \bibinfo{journal}{Phys. Rev. D} \textbf{\bibinfo{volume}{29}},
  \bibinfo{pages}{2423} (\bibinfo{year}{1984}),
  \urlprefix\url{https://link.aps.org/doi/10.1103/PhysRevD.29.2423}.

\bibitem[{\citenamefont{Vafa and Witten}(1984)}]{vafa_1984_eigenvalue}
\bibinfo{author}{\bibfnamefont{C.}~\bibnamefont{Vafa}} \bibnamefont{and}
  \bibinfo{author}{\bibfnamefont{E.}~\bibnamefont{Witten}},
  \bibinfo{journal}{Communications in Mathematical Physics}
  \textbf{\bibinfo{volume}{95}}, \bibinfo{pages}{257} (\bibinfo{year}{1984}),
  ISSN \bibinfo{issn}{1432-0916},
  \urlprefix\url{https://doi.org/10.1007/BF01212397}.

\bibitem[{\citenamefont{Appelquist et~al.}(1988)\citenamefont{Appelquist, Nash,
  and Wijewardhana}}]{appelquist_critical_1988}
\bibinfo{author}{\bibfnamefont{T.}~\bibnamefont{Appelquist}},
  \bibinfo{author}{\bibfnamefont{D.}~\bibnamefont{Nash}}, \bibnamefont{and}
  \bibinfo{author}{\bibfnamefont{L.~C.~R.} \bibnamefont{Wijewardhana}},
  \bibinfo{journal}{Phys. Rev. Lett.} \textbf{\bibinfo{volume}{60}},
  \bibinfo{pages}{2575} (\bibinfo{year}{1988}),
  \urlprefix\url{https://link.aps.org/doi/10.1103/PhysRevLett.60.2575}.

\bibitem[{\citenamefont{Appelquist and
  Wijewardhana}(2004)}]{appelquist_phase_2004}
\bibinfo{author}{\bibfnamefont{T.}~\bibnamefont{Appelquist}} \bibnamefont{and}
  \bibinfo{author}{\bibfnamefont{L.~C.~R.} \bibnamefont{Wijewardhana}},
  \bibinfo{type}{Tech. Rep.} \bibinfo{number}{hep-ph/0403250. UCTP-106-04},
  \bibinfo{institution}{Cincinnati Univ. Dept. Phys.},
  \bibinfo{address}{Cincinnati, OH} (\bibinfo{year}{2004}),
  \urlprefix\url{http://cds.cern.ch/record/726218}.

\bibitem[{\citenamefont{Braun et~al.}(2014)\citenamefont{Braun, Gies, Janssen,
  and Roscher}}]{braun_phase_2014}
\bibinfo{author}{\bibfnamefont{J.}~\bibnamefont{Braun}},
  \bibinfo{author}{\bibfnamefont{H.}~\bibnamefont{Gies}},
  \bibinfo{author}{\bibfnamefont{L.}~\bibnamefont{Janssen}}, \bibnamefont{and}
  \bibinfo{author}{\bibfnamefont{D.}~\bibnamefont{Roscher}},
  \bibinfo{journal}{Phys. Rev. D} \textbf{\bibinfo{volume}{90}},
  \bibinfo{pages}{036002} (\bibinfo{year}{2014}),
  \urlprefix\url{https://link.aps.org/doi/10.1103/PhysRevD.90.036002}.

\bibitem[{\citenamefont{Giombi et~al.}(2016{\natexlab{a}})\citenamefont{Giombi,
  Klebanov, and Tarnopolsky}}]{giombi_conformal_2016}
\bibinfo{author}{\bibfnamefont{S.}~\bibnamefont{Giombi}},
  \bibinfo{author}{\bibfnamefont{I.~R.} \bibnamefont{Klebanov}},
  \bibnamefont{and}
  \bibinfo{author}{\bibfnamefont{G.}~\bibnamefont{Tarnopolsky}},
  \bibinfo{journal}{Journal of Physics A: Mathematical and Theoretical}
  \textbf{\bibinfo{volume}{49}}, \bibinfo{pages}{135403}
  (\bibinfo{year}{2016}{\natexlab{a}}),
  \urlprefix\url{https://doi.org/10.1088%2F1751-8113%2F49%2F13%2F135403}.

\bibitem[{\citenamefont{Karthik and Narayanan}(2016)}]{karthik_evidence_2016}
\bibinfo{author}{\bibfnamefont{N.}~\bibnamefont{Karthik}} \bibnamefont{and}
  \bibinfo{author}{\bibfnamefont{R.}~\bibnamefont{Narayanan}},
  \bibinfo{journal}{Phys. Rev. D} \textbf{\bibinfo{volume}{93}},
  \bibinfo{pages}{045020} (\bibinfo{year}{2016}),
  \urlprefix\url{https://link.aps.org/doi/10.1103/PhysRevD.93.045020}.

\bibitem[{\citenamefont{Di~Pietro et~al.}(2016)\citenamefont{Di~Pietro,
  Komargodski, Shamir, and Stamou}}]{dipietro_quantum_2016}
\bibinfo{author}{\bibfnamefont{L.}~\bibnamefont{Di~Pietro}},
  \bibinfo{author}{\bibfnamefont{Z.}~\bibnamefont{Komargodski}},
  \bibinfo{author}{\bibfnamefont{I.}~\bibnamefont{Shamir}}, \bibnamefont{and}
  \bibinfo{author}{\bibfnamefont{E.}~\bibnamefont{Stamou}},
  \bibinfo{journal}{Phys. Rev. Lett.} \textbf{\bibinfo{volume}{116}},
  \bibinfo{pages}{131601} (\bibinfo{year}{2016}),
  \urlprefix\url{https://link.aps.org/doi/10.1103/PhysRevLett.116.131601}.

\bibitem[{\citenamefont{Chester and
  Pufu}(2016{\natexlab{a}})}]{chester_anomalous_2016}
\bibinfo{author}{\bibfnamefont{S.~M.} \bibnamefont{Chester}} \bibnamefont{and}
  \bibinfo{author}{\bibfnamefont{S.~S.} \bibnamefont{Pufu}},
  \bibinfo{journal}{Journal of High Energy Physics}
  \textbf{\bibinfo{volume}{2016}}, \bibinfo{pages}{69}
  (\bibinfo{year}{2016}{\natexlab{a}}), ISSN \bibinfo{issn}{1029-8479},
  \urlprefix\url{https://doi.org/10.1007/JHEP08(2016)069}.

\bibitem[{\citenamefont{Giombi et~al.}(2016{\natexlab{b}})\citenamefont{Giombi,
  Tarnopolsky, and Klebanov}}]{giombi_CJ_2016}
\bibinfo{author}{\bibfnamefont{S.}~\bibnamefont{Giombi}},
  \bibinfo{author}{\bibfnamefont{G.}~\bibnamefont{Tarnopolsky}},
  \bibnamefont{and} \bibinfo{author}{\bibfnamefont{I.~R.}
  \bibnamefont{Klebanov}}, \bibinfo{journal}{Journal of High Energy Physics}
  \textbf{\bibinfo{volume}{2016}}, \bibinfo{pages}{156}
  (\bibinfo{year}{2016}{\natexlab{b}}), ISSN \bibinfo{issn}{1029-8479},
  \urlprefix\url{https://doi.org/10.1007/JHEP08(2016)156}.

\bibitem[{\citenamefont{Kotikov et~al.}(2016)\citenamefont{Kotikov, Shilin, and
  Teber}}]{kotikov_critical_2016}
\bibinfo{author}{\bibfnamefont{A.~V.} \bibnamefont{Kotikov}},
  \bibinfo{author}{\bibfnamefont{V.~I.} \bibnamefont{Shilin}},
  \bibnamefont{and} \bibinfo{author}{\bibfnamefont{S.}~\bibnamefont{Teber}},
  \bibinfo{journal}{Phys. Rev. D} \textbf{\bibinfo{volume}{94}},
  \bibinfo{pages}{056009} (\bibinfo{year}{2016}),
  \urlprefix\url{https://link.aps.org/doi/10.1103/PhysRevD.94.056009}.

\bibitem[{\citenamefont{Kotikov and Teber}(2016)}]{kotikov_critical2_2016}
\bibinfo{author}{\bibfnamefont{A.~V.} \bibnamefont{Kotikov}} \bibnamefont{and}
  \bibinfo{author}{\bibfnamefont{S.}~\bibnamefont{Teber}},
  \bibinfo{journal}{Phys. Rev. D} \textbf{\bibinfo{volume}{94}},
  \bibinfo{pages}{114011} (\bibinfo{year}{2016}),
  \urlprefix\url{https://link.aps.org/doi/10.1103/PhysRevD.94.114011}.

\bibitem[{\citenamefont{Borokhov
  et~al.}(2002{\natexlab{a}})\citenamefont{Borokhov, Kapustin, and
  Wu}}]{borokhov_topological_2003}
\bibinfo{author}{\bibfnamefont{V.}~\bibnamefont{Borokhov}},
  \bibinfo{author}{\bibfnamefont{A.}~\bibnamefont{Kapustin}}, \bibnamefont{and}
  \bibinfo{author}{\bibfnamefont{X.}~\bibnamefont{Wu}},
  \bibinfo{journal}{Journal of High Energy Physics}
  \textbf{\bibinfo{volume}{2002}}, \bibinfo{pages}{049}
  (\bibinfo{year}{2002}{\natexlab{a}}),
  \urlprefix\url{https://doi.org/10.1088%2F1126-6708%2F2002%2F11%2F049}.

\bibitem[{\citenamefont{Polyakov}(1977)}]{polyakov_quark_1977}
\bibinfo{author}{\bibfnamefont{A.~M.} \bibnamefont{Polyakov}},
  \bibinfo{journal}{Nuclear Physics B} \textbf{\bibinfo{volume}{120}},
  \bibinfo{pages}{429} (\bibinfo{year}{1977}), ISSN \bibinfo{issn}{0550-3213},
  \urlprefix\url{http://www.sciencedirect.com/science/article/pii/0550321377900864}.

\bibitem[{\citenamefont{Pufu}(2014)}]{pufu_anomalous_2014}
\bibinfo{author}{\bibfnamefont{S.~S.} \bibnamefont{Pufu}},
  \bibinfo{journal}{Physical Review D} \textbf{\bibinfo{volume}{89}}
  (\bibinfo{year}{2014}), ISSN \bibinfo{issn}{1550-7998, 1550-2368},
  \urlprefix\url{https://link.aps.org/doi/10.1103/PhysRevD.89.065016}.

\bibitem[{\citenamefont{Karthik}(2018)}]{karthik_monopole_2018}
\bibinfo{author}{\bibfnamefont{N.}~\bibnamefont{Karthik}},
  \bibinfo{journal}{Phys. Rev. D} \textbf{\bibinfo{volume}{98}},
  \bibinfo{pages}{074513} (\bibinfo{year}{2018}),
  \urlprefix\url{https://link.aps.org/doi/10.1103/PhysRevD.98.074513}.

\bibitem[{\citenamefont{Karthik and Narayanan}(2019)}]{karthik_numerical_2019}
\bibinfo{author}{\bibfnamefont{N.}~\bibnamefont{Karthik}} \bibnamefont{and}
  \bibinfo{author}{\bibfnamefont{R.}~\bibnamefont{Narayanan}},
  \bibinfo{journal}{Phys. Rev. D} \textbf{\bibinfo{volume}{100}},
  \bibinfo{pages}{054514} (\bibinfo{year}{2019}),
  \urlprefix\url{https://link.aps.org/doi/10.1103/PhysRevD.100.054514}.

\bibitem[{\citenamefont{Chester and
  Pufu}(2016{\natexlab{b}})}]{chester_towards_2016}
\bibinfo{author}{\bibfnamefont{S.~M.} \bibnamefont{Chester}} \bibnamefont{and}
  \bibinfo{author}{\bibfnamefont{S.~S.} \bibnamefont{Pufu}},
  \bibinfo{journal}{Journal of High Energy Physics}
  \textbf{\bibinfo{volume}{2016}}, \bibinfo{pages}{19}
  (\bibinfo{year}{2016}{\natexlab{b}}), ISSN \bibinfo{issn}{1029-8479},
  \urlprefix\url{https://doi.org/10.1007/JHEP08(2016)019}.

\bibitem[{\citenamefont{Chester et~al.}(2018)\citenamefont{Chester, Iliesiu,
  Mezei, and Pufu}}]{chester_monopole_2018}
\bibinfo{author}{\bibfnamefont{S.~M.} \bibnamefont{Chester}},
  \bibinfo{author}{\bibfnamefont{L.~V.} \bibnamefont{Iliesiu}},
  \bibinfo{author}{\bibfnamefont{M.}~\bibnamefont{Mezei}}, \bibnamefont{and}
  \bibinfo{author}{\bibfnamefont{S.~S.} \bibnamefont{Pufu}},
  \bibinfo{journal}{Journal of High Energy Physics}
  \textbf{\bibinfo{volume}{2018}} (\bibinfo{year}{2018}), ISSN
  \bibinfo{issn}{1029-8479},
  \urlprefix\url{http://link.springer.com/10.1007/JHEP05(2018)157}.

\bibitem[{\citenamefont{Alicea}(2008)}]{alicea_monopole_2008}
\bibinfo{author}{\bibfnamefont{J.}~\bibnamefont{Alicea}},
  \bibinfo{journal}{Physical Review B} \textbf{\bibinfo{volume}{78}}
  (\bibinfo{year}{2008}), ISSN \bibinfo{issn}{1098-0121, 1550-235X},
  \urlprefix\url{https://link.aps.org/doi/10.1103/PhysRevB.78.035126}.

\bibitem[{\citenamefont{Hermele et~al.}(2008)\citenamefont{Hermele, Ran, Lee,
  and Wen}}]{hermele_properties_2008}
\bibinfo{author}{\bibfnamefont{M.}~\bibnamefont{Hermele}},
  \bibinfo{author}{\bibfnamefont{Y.}~\bibnamefont{Ran}},
  \bibinfo{author}{\bibfnamefont{P.~A.} \bibnamefont{Lee}}, \bibnamefont{and}
  \bibinfo{author}{\bibfnamefont{X.-G.} \bibnamefont{Wen}},
  \bibinfo{journal}{Physical Review B} \textbf{\bibinfo{volume}{77}}
  (\bibinfo{year}{2008}), ISSN \bibinfo{issn}{1098-0121, 1550-235X},
  \bibinfo{note}{arXiv: 0803.1150},
  \urlprefix\url{http://arxiv.org/abs/0803.1150}.

\bibitem[{\citenamefont{Song et~al.}(2019)\citenamefont{Song, Wang, Vishwanath,
  and He}}]{song_unifying_2019}
\bibinfo{author}{\bibfnamefont{X.-Y.} \bibnamefont{Song}},
  \bibinfo{author}{\bibfnamefont{C.}~\bibnamefont{Wang}},
  \bibinfo{author}{\bibfnamefont{A.}~\bibnamefont{Vishwanath}},
  \bibnamefont{and} \bibinfo{author}{\bibfnamefont{Y.-C.} \bibnamefont{He}},
  \bibinfo{journal}{Nature Communications} \textbf{\bibinfo{volume}{10}},
  \bibinfo{pages}{4254} (\bibinfo{year}{2019}), ISSN \bibinfo{issn}{2041-1723},
  \urlprefix\url{https://doi.org/10.1038/s41467-019-11727-3}.

\bibitem[{\citenamefont{Song et~al.}(2018)\citenamefont{Song, He, Vishwanath,
  and Wang}}]{song_spinon_2018}
\bibinfo{author}{\bibfnamefont{X.-Y.} \bibnamefont{Song}},
  \bibinfo{author}{\bibfnamefont{Y.-C.} \bibnamefont{He}},
  \bibinfo{author}{\bibfnamefont{A.}~\bibnamefont{Vishwanath}},
  \bibnamefont{and} \bibinfo{author}{\bibfnamefont{C.}~\bibnamefont{Wang}},
  \bibinfo{journal}{arXiv:1811.11182 [cond-mat, physics:hep-lat,
  physics:hep-th]}  (\bibinfo{year}{2018}), \bibinfo{note}{arXiv: 1811.11182},
  \urlprefix\url{http://arxiv.org/abs/1811.11182}.

\bibitem[{\citenamefont{Borokhov
  et~al.}(2002{\natexlab{b}})\citenamefont{Borokhov, Kapustin, and
  Wu}}]{borokhov_monopole_2002}
\bibinfo{author}{\bibfnamefont{V.}~\bibnamefont{Borokhov}},
  \bibinfo{author}{\bibfnamefont{A.}~\bibnamefont{Kapustin}}, \bibnamefont{and}
  \bibinfo{author}{\bibfnamefont{X.}~\bibnamefont{Wu}},
  \bibinfo{journal}{Journal of High Energy Physics}
  \textbf{\bibinfo{volume}{2002}}, \bibinfo{pages}{044}
  (\bibinfo{year}{2002}{\natexlab{b}}), ISSN \bibinfo{issn}{1029-8479},
  \urlprefix\url{http://stacks.iop.org/1126-6708/2002/i=12/a=044?key=crossref.62e8903454b6a90cca5db93bdc6fbd62}.

\bibitem[{\citenamefont{Borokhov}(2004)}]{borokhov_monopole_2004}
\bibinfo{author}{\bibfnamefont{V.}~\bibnamefont{Borokhov}},
  \bibinfo{journal}{Journal of High Energy Physics}
  \textbf{\bibinfo{volume}{2004}}, \bibinfo{pages}{008} (\bibinfo{year}{2004}),
  \urlprefix\url{https://doi.org/10.1088%2F1126-6708%2F2004%2F03%2F008}.

\bibitem[{\citenamefont{Dyer et~al.}(2013)\citenamefont{Dyer, Mezei, and
  Pufu}}]{dyer_monopole_2013}
\bibinfo{author}{\bibfnamefont{E.}~\bibnamefont{Dyer}},
  \bibinfo{author}{\bibfnamefont{M.}~\bibnamefont{Mezei}}, \bibnamefont{and}
  \bibinfo{author}{\bibfnamefont{S.~S.} \bibnamefont{Pufu}},
  \emph{\bibinfo{title}{Monopole taxonomy in three-dimensional conformal field
  theories}} (\bibinfo{year}{2013}), \eprint{arXiv:1309.1160}.

\bibitem[{\citenamefont{Ra{\dj}i{\v{c}}evi{\'{c}}}(2016)}]{radicevic_disorder_2016}
\bibinfo{author}{\bibfnamefont{{\DJ}.}~\bibnamefont{Ra{\dj}i{\v{c}}evi{\'{c}}}},
  \bibinfo{journal}{Journal of High Energy Physics}
  \textbf{\bibinfo{volume}{2016}}, \bibinfo{pages}{131} (\bibinfo{year}{2016}),
  ISSN \bibinfo{issn}{1029-8479},
  \urlprefix\url{https://doi.org/10.1007/JHEP03(2016)131}.

\bibitem[{\citenamefont{Assel}(2019)}]{assel_note_2019}
\bibinfo{author}{\bibfnamefont{B.}~\bibnamefont{Assel}},
  \bibinfo{journal}{Journal of High Energy Physics}
  \textbf{\bibinfo{volume}{2019}}, \bibinfo{pages}{74} (\bibinfo{year}{2019}),
  ISSN \bibinfo{issn}{1029-8479},
  \urlprefix\url{https://doi.org/10.1007/JHEP03(2019)074}.

\bibitem[{\citenamefont{Senthil et~al.}(2004)\citenamefont{Senthil, Balents,
  Sachdev, Vishwanath, and Fisher}}]{senthil_quantum_2004}
\bibinfo{author}{\bibfnamefont{T.}~\bibnamefont{Senthil}},
  \bibinfo{author}{\bibfnamefont{L.}~\bibnamefont{Balents}},
  \bibinfo{author}{\bibfnamefont{S.}~\bibnamefont{Sachdev}},
  \bibinfo{author}{\bibfnamefont{A.}~\bibnamefont{Vishwanath}},
  \bibnamefont{and} \bibinfo{author}{\bibfnamefont{M.~P.~A.}
  \bibnamefont{Fisher}}, \bibinfo{journal}{Phys. Rev. B}
  \textbf{\bibinfo{volume}{70}}, \bibinfo{pages}{144407}
  (\bibinfo{year}{2004}),
  \urlprefix\url{https://link.aps.org/doi/10.1103/PhysRevB.70.144407}.

\bibitem[{\citenamefont{Senthil et~al.}(2005)\citenamefont{Senthil, Balents,
  Sachdev, Vishwanath, and P.~A.~Fisher}}]{senthil_deconfined_2005}
\bibinfo{author}{\bibfnamefont{T.}~\bibnamefont{Senthil}},
  \bibinfo{author}{\bibfnamefont{L.}~\bibnamefont{Balents}},
  \bibinfo{author}{\bibfnamefont{S.}~\bibnamefont{Sachdev}},
  \bibinfo{author}{\bibfnamefont{A.}~\bibnamefont{Vishwanath}},
  \bibnamefont{and}
  \bibinfo{author}{\bibfnamefont{M.}~\bibnamefont{P.~A.~Fisher}},
  \bibinfo{journal}{Journal of the Physical Society of Japan}
  \textbf{\bibinfo{volume}{74}}, \bibinfo{pages}{1} (\bibinfo{year}{2005}),
  \eprint{https://doi.org/10.1143/JPSJS.74S.1},
  \urlprefix\url{https://doi.org/10.1143/JPSJS.74S.1}.

\bibitem[{\citenamefont{Metlitski and
  Thorngren}(2018)}]{Metlitski_intrinsic_2018}
\bibinfo{author}{\bibfnamefont{M.~A.} \bibnamefont{Metlitski}}
  \bibnamefont{and}
  \bibinfo{author}{\bibfnamefont{R.}~\bibnamefont{Thorngren}},
  \bibinfo{journal}{Phys. Rev. B} \textbf{\bibinfo{volume}{98}},
  \bibinfo{pages}{085140} (\bibinfo{year}{2018}),
  \urlprefix\url{https://link.aps.org/doi/10.1103/PhysRevB.98.085140}.

\bibitem[{\citenamefont{Lee et~al.}(2019)\citenamefont{Lee, You, Sachdev, and
  Vishwanath}}]{lee_signatures_2019}
\bibinfo{author}{\bibfnamefont{J.~Y.} \bibnamefont{Lee}},
  \bibinfo{author}{\bibfnamefont{Y.-Z.} \bibnamefont{You}},
  \bibinfo{author}{\bibfnamefont{S.}~\bibnamefont{Sachdev}}, \bibnamefont{and}
  \bibinfo{author}{\bibfnamefont{A.}~\bibnamefont{Vishwanath}},
  \bibinfo{journal}{Phys. Rev. X} \textbf{\bibinfo{volume}{9}},
  \bibinfo{pages}{041037} (\bibinfo{year}{2019}),
  \urlprefix\url{https://link.aps.org/doi/10.1103/PhysRevX.9.041037}.

\bibitem[{\citenamefont{Murthy and Sachdev}(1990)}]{murthy_action_1990}
\bibinfo{author}{\bibfnamefont{G.}~\bibnamefont{Murthy}} \bibnamefont{and}
  \bibinfo{author}{\bibfnamefont{S.}~\bibnamefont{Sachdev}},
  \bibinfo{journal}{Nuclear Physics B} \textbf{\bibinfo{volume}{344}},
  \bibinfo{pages}{557} (\bibinfo{year}{1990}).

\bibitem[{\citenamefont{Metlitski et~al.}(2008)\citenamefont{Metlitski,
  Hermele, Senthil, and Fisher}}]{metlitski_monopoles_2008}
\bibinfo{author}{\bibfnamefont{M.~A.} \bibnamefont{Metlitski}},
  \bibinfo{author}{\bibfnamefont{M.}~\bibnamefont{Hermele}},
  \bibinfo{author}{\bibfnamefont{T.}~\bibnamefont{Senthil}}, \bibnamefont{and}
  \bibinfo{author}{\bibfnamefont{M.~P.~A.} \bibnamefont{Fisher}},
  \bibinfo{journal}{Physical Review B} \textbf{\bibinfo{volume}{78}}
  (\bibinfo{year}{2008}), ISSN \bibinfo{issn}{1098-0121, 1550-235X},
  \urlprefix\url{https://link.aps.org/doi/10.1103/PhysRevB.78.214418}.

\bibitem[{\citenamefont{Dyer et~al.}(2015)\citenamefont{Dyer, Mezei, Pufu, and
  Sachdev}}]{dyer_scaling_2015}
\bibinfo{author}{\bibfnamefont{E.}~\bibnamefont{Dyer}},
  \bibinfo{author}{\bibfnamefont{M.}~\bibnamefont{Mezei}},
  \bibinfo{author}{\bibfnamefont{S.~S.} \bibnamefont{Pufu}}, \bibnamefont{and}
  \bibinfo{author}{\bibfnamefont{S.}~\bibnamefont{Sachdev}},
  \bibinfo{journal}{Journal of High Energy Physics}
  \textbf{\bibinfo{volume}{2015}} (\bibinfo{year}{2015}), ISSN
  \bibinfo{issn}{1029-8479},
  \urlprefix\url{http://link.springer.com/10.1007/JHEP06(2015)037}.

\bibitem[{\citenamefont{Dyer et~al.}(2016)\citenamefont{Dyer, Mezei, Pufu, and
  Sachdev}}]{dyer_erratum_2016}
\bibinfo{author}{\bibfnamefont{E.}~\bibnamefont{Dyer}},
  \bibinfo{author}{\bibfnamefont{M.}~\bibnamefont{Mezei}},
  \bibinfo{author}{\bibfnamefont{S.~S.} \bibnamefont{Pufu}}, \bibnamefont{and}
  \bibinfo{author}{\bibfnamefont{S.}~\bibnamefont{Sachdev}},
  \bibinfo{journal}{Journal of High Energy Physics}
  \textbf{\bibinfo{volume}{2016}}, \bibinfo{pages}{111} (\bibinfo{year}{2016}),
  ISSN \bibinfo{issn}{1029-8479},
  \urlprefix\url{https://doi.org/10.1007/JHEP03(2016)111}.

\bibitem[{\citenamefont{de~la Fuente}(2018)}]{delaFuente_large_2018}
\bibinfo{author}{\bibfnamefont{A.}~\bibnamefont{de~la Fuente}},
  \bibinfo{journal}{Journal of High Energy Physics}
  \textbf{\bibinfo{volume}{2018}}, \bibinfo{pages}{41} (\bibinfo{year}{2018}),
  ISSN \bibinfo{issn}{1029-8479},
  \urlprefix\url{https://doi.org/10.1007/JHEP08(2018)041}.

\bibitem[{\citenamefont{Block et~al.}(2013)\citenamefont{Block, Melko, and
  Kaul}}]{ribhu_fate_2013}
\bibinfo{author}{\bibfnamefont{M.~S.} \bibnamefont{Block}},
  \bibinfo{author}{\bibfnamefont{R.~G.} \bibnamefont{Melko}}, \bibnamefont{and}
  \bibinfo{author}{\bibfnamefont{R.~K.} \bibnamefont{Kaul}},
  \bibinfo{journal}{Phys. Rev. Lett.} \textbf{\bibinfo{volume}{111}},
  \bibinfo{pages}{137202} (\bibinfo{year}{2013}),
  \urlprefix\url{https://link.aps.org/doi/10.1103/PhysRevLett.111.137202}.

\bibitem[{\citenamefont{Sreejith and Powell}(2015)}]{sreejith_scaling_2015}
\bibinfo{author}{\bibfnamefont{G.~J.} \bibnamefont{Sreejith}} \bibnamefont{and}
  \bibinfo{author}{\bibfnamefont{S.}~\bibnamefont{Powell}},
  \bibinfo{journal}{Phys. Rev. B} \textbf{\bibinfo{volume}{92}},
  \bibinfo{pages}{184413} (\bibinfo{year}{2015}),
  \urlprefix\url{https://link.aps.org/doi/10.1103/PhysRevB.92.184413}.

\bibitem[{\citenamefont{Pujari et~al.}(2015)\citenamefont{Pujari, Alet, and
  Damle}}]{pujari_transitions_2015}
\bibinfo{author}{\bibfnamefont{S.}~\bibnamefont{Pujari}},
  \bibinfo{author}{\bibfnamefont{F.}~\bibnamefont{Alet}}, \bibnamefont{and}
  \bibinfo{author}{\bibfnamefont{K.}~\bibnamefont{Damle}},
  \bibinfo{journal}{Phys. Rev. B} \textbf{\bibinfo{volume}{91}},
  \bibinfo{pages}{104411} (\bibinfo{year}{2015}),
  \urlprefix\url{https://link.aps.org/doi/10.1103/PhysRevB.91.104411}.

\bibitem[{\citenamefont{Wang et~al.}(2017)\citenamefont{Wang, Nahum, Metlitski,
  Xu, and Senthil}}]{wang_deconfined_2018}
\bibinfo{author}{\bibfnamefont{C.}~\bibnamefont{Wang}},
  \bibinfo{author}{\bibfnamefont{A.}~\bibnamefont{Nahum}},
  \bibinfo{author}{\bibfnamefont{M.~A.} \bibnamefont{Metlitski}},
  \bibinfo{author}{\bibfnamefont{C.}~\bibnamefont{Xu}}, \bibnamefont{and}
  \bibinfo{author}{\bibfnamefont{T.}~\bibnamefont{Senthil}},
  \bibinfo{journal}{Phys. Rev. X} \textbf{\bibinfo{volume}{7}},
  \bibinfo{pages}{031051} (\bibinfo{year}{2017}),
  \urlprefix\url{https://link.aps.org/doi/10.1103/PhysRevX.7.031051}.

\bibitem[{\citenamefont{Janssen and He}(2017)}]{janssen_critical_2017}
\bibinfo{author}{\bibfnamefont{L.}~\bibnamefont{Janssen}} \bibnamefont{and}
  \bibinfo{author}{\bibfnamefont{Y.-C.} \bibnamefont{He}},
  \bibinfo{journal}{Phys. Rev. B} \textbf{\bibinfo{volume}{96}},
  \bibinfo{pages}{205113} (\bibinfo{year}{2017}),
  \urlprefix\url{https://link.aps.org/doi/10.1103/PhysRevB.96.205113}.

\bibitem[{\citenamefont{Ihrig et~al.}(2018)\citenamefont{Ihrig, Janssen,
  Mihaila, and Scherer}}]{bernhard_deconfined_2018}
\bibinfo{author}{\bibfnamefont{B.}~\bibnamefont{Ihrig}},
  \bibinfo{author}{\bibfnamefont{L.}~\bibnamefont{Janssen}},
  \bibinfo{author}{\bibfnamefont{L.~N.} \bibnamefont{Mihaila}},
  \bibnamefont{and} \bibinfo{author}{\bibfnamefont{M.~M.}
  \bibnamefont{Scherer}}, \bibinfo{journal}{Phys. Rev. B}
  \textbf{\bibinfo{volume}{98}}, \bibinfo{pages}{115163}
  (\bibinfo{year}{2018}),
  \urlprefix\url{https://link.aps.org/doi/10.1103/PhysRevB.98.115163}.

\bibitem[{\citenamefont{Gracey}(2018)}]{gracey_fermion_2018}
\bibinfo{author}{\bibfnamefont{J.~A.} \bibnamefont{Gracey}},
  \bibinfo{journal}{Phys. Rev. D} \textbf{\bibinfo{volume}{98}},
  \bibinfo{pages}{085012} (\bibinfo{year}{2018}),
  \urlprefix\url{https://link.aps.org/doi/10.1103/PhysRevD.98.085012}.

\bibitem[{\citenamefont{Zerf et~al.}(2018)\citenamefont{Zerf, Marquard, Boyack,
  and Maciejko}}]{zerf_critical_2018}
\bibinfo{author}{\bibfnamefont{N.}~\bibnamefont{Zerf}},
  \bibinfo{author}{\bibfnamefont{P.}~\bibnamefont{Marquard}},
  \bibinfo{author}{\bibfnamefont{R.}~\bibnamefont{Boyack}}, \bibnamefont{and}
  \bibinfo{author}{\bibfnamefont{J.}~\bibnamefont{Maciejko}},
  \bibinfo{journal}{Phys. Rev. B} \textbf{\bibinfo{volume}{98}},
  \bibinfo{pages}{165125} (\bibinfo{year}{2018}),
  \urlprefix\url{https://link.aps.org/doi/10.1103/PhysRevB.98.165125}.

\bibitem[{\citenamefont{Boyack et~al.}(2019)\citenamefont{Boyack, Rayyan, and
  Maciejko}}]{boyack_deconfined_2019}
\bibinfo{author}{\bibfnamefont{R.}~\bibnamefont{Boyack}},
  \bibinfo{author}{\bibfnamefont{A.}~\bibnamefont{Rayyan}}, \bibnamefont{and}
  \bibinfo{author}{\bibfnamefont{J.}~\bibnamefont{Maciejko}},
  \bibinfo{journal}{Phys. Rev. B} \textbf{\bibinfo{volume}{99}},
  \bibinfo{pages}{195135} (\bibinfo{year}{2019}),
  \urlprefix\url{https://link.aps.org/doi/10.1103/PhysRevB.99.195135}.

\bibitem[{\citenamefont{Benvenuti and Khachatryan}(2019)}]{benvenuti_easy_2019}
\bibinfo{author}{\bibfnamefont{S.}~\bibnamefont{Benvenuti}} \bibnamefont{and}
  \bibinfo{author}{\bibfnamefont{H.}~\bibnamefont{Khachatryan}},
  \bibinfo{journal}{Journal of High Energy Physics}
  \textbf{\bibinfo{volume}{2019}}, \bibinfo{pages}{214} (\bibinfo{year}{2019}),
  ISSN \bibinfo{issn}{1029-8479},
  \urlprefix\url{https://doi.org/10.1007/JHEP05(2019)214}.

\bibitem[{\citenamefont{Dupuis et~al.}(2019)\citenamefont{Dupuis, Paranjape,
  and Witczak-Krempa}}]{dupuis_transition_2019}
\bibinfo{author}{\bibfnamefont{E.}~\bibnamefont{Dupuis}},
  \bibinfo{author}{\bibfnamefont{M.~B.} \bibnamefont{Paranjape}},
  \bibnamefont{and}
  \bibinfo{author}{\bibfnamefont{W.}~\bibnamefont{Witczak-Krempa}},
  \bibinfo{journal}{Phys. Rev. B} \textbf{\bibinfo{volume}{100}},
  \bibinfo{pages}{094443} (\bibinfo{year}{2019}),
  \urlprefix\url{https://link.aps.org/doi/10.1103/PhysRevB.100.094443}.

\bibitem[{\citenamefont{Dupuis et~al.}(To appear, 2021)\citenamefont{Dupuis,
  Paranjape, and Witczak-Krempa}}]{proc}
\bibinfo{author}{\bibfnamefont{E.}~\bibnamefont{Dupuis}},
  \bibinfo{author}{\bibfnamefont{M.~B.} \bibnamefont{Paranjape}},
  \bibnamefont{and}
  \bibinfo{author}{\bibfnamefont{W.}~\bibnamefont{Witczak-Krempa}}, in
  \emph{\bibinfo{booktitle}{Quantum Theory and Symmetries : Proceedings of the
  11th International Symposium, Montreal, Canada}}, edited by
  \bibinfo{editor}{\bibfnamefont{M.}~\bibnamefont{Paranjape}},
  \bibinfo{editor}{\bibfnamefont{R.}~\bibnamefont{MacKenzie}},
  \bibinfo{editor}{\bibfnamefont{Z.}~\bibnamefont{Thomova}},
  \bibinfo{editor}{\bibfnamefont{P.}~\bibnamefont{Winternitz}},
  \bibnamefont{and}
  \bibinfo{editor}{\bibfnamefont{W.}~\bibnamefont{Witczak-Krempa}}
  (\bibinfo{year}{To appear, 2021}).

\bibitem[{\citenamefont{Hermele et~al.}(2005)\citenamefont{Hermele, Senthil,
  and Fisher}}]{hermele_algebraic_2005}
\bibinfo{author}{\bibfnamefont{M.}~\bibnamefont{Hermele}},
  \bibinfo{author}{\bibfnamefont{T.}~\bibnamefont{Senthil}}, \bibnamefont{and}
  \bibinfo{author}{\bibfnamefont{M.~P.~A.} \bibnamefont{Fisher}},
  \bibinfo{journal}{Phys. Rev. B} \textbf{\bibinfo{volume}{72}},
  \bibinfo{pages}{104404} (\bibinfo{year}{2005}),
  \urlprefix\url{https://link.aps.org/doi/10.1103/PhysRevB.72.104404}.

\bibitem[{\citenamefont{He et~al.}(2015)\citenamefont{He, Fuji, and
  Bhattacharjee}}]{bhattacharjee_kagome_2015}
\bibinfo{author}{\bibfnamefont{Y.-C.} \bibnamefont{He}},
  \bibinfo{author}{\bibfnamefont{Y.}~\bibnamefont{Fuji}}, \bibnamefont{and}
  \bibinfo{author}{\bibfnamefont{S.}~\bibnamefont{Bhattacharjee}},
  \emph{\bibinfo{title}{Kagome spin liquid: a deconfined critical phase driven
  by $u(1)$ gauge fluctuation}} (\bibinfo{year}{2015}),
  \eprint{arXiv:1512.05381}.

\bibitem[{\citenamefont{Ghaemi and Senthil}(2006)}]{ghaemi_neel_2006}
\bibinfo{author}{\bibfnamefont{P.}~\bibnamefont{Ghaemi}} \bibnamefont{and}
  \bibinfo{author}{\bibfnamefont{T.}~\bibnamefont{Senthil}},
  \bibinfo{journal}{Phys. Rev. B} \textbf{\bibinfo{volume}{73}},
  \bibinfo{pages}{054415} (\bibinfo{year}{2006}),
  \urlprefix\url{https://link.aps.org/doi/10.1103/PhysRevB.73.054415}.

\bibitem[{\citenamefont{Lu et~al.}(2017)\citenamefont{Lu, Cho, and
  Vishwanath}}]{lu_unification_2017}
\bibinfo{author}{\bibfnamefont{Y.-M.} \bibnamefont{Lu}},
  \bibinfo{author}{\bibfnamefont{G.~Y.} \bibnamefont{Cho}}, \bibnamefont{and}
  \bibinfo{author}{\bibfnamefont{A.}~\bibnamefont{Vishwanath}},
  \bibinfo{journal}{Phys. Rev. B} \textbf{\bibinfo{volume}{96}},
  \bibinfo{pages}{205150} (\bibinfo{year}{2017}),
  \urlprefix\url{https://link.aps.org/doi/10.1103/PhysRevB.96.205150}.

\bibitem[{\citenamefont{Hastings}(2000)}]{hastings_dirac_2000}
\bibinfo{author}{\bibfnamefont{M.~B.} \bibnamefont{Hastings}},
  \bibinfo{journal}{Phys. Rev. B} \textbf{\bibinfo{volume}{63}},
  \bibinfo{pages}{014413} (\bibinfo{year}{2000}),
  \urlprefix\url{https://link.aps.org/doi/10.1103/PhysRevB.63.014413}.

\bibitem[{\citenamefont{Baskaran and Anderson}(1988)}]{baskaran_gauge_1988}
\bibinfo{author}{\bibfnamefont{G.}~\bibnamefont{Baskaran}} \bibnamefont{and}
  \bibinfo{author}{\bibfnamefont{P.~W.} \bibnamefont{Anderson}},
  \bibinfo{journal}{Phys. Rev. B} \textbf{\bibinfo{volume}{37}},
  \bibinfo{pages}{580} (\bibinfo{year}{1988}),
  \urlprefix\url{https://link.aps.org/doi/10.1103/PhysRevB.37.580}.

\bibitem[{\citenamefont{Polyakov}(1975)}]{polyakov_compact_1975}
\bibinfo{author}{\bibfnamefont{A.~M.} \bibnamefont{Polyakov}},
  \bibinfo{journal}{Physics Letters B} \textbf{\bibinfo{volume}{59}},
  \bibinfo{pages}{82} (\bibinfo{year}{1975}), ISSN \bibinfo{issn}{0370-2693},
  \urlprefix\url{http://www.sciencedirect.com/science/article/pii/0370269375901628}.

\bibitem[{\citenamefont{Rychkov}(2017)}]{rychkov_epfl_2017}
\bibinfo{author}{\bibfnamefont{S.}~\bibnamefont{Rychkov}},
  \emph{\bibinfo{title}{{EPFL} {Lectures} on {Conformal} {Field} {Theory} in
  {D} ≥ 3 {Dimensions}}}, {SpringerBriefs} in {Physics}
  (\bibinfo{publisher}{Springer International Publishing},
  \bibinfo{address}{Cham}, \bibinfo{year}{2017}), ISBN
  \bibinfo{isbn}{978-3-319-43625-8},
  \urlprefix\url{//www.springer.com/gp/book/9783319436258}.

\bibitem[{\citenamefont{Atiyah and Singer}(1963)}]{atiyah_index_1963}
\bibinfo{author}{\bibfnamefont{M.~F.} \bibnamefont{Atiyah}} \bibnamefont{and}
  \bibinfo{author}{\bibfnamefont{I.~M.} \bibnamefont{Singer}},
  \bibinfo{journal}{Bulletin of the American Mathematical Society}
  \textbf{\bibinfo{volume}{69}}, \bibinfo{pages}{422} (\bibinfo{year}{1963}),
  ISSN \bibinfo{issn}{0002-9904, 1936-881X},
  \urlprefix\url{https://www.ams.org/home/page/}.

\bibitem[{\citenamefont{Wu and Yang}(1976)}]{wu_dirac_1976}
\bibinfo{author}{\bibfnamefont{T.~T.} \bibnamefont{Wu}} \bibnamefont{and}
  \bibinfo{author}{\bibfnamefont{C.~N.} \bibnamefont{Yang}},
  \bibinfo{journal}{Nuclear Physics B} \textbf{\bibinfo{volume}{107}},
  \bibinfo{pages}{365} (\bibinfo{year}{1976}).

\bibitem[{\citenamefont{Hellerman et~al.}(2015)\citenamefont{Hellerman,
  Orlando, Reffert, and Watanabe}}]{Hellerman_on_2015}
\bibinfo{author}{\bibfnamefont{S.}~\bibnamefont{Hellerman}},
  \bibinfo{author}{\bibfnamefont{D.}~\bibnamefont{Orlando}},
  \bibinfo{author}{\bibfnamefont{S.}~\bibnamefont{Reffert}}, \bibnamefont{and}
  \bibinfo{author}{\bibfnamefont{M.}~\bibnamefont{Watanabe}},
  \bibinfo{journal}{Journal of High Energy Physics}
  \textbf{\bibinfo{volume}{2015}}, \bibinfo{pages}{1} (\bibinfo{year}{2015}),
  ISSN \bibinfo{issn}{1029-8479},
  \urlprefix\url{https://doi.org/10.1007/JHEP12(2015)071}.

\bibitem[{\citenamefont{Itzykson and Nauenberg}(1966)}]{itzykson_unitary_1966}
\bibinfo{author}{\bibfnamefont{C.}~\bibnamefont{Itzykson}} \bibnamefont{and}
  \bibinfo{author}{\bibfnamefont{M.}~\bibnamefont{Nauenberg}},
  \bibinfo{journal}{Rev. Mod. Phys.} \textbf{\bibinfo{volume}{38}},
  \bibinfo{pages}{95} (\bibinfo{year}{1966}),
  \urlprefix\url{https://link.aps.org/doi/10.1103/RevModPhys.38.95}.

\bibitem[{\citenamefont{Feger et~al.}(2019)\citenamefont{Feger, Kephart, and
  Saskowski}}]{feger_lieart_2019}
\bibinfo{author}{\bibfnamefont{R.}~\bibnamefont{Feger}},
  \bibinfo{author}{\bibfnamefont{T.~W.} \bibnamefont{Kephart}},
  \bibnamefont{and} \bibinfo{author}{\bibfnamefont{R.~J.}
  \bibnamefont{Saskowski}}, \emph{\bibinfo{title}{Lieart 2.0 -- a mathematica
  application for lie algebras and representation theory}}
  (\bibinfo{year}{2019}), \eprint{arXiv:1912.10969}.

\bibitem[{\citenamefont{Xu et~al.}(2018)\citenamefont{Xu, Qi, Zhang, Assaad,
  Xu, and Meng}}]{yang-xu_monte_2018}
\bibinfo{author}{\bibfnamefont{X.~Y.} \bibnamefont{Xu}},
  \bibinfo{author}{\bibfnamefont{Y.}~\bibnamefont{Qi}},
  \bibinfo{author}{\bibfnamefont{L.}~\bibnamefont{Zhang}},
  \bibinfo{author}{\bibfnamefont{F.~F.} \bibnamefont{Assaad}},
  \bibinfo{author}{\bibfnamefont{C.}~\bibnamefont{Xu}}, \bibnamefont{and}
  \bibinfo{author}{\bibfnamefont{Z.~Y.} \bibnamefont{Meng}}
  (\bibinfo{year}{2018}), \eprint{arXiv:1807.07574}.

\bibitem[{\citenamefont{Xu et~al.}(2019)\citenamefont{Xu, Qi, Zhang, Assaad,
  Xu, and Meng}}]{meng_monte_2019}
\bibinfo{author}{\bibfnamefont{X.~Y.} \bibnamefont{Xu}},
  \bibinfo{author}{\bibfnamefont{Y.}~\bibnamefont{Qi}},
  \bibinfo{author}{\bibfnamefont{L.}~\bibnamefont{Zhang}},
  \bibinfo{author}{\bibfnamefont{F.~F.} \bibnamefont{Assaad}},
  \bibinfo{author}{\bibfnamefont{C.}~\bibnamefont{Xu}}, \bibnamefont{and}
  \bibinfo{author}{\bibfnamefont{Z.~Y.} \bibnamefont{Meng}},
  \bibinfo{journal}{Phys. Rev. X} \textbf{\bibinfo{volume}{9}},
  \bibinfo{pages}{021022} (\bibinfo{year}{2019}),
  \urlprefix\url{https://link.aps.org/doi/10.1103/PhysRevX.9.021022}.

\bibitem[{{\relax DLMF}()}]{NIST:DLMF}
{\relax DLMF}, \emph{\bibinfo{title}{{\it NIST Digital Library of Mathematical
  Functions}}}, \bibinfo{howpublished}{http://dlmf.nist.gov/, Release 1.0.21 of
  2018-12-15}, \bibinfo{note}{f.~W.~J. Olver, A.~B. {Olde Daalhuis}, D.~W.
  Lozier, B.~I. Schneider, R.~F. Boisvert, C.~W. Clark, B.~R. Miller and B.~V.
  Saunders, eds.}, \urlprefix\url{http://dlmf.nist.gov/}.

\bibitem[{\citenamefont{Georgi}(1999)}]{georgi_lie_1999}
\bibinfo{author}{\bibfnamefont{H.~M.} \bibnamefont{Georgi}},
  \emph{\bibinfo{title}{{Lie algebras in particle physics; 2nd ed.}}},
  Frontiers in Physics (\bibinfo{publisher}{Perseus},
  \bibinfo{address}{Cambridge}, \bibinfo{year}{1999}).

\end{thebibliography}

\end{document}